\documentclass[fleqn,usenatbib,useAMS]{mnras}
\pdfoutput=1

\usepackage{amssymb}
 \usepackage{color}
\usepackage{graphicx,mathrsfs,amsmath}
\usepackage[caption=false]{subfig}    

\usepackage[T1]{fontenc}
\usepackage{graphicx}

\newcommand{\MgII}{{\ion{Mg}{ii}}}
\newcommand{\FeII}{{\ion{Fe}{ii}}}
\newcommand{\CIV}{{\ion{C}{iv}}}
\newcommand{\SiII}{{\ion{Si}{ii}}}
\newcommand\kms{$\rm{kms^{-1}}$}
\newcommand{\arcnamelong}{RCSGA 032727-132609}
\newcommand{\clustername}{RCS2 032727-132623}
\newcommand{\arcname}{RCS0327}

\usepackage{ae,aecompl}
 \usepackage{textcomp, libertine}
 
\usepackage{mathptmx}
\usepackage{txfonts}

\title[Resolved outflows @ $z=1.70$]{Spatially Resolved Galactic Wind in Lensed Galaxy RCSGA 032727-132609}
\author[R. Bordoloi et al.]
{\parbox{\textwidth}{Rongmon~Bordoloi$^{1,9,10}$\thanks{Contact e-mail: \href{mailto:bordoloi@stsci.edu}{bordoloi@mit.edu}}, 
{Jane~R.~Rigby$^2$}, 
{Jason~Tumlinson$^1$},  
{Matthew~B.~Bayliss$^{3,4}$}, 
{Keren~Sharon$^5$}, 
{Michael~G.~Gladders$^{6,7}$} \&
{Eva~Wuyts$^8$}}\vspace{0.4cm}\\
 $^1$Space Telescope Science Institute, 3700 San Martin Drive, 21218, Baltimore, MD\\
 $^2$NASA Goddard Space Flight Center, Code 665, Greenbelt MD 20771\\
 $^3$Dept. of Physics, Harvard University, 17 Oxford St., Cambridge, MA, 02138\\
 $^4$Harvard-Smithsonian Center for Astrophysics, 60 Garden Street, Cambridge, MA 02138, USA\\
 $^5$ Department of Astronomy and Astrophysics, University of Michigan, 500 Church Street, Ann Arbor, MI 48109, USA\\
 $^6$Department of Astronomy and Astrophysics, University of Chicago, 5640 South Ellis Avenue, Chicago, IL 60637, USA\\
 $^7$Kavli Institute for Cosmological Physics at the University of Chicago\\
 $^8$Max-Planck-Institut f\''{u}r extraterrestrische Physik, Postfach 1312, Giessenbachstr., D-85741 Garching, Germany\\
 $^9$MIT-Kavli Center for Astrophysics and Space Re- search, 77 Massachusetts Avenue, Cambridge, MA, 02139, USA\\
$^{10}$ Hubble Fellow }

\date{Accepted for publication February 24, 2016}
\pagerange{\pageref{firstpage}--\pageref{lastpage}} \pubyear{}
\begin{document}

\label{firstpage}

\maketitle

\begin{abstract}
We probe the spatial distribution of outflowing gas along four lines of sight separated by up to 6 kpc in a gravitationally-lensed star-forming galaxy at $z=1.70$.  Using {\MgII} and {\FeII} emission and absorption as tracers, we find that the clumps of star formation are driving galactic outflows with velocities of  -170 to -250 km/sec.  The velocities of {\MgII} emission are redshifted with respect to the systemic velocities of the galaxy, consistent with being back-scattered.  By contrast, the {\FeII} fluorescent emission lines are either slightly blueshifted or at the systemic velocity of the galaxy.  Taken together, the velocity structure of the {\MgII} and {\FeII} emission is consistent with arising through scattering in galactic winds.  Assuming a thin shell geometry for the outflowing gas, the estimated masses carried out by these outflows are large ($\gtrsim$ 30 - 50 $\rm{M_{\odot}\; yr^{-1}}$), with mass loading factors several times the star-formation rate. Almost 20\% to 50\% of the blueshifted absorption probably escapes the gravitational potential of the galaxy. In this galaxy, the outflow is ``locally sourced'', that is, the properties of the outflow in each line of sight are dominated by the properties of the nearest clump of star formation; the wind is not global to the galaxy. The mass outflow rates and the momentum flux carried out by outflows in individual star forming knots of this object are comparable to that of starburst galaxies in the local Universe. 
\end{abstract}


\section{Introduction}
In the last decade, our understanding of the formation and evolution of galaxies over cosmic time has been significantly enhanced by large scale spectroscopic galaxy surveys. These surveys have unearthed a detailed picture of the properties of the galaxies, establishing their bi-modal distribution in the color-magnitude diagram \citep{Blanton2003}, mass-metalicity relation \citep{Tremonti2004} and the galaxy stellar mass function to high precision \citep{Bell2003}. Surveys observing galaxies over a broad range of cosmic time have identified that the average star-formation rates (SFR) drop several-fold \citep{Daddi2007, Noeske2007}, the fraction of galaxies on the red sequence more than doubles \citep{Faber2007, Brammer2011}, and red galaxies grow passively \citep{Patel2013} while star-forming disks become more ordered \citep{Kassin2012}. 

Theoretical models explaining these galaxy properties as well as the observed metal enrichment of the intergalactic medium (IGM; \citealt{Cooksey2010, Adelberger2005}), consistently invoke feedback mechanisms to account for the discrepancies between theory and observations. These processes, and in particular galactic outflows are thought to drive the mass-metallicity relation, enrich the IGM \citep{Bordoloi2011a, Tumlinson2011a, Werk2014, zhu2013a, Bordoloi2014c}, and regulate star formation and black hole growth \citep{Gabor2011}.

In the local Universe, hot outflowing galactic winds are observed with X-ray emission, and cooler phases of the outflowing gas are detected via optical emission lines such as H$\alpha$ (e.g. \citealt{Lehnert1996}), and in ``down-the-barrel'' observations of  absorption lines against the galaxy's stellar continuum (see \citealt{Veilleux2005} for review). The kinematics and column density of the cool (100-1000 K) gas entrained in the  hot outflowing gas of the local dwarf starbursts and luminous infrared galaxies (LIRGs) up to z $\sim$ 0.5 are detected using the blueshifed Na I D $\lambda \lambda$ 5890, 5896 doublet \citep{Heckman2000, Martin2005, Rupke2005b}. 

At  $z\sim$ 3, using UV transitions such as {\SiII} $\lambda$ 1206 and {\CIV} $\lambda \lambda$ 1548, 1550 blueshifed outflowing gas with velocities of hundreds of $\rm{km s^{-1}}$ have been observed in the spectra of Lyman Break Galaxies (LBGs) \citep{Shapley2003}. Outflows have also been detected at $z\sim$ 2 using extended $H\alpha$ emission wings \citep{Newman2012,ForsterSchreiber2014,Genzel2014}.  At $z \sim$ 1, outflowing gas has been found by several studies \citep{Weiner2009, Rubin2011, Martin2012,Bordoloi2014b} to be largely ubiquitous. These large scale outflows are biconical in morphology \citep{Bordoloi2011a, Martin2012, Bordoloi2012a, Bordoloi2014b, Rubin2014a} and exhibit resonant {\MgII} and fine-structure {\FeII} emission lines \citep{Erb2012, Kornei2013}. 

Moreover, studies of down-the-barrel spectra of individual galaxies have also yielded evidence of redshifted inflowing absorption in {\MgII} and {\FeII} transitions \citep{Rubin2011a, Kornei2012, Martin2012}. The covering fraction of inflowing gas (not masked by outflows) around these galaxies has been found to be small ($\approx$ 6\%).  These studies have identified galaxies with large concentration of massive stars to be driving strong outflows \citep{Kornei2012,Kornei2013, Bordoloi2014b}, consistent with theoretical studies requiring massive star clusters to drive galactic outflows \citep{Murray2011}.  To understand these outflows, it is particularly important to understand the spatial scales involved and the mass outflow rate; the latter is a crucial, as-yet-poorly-constrained parameter in galaxy evolution models \citep{Dave2012}.

Previous studies have provided constraints on the spatial scales of the outflows, particularly as traced by emission from the resonant doublet {\MgII} 2796, 2803 Angstrom.  \cite{Erb2012} constructed a two-dimensional composite spectrum of 33 galaxies that show {\MgII} emission; the stacked {\MgII} is slightly more extended than the continuum at distances of 0.8 arcsec (7 kpc). {\MgII} emission has been spatially resolved in at least two individual galaxies.  \cite{Rubin2011b} found that {\MgII} emission from a $z$=0.69 galaxy extended to radii of $\ge 6.5$ kpc; the mass outflow rate estimated by \cite{Martin2013} of 35-40 $\rm{M_{\odot} yr^{-1}}$ is about half the star formation rate.  For a  z=0.93 galaxy, \cite{Martin2013} found {\MgII} emission extending out to 11~kpc; this {\MgII} emission shows  different spatial distribution and kinematics than the [O II] emission, indicating that the {\MgII} is scattered rather than coming directly from the H II regions.  They inferred a mass flux of  330-500 $\rm{M_{\odot} yr^{-1}}$, which is 4-6 times larger than the star formation rate. \cite{Martin2013} caution that this galaxy is exceptional in their sample: of the 145 galaxy spectra from \cite{Martin2012} that covered the {\MgII} doublet and {\FeII}�absorption, 22 showed strong {\MgII} emission, but for only three galaxies was the emission clearly spatially resolved.

{\FeII} fine structure emission is generally not seen in local star-forming and starburst galaxies \citep{Leitherer2011}. However, both {\FeII} and MgII emission features are seen at higher redshift star-forming galaxies. Some authors have argued that if {\MgII} emission typically occurs on large spatial scales, compared to the stellar continuum, then slit losses may explain why {\MgII} emission is rarely seen in observations of $z\sim$0 galaxies; as the slits used to observe such objects are generally placed to cover  only the central regions of the galaxies and capture little emission from the spatially extended outflows \citep{Giavalisco2011, Erb2012}. \cite{Prochaska2011} modeled this effects of slit loss, finding in their models that the emission is sufficiently centrally concentrated, and the emission-to-absorption ratio is unaffected for slit widths that cover more than 4-5 kpc of the galaxies.

The limited spatial resolution typical of observations of {\MgII} and {\FeII} emission translates to as-yet unsatisfying constraints on the location, maximum velocity, and total column density of the warm outflowing gas. Absorbing clouds may lie anywhere along the line of sight to the galaxy, and the decline in gas covering fraction with increasing blueshift makes it challenging to detect the highest velocity gas \citep{Martin2009}. Gravitationally lensed galaxies provide access to smaller spatial scales, and higher signal-to-noise ratios, than are possible for field galaxies. One approach would be to map the outflows via {\MgII} and {\FeII} emission using large-format integral field spectrographs like MUSE on the VLT.  Such an approach would cover both large spatial scales (from the large format of the IFU) as well as sub-kiloparsec scales (accessible through lensing magnification).

Another approach, employed in this paper, is to examine the kinematics of the outflow at multiple positions within a lensed galaxy, and examine how the properties of the outflow may depend on the properties of the star-forming complexes that ultimately should drive the outflow.  In this work, we perform the first spatially-resolved down-the-barrel study of galactic outflows using the lensed,  $z$ = 1.7 spatially resolved galaxy RCSGA 032727-132609 as the background source \citep{Wuyts2010}. We observe the {\MgII} and {\FeII} absorption and emission lines for four spatially resolved star-forming knots in this galaxy. These knots are physically separated by 2 to 6 kpc \citep{Wuyts2014}.  As such, this work probes smaller spatial scales,  with more detailed kinematics, than have been accessed previously with observations of field galaxies. We compare the outflow absorption and emission strength and kinematics to the properties of the star forming regions, namely star formation rates and stellar masses estimated from HST/WFC3 photometry and grism spectra.

The paper is organized as follows. In section 2, we describe the observations and the methods used to estimate the properties of these star forming knots. In section 3, we describe the method used to estimate the outflowing absorption and emission gas properties. In sections 4.1 and 4.2, we compare the estimated outflowing absorption and emission strength and kinematics with the properties of the star forming knots. In section 4.3, we study the pixel velocity distribution of the outflowing gas. In section 4.4 we estimate the spatial extent of the {\MgII} emission region. In section 4.5 we estimate the minimum mass outflow rates in each individual knot. In section 4.6 we compare our findings with the outflow properties observed in local star-burst galaxies. In section 5 we summarize our findings. 

Throughout this paper, we assume a cosmology of $\Omega_m = 0.3$, $\Omega_{\Lambda} = 0.7$,
$H_0= 70$~\kms~Mpc$^{-1}$, and $h \equiv H_0/100 =0.7$.
 Solar abundances are taken from Table~1 of \citet{Asplund09}.  
The initial mass function (IMF) is \citet{Chabrier03}.

\section{Methods}

\subsection{Observations}

The lensed galaxy \arcnamelong, hereafter \arcname, is lensed by the galaxy cluster \clustername.  It was discovered in the Second Red Sequence Cluster Survey \citep{Gilbank2011}; the discovery paper is \citet{Wuyts2010}. Images obtained with the \textit{Hubble Space Telescope} (Program 12267, PI Rigby) have enabled a high-fidelity lensing model and source--plane reconstruction (\citealt{Sharon2012}; see their Figure~4).  We adopt this lensing model. Figure~1 of \citet{Sharon2012} named the regions within \arcname; we adopt this nomenclature, but capitalize the named regions for readability.

Rest-ultraviolet spectra of \arcname\ were obtained using the MagE spectrograph \citep{magespie} on the 6.5~m Magellan II telescope at Las Campanas Observatory in Chile.  Observations, mostly at low airmass (sec(z)$<$ 1.4), were obtained in the course of seven nights from 2008 to 2013.  The atmospheric seeing, as measured by the guide camera during observations, was sub-arcsecond for all observations. The observations are part of a larger survey of the rest-ultraviolet spectra of bright lensed galaxies; a future paper will present the full spectra and complete details of the observations for the full sample (Rigby et al.\ in prep).

To summarize the MagE observations of RCS0327 for the purposes of this paper, three different slits were used, all 10\arcsec\ length, with the following widths: 
\begin{itemize}
\item 2\arcsec\  wide, with resolution R$=$2050 (145 \kms\ per resolution element);
\item 1.5\arcsec\ wide, with R$=$2730 (108 \kms); 
\item 1\arcsec\ wide, with R$=$4100 (72 \kms).
\end{itemize}
No detector binning was used. The slit position angle was updated hourly to keep it close to the parallactic angle, thus preserving relative fluxing of the spectrum; this is especially important given the wide wavelength coverage of MagE. For a few integrations, a non-parallactic slit position was chosen to keep potentially contaminating objects out of the slit.

Since the RCS0327 arc at 38\arcsec\ is much longer than the 10\arcsec\ MagE slit, we could not target the entire arc. Instead we targeted four physically distinct bright physical regions within \arcname : Knots B, E, G, and U. Figure~\ref{fig:finder_chart} shows how the MagE slit was positioned in the image plane.  The total integration time was 10.00~hr on Knot E, 7.94~hr on Knot U,  2.83~hr on Knot B, and 1.83~hr on Knot G. Table~\ref{table:Knot_properties} tabulates the effective spectral resolution for each knot spectrum weighted by the integration time. Figure~\ref{fig:finder_chart} shows that in the image plane, knots U, E, B, and G are separated from each other by more than 2\arcsec. Given the sub-arcsecond seeing conditions during the observations, and the slit position angles shown in Figure~\ref{fig:finder_chart}, we expect minimal cross-contamination of each slit by emission from adjacent regions.

\subsection{Calibration and data reduction}
Full details of the calibration and reduction of the MagE data will be given in the MagE sample paper (Rigby et al.\ in prep). To summarize, wavelength calibration was accomplished by internal thorium argon lamp spectra, obtained at each pointing of the telescope, to guard against instrument flexure. Flat-fielding was accomplished using an internal Xe lamp for the blue orders, and a quartz lamp for the redder orders. The MagE spectra for each night were reduced separately using the LCO MagE pipeline written by D.~Kelson. The spectra for each night were then fluxed using spectrophotometric standard stars observed nightly. Overlapping orders were combined with a weighted average, which was then corrected to vacuum barycentric wavelength. Telluric absorption, particularly the A-band of molecular absorption at 7590{\AA} to 7660 {\AA}  \citep{WarkMercer}, was corrected using the featureless standard star EG~131 and the IRAF task \textit{noao.onedspec.telluric}. Since the A-band begins 10~\AA\ redward of the reddest part of the {\MgII} doublet in \arcname, the telluric correction is largely cosmetic and does not affect the results presented here. For each knot, spectra from multiple nights were then combined with a weighted average.

\subsection{Knot properties}
Knots B, D, E, F, G, and U of \arcname\ have complementary rest-frame optical (observed-frame near-infrared) spectra from NIRSPEC on the Keck II telescope and OSIRIS on Keck II \citep{Rigby11,Wuyts2014}. We list the redshifts of these knots in Table~\ref{table:Knot_properties}, derived from H$\alpha$ in the NIRSPEC or OSIRIS spectra. 

Multiple measurements exist for the star formation rates of the RCS0327 knots. We choose not to use the H$\alpha$ measurements due to challenges of flux calibration across multiple telescopes, instruments with different slit widths, and variation in seeing conditions. Instead, we take as our star formation rate indicator, the extinction-corrected H$\beta$ luminosity, as measured by the WFC3 G141 grism and published in \cite{Whitaker2014}. For the extinction, we use $E(B-V)=0.4$, which was found by \cite{Whitaker2014} to be the reddening that best fit the H$\beta$/H$\gamma$ ratio in the grism spectra, and the conversion from reddening to extinction of R$=$4.05 \citep{1994ApJ...429..582C}. We correct the H$\beta$ flux for the outflow component, which contributes 0 to 50$\%$ of the total H$\alpha$ line flux, as measured by \cite{Wuyts2014} from the velocity profile of H$\alpha$.  These outflow fractions are taken from column 7 of Table 3 of \citep{Wuyts2014}. To convert from H$\beta$ luminosity to star formation rate, we use equation 2 of \cite{Kennicutt1998}, modified to use the Chabrier IMF \citep{Chabrier03}, and an H$\alpha$ to H$\beta$ flux ratio of 2.863 which is true for Case B, $T=10^4$K, $n=100$ cm$^{-3}$.  

When grism H$\beta$ measurements are available for multiple lensed images of the same knot, we quote the weighted mean and the error in the weighted mean. For cases with only one measurement, we quote the statistical  uncertainty, which mostly comes from the line flux uncertainty, and ignores the magnification uncertainty which probably dominates. These derived star formation rates are tabulated in Table \ref{table:Knot_properties}. Stellar masses for each knot were measured by spectral energy distribution (SED) fitting, as described in \cite{Wuyts2014}. SED fitting is highly uncertain for Knot U because it sits on the caustic and therefore has an extremely high magnification whose value is quite uncertain, as well as a highly distorted appearance in the image plane that complicates aperture photometry. Therefore, to derive an approximate stellar mass for Knot U, we assume that the knots E and U have the same mass-to-light ratio, and scale their H$\beta$ fluxes, resulting in an estimated stellar mass for Knot U of $\log M*=7.5\; M_{\odot}$. 

We estimate the halo mass of the galaxy from mock catalogues generated from the Bolshoi simulation \citep{Klypin2011}. Halos were found using \texttt{ROCKSTAR} halo finder \citep{Behroozi2013}. RCS0327 is a merger with stellar masses of the two merging galaxies being 5.1 $\times\; 10^{9}  M_{\odot}$ and 1.2 $\times\; 10^{9} M_{\odot}$ respectively \citep{Wuyts2014}. We search for galaxy pairs in the mocks with similar stellar masses and physical separations $<$100 kpc from each other. This gives us a distribution of halo masses for the higher mass galaxy and we adopt the mean halo mass as the halo mass of the system. Thus we assign a halo mass for the galaxy as $\rm{\log\; M_{halo}= 11.6 \;\pm 0.3 \; M_{\odot}}$, the maximum circular velocity $v_{circ} = 156 \pm 28$ km/sec.    For a halo with outer radius $r_h$ the escape velocity at radius R (for a flat rotation curve), is given as $\rm{v^{2}_{esc}(R) \;=\; 2 v^{2}_{circ} \ln (1 + r_{h}/R)}$ \citep{Galactic_Dynamics}. The escape velocities ($v_{esc}$) for this galaxy at R = 5 kpc, 20 kpc and 50 kpc are given as 411 $\pm$ 74 {\kms}, 324 $\pm$ 59 {\kms} and 262 $\pm$ 48 {\kms}, respectively.

\begin{figure*}
\centering
    \includegraphics[height=15.25cm,width=12.2cm,angle=270]{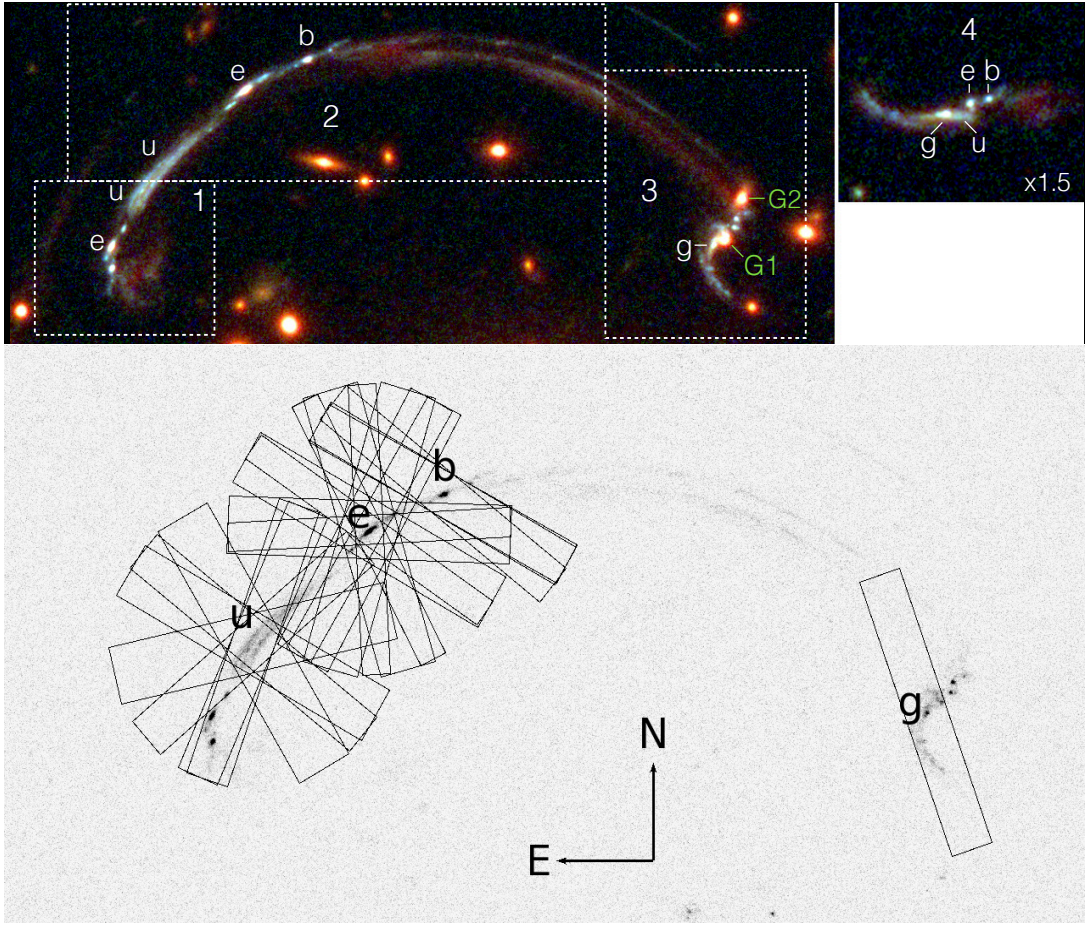}        
    \caption{ Lensed galaxy RCSGA 032727-132609.  
TOP: Color rendition composed of
HST/WFC3   F160W, F125W, F098M (red); 
F814W, F606W (green); and F390W (blue).  The dashed lines approximately
mark the multiple images, indicated by numbers.  The emission knots 
with MagE spectroscopy are labeled as ``e", ``u", ``b", and ``g".  
Two cluster galaxies ``G1" and ``G2" are in image 3.
The right panel shows a zoomed-in view of the counter-image, which is a 
relatively undistorted image of the source-plane galaxy.  
Figure adapted from \citet{Sharon2012}. BOTTOM: HST/WFC3 F390W image, with MagE observations overplotted.  One
slit is drawn for each integration, using our best estimate 
of the pointing, the actual slit position angle, and the
slit width used during that observation.  
Since the slit was regularly rotated to track the parallactic angle, 
knots U and E were observed using a wide range of slit position angles.  
The length of the MagE slit is a fixed 10 arcsec.   
North is up and East is left.}
\label{fig:finder_chart}
\end{figure*}

\begin{figure*}
 \includegraphics[height=6.cm,width=12.5cm]{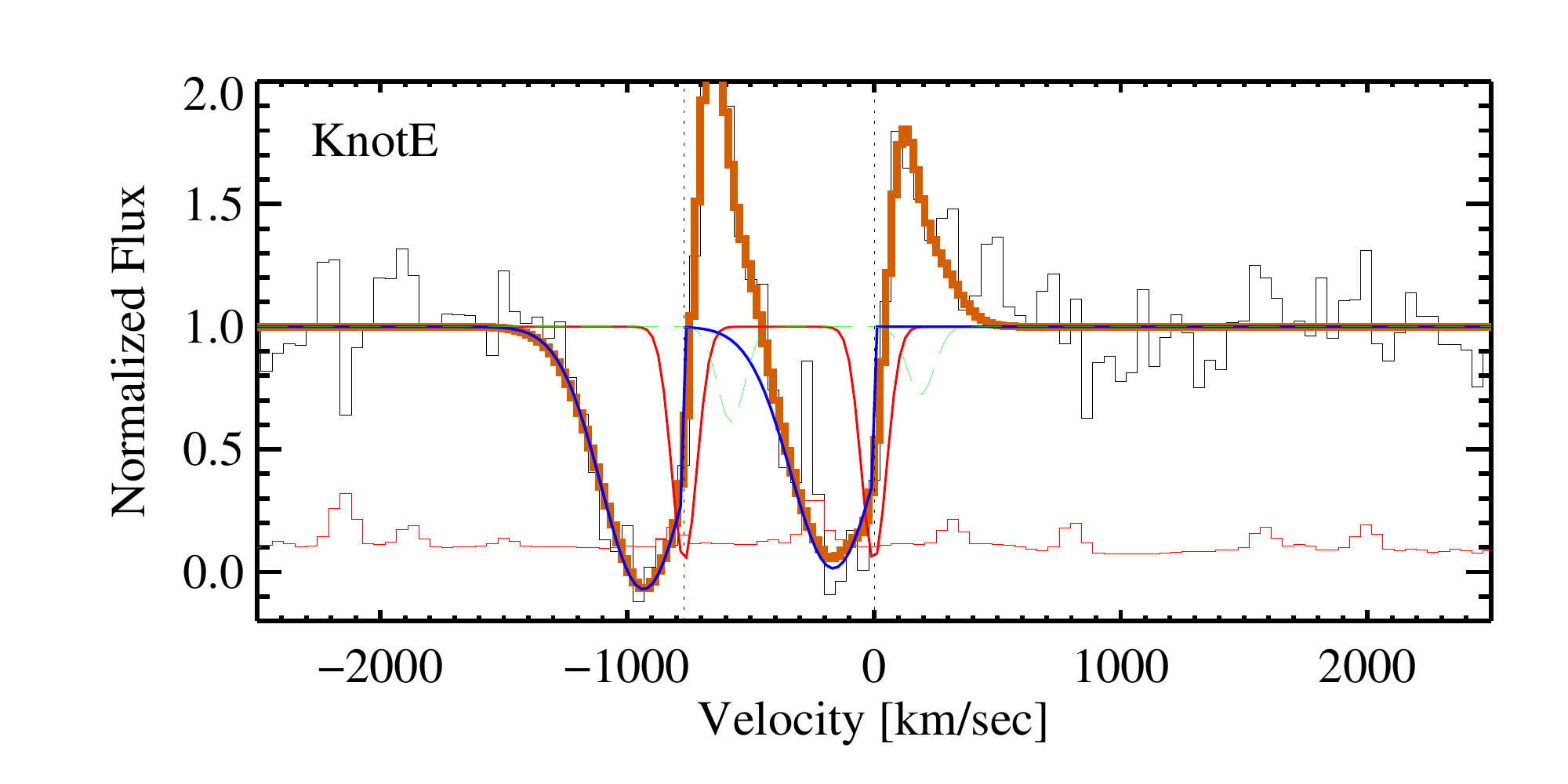}     
\includegraphics[height=9.5cm,width=10.5cm]{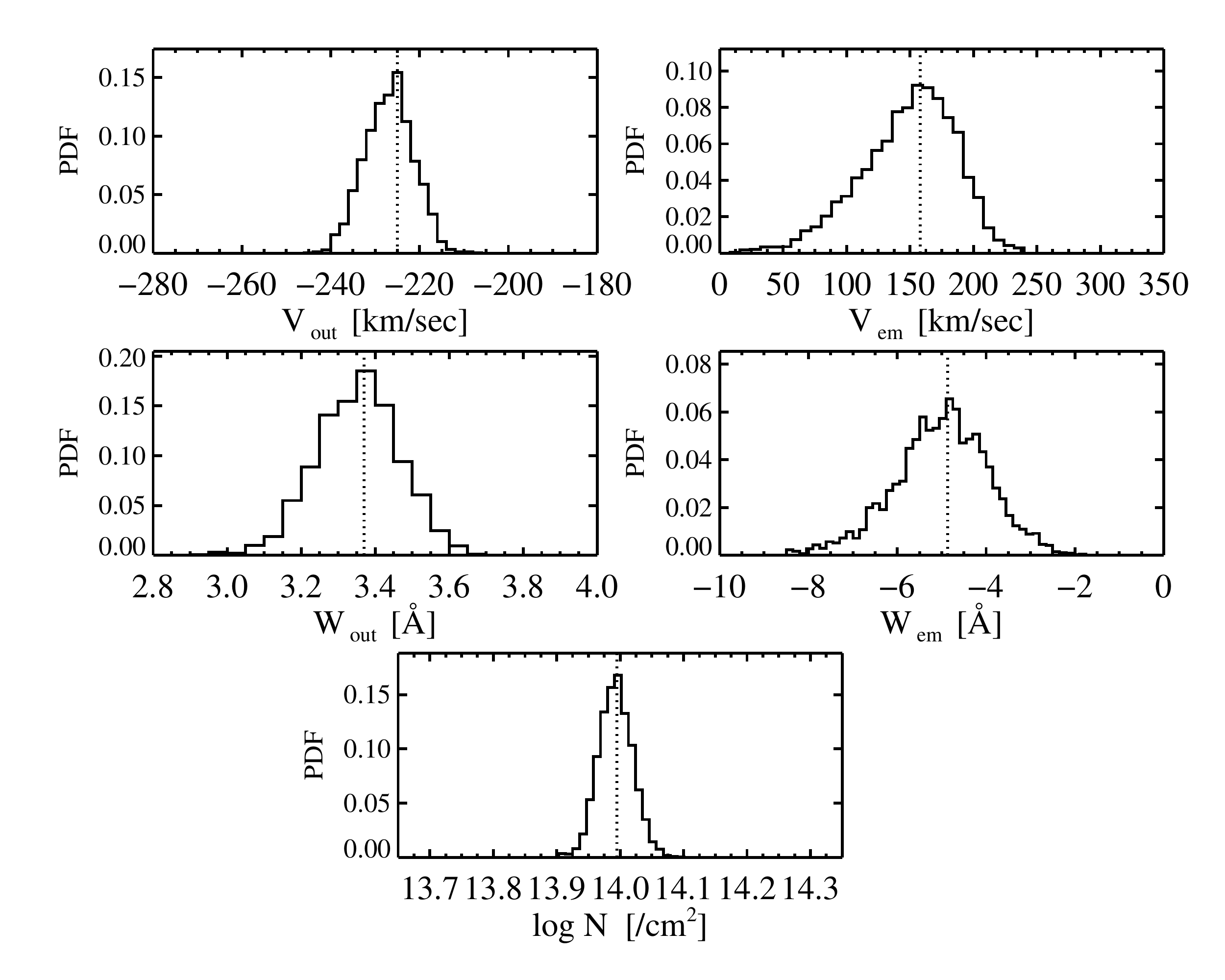}
  \caption{Top:-Observations of {\MgII} 2796, 2803 transitions with the model fits for Knot E. The black stairs represent the observer spectra and the red stairs represent the error spectra. The brown lines show the best fit model profile to the data, after convolving with the instrumental spread function. The blue lines show the outflowing components, and the red smooth line shows the systemic component. The dashed green line represents the inflowing component, which is very small. Lower Panels:- The marginalized PPDFs for the outflow velocity ($V_{out}$), emission velocity ($V_{em}$), rest frame outflow absorption equivalent width ($W_{out}$), rest frame emission equivalent width ($W_{em}$) and {\MgII} column density estimate ($\log\; N$) respectively. The vertical dashed lines show the mode of the distributions which are taken as the best model estimates for each case respectively.}
\label{fig:KnotE_MgII_estimate}
\end{figure*}

\begin{figure*}
    \includegraphics[height= 4.5cm,width=8.5cm]{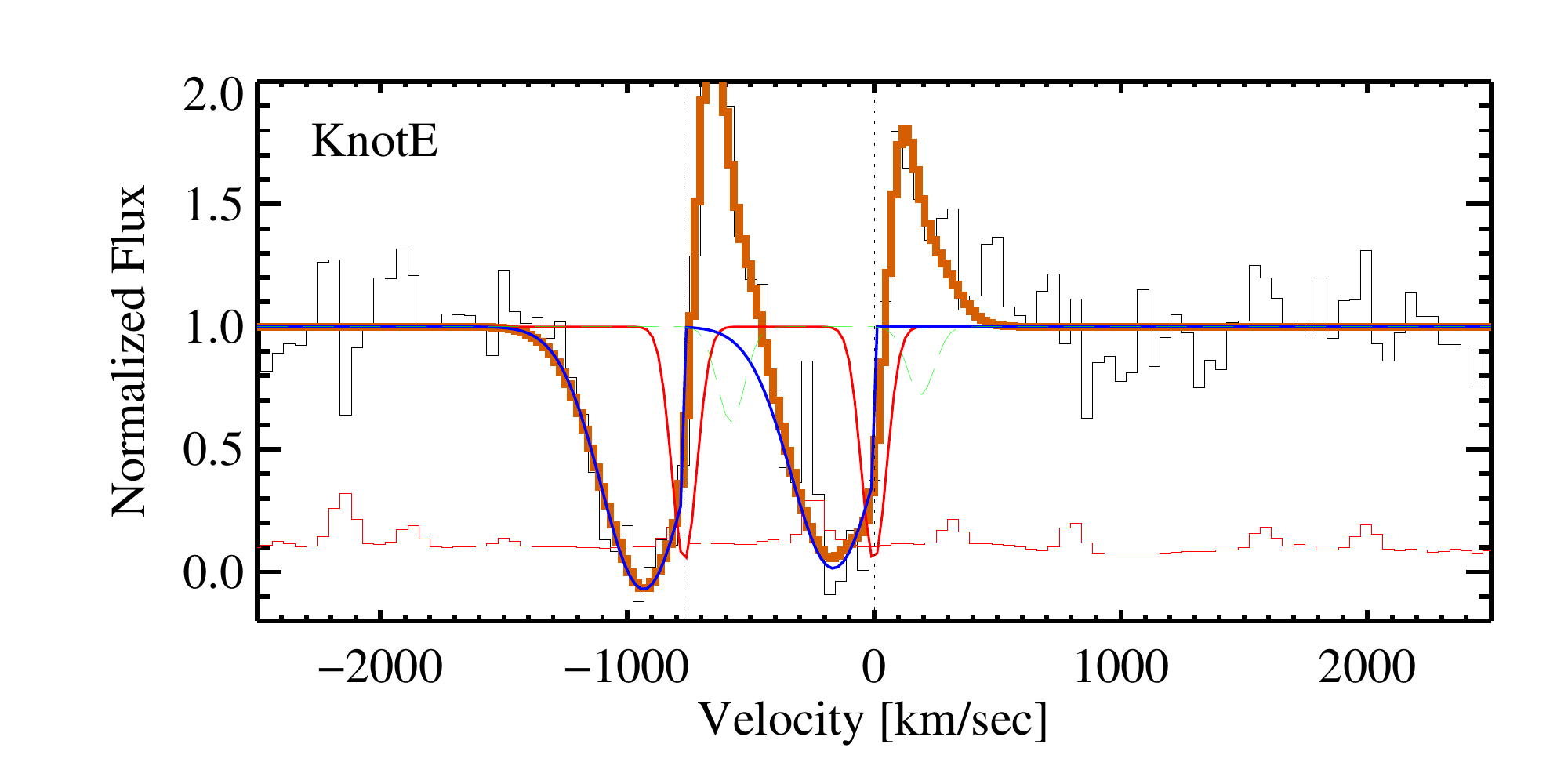}     
     \includegraphics[height=4.5cm,width=8.5cm]{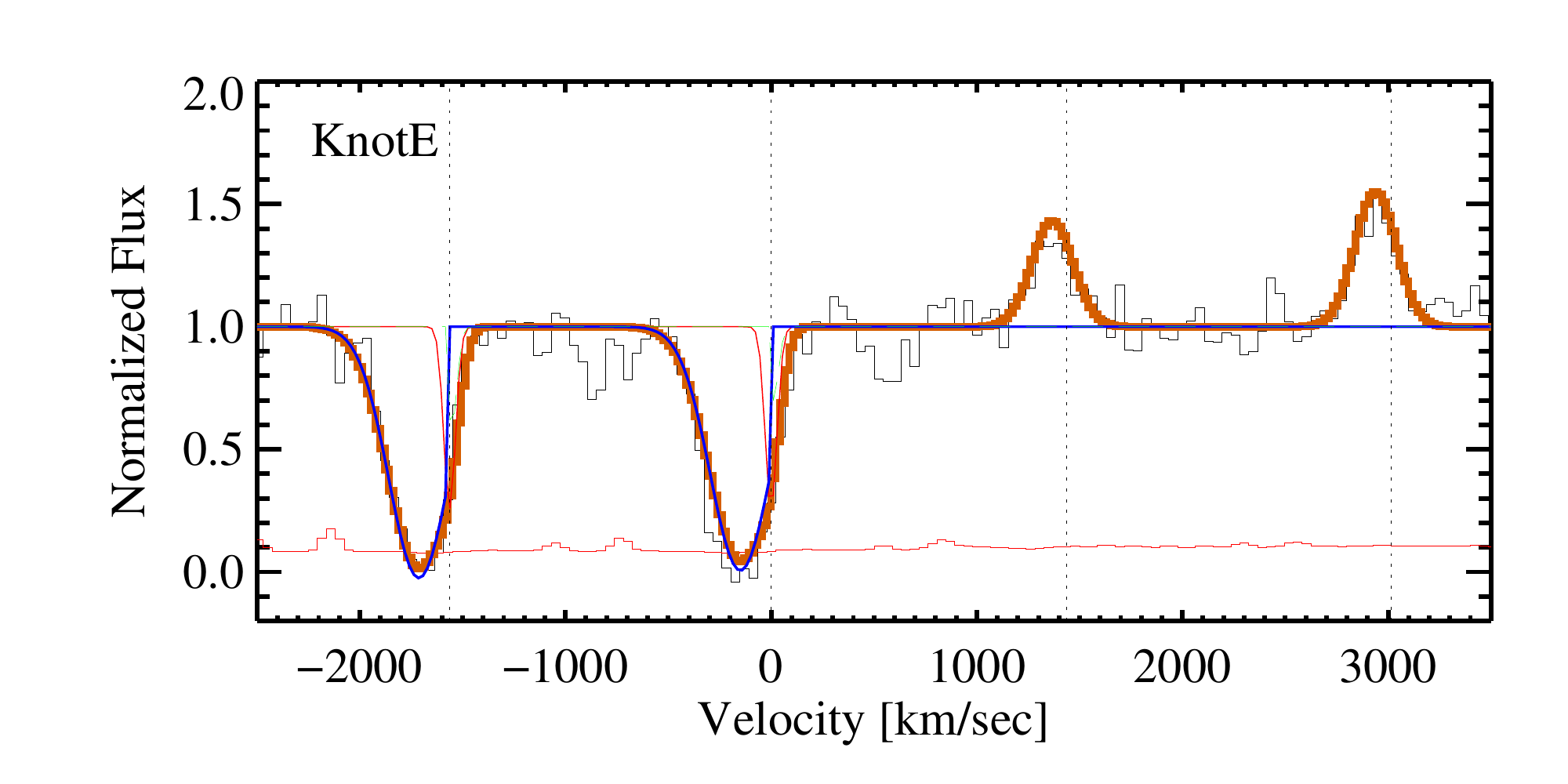}     

    \includegraphics[height=4.5cm,width=8.5cm]{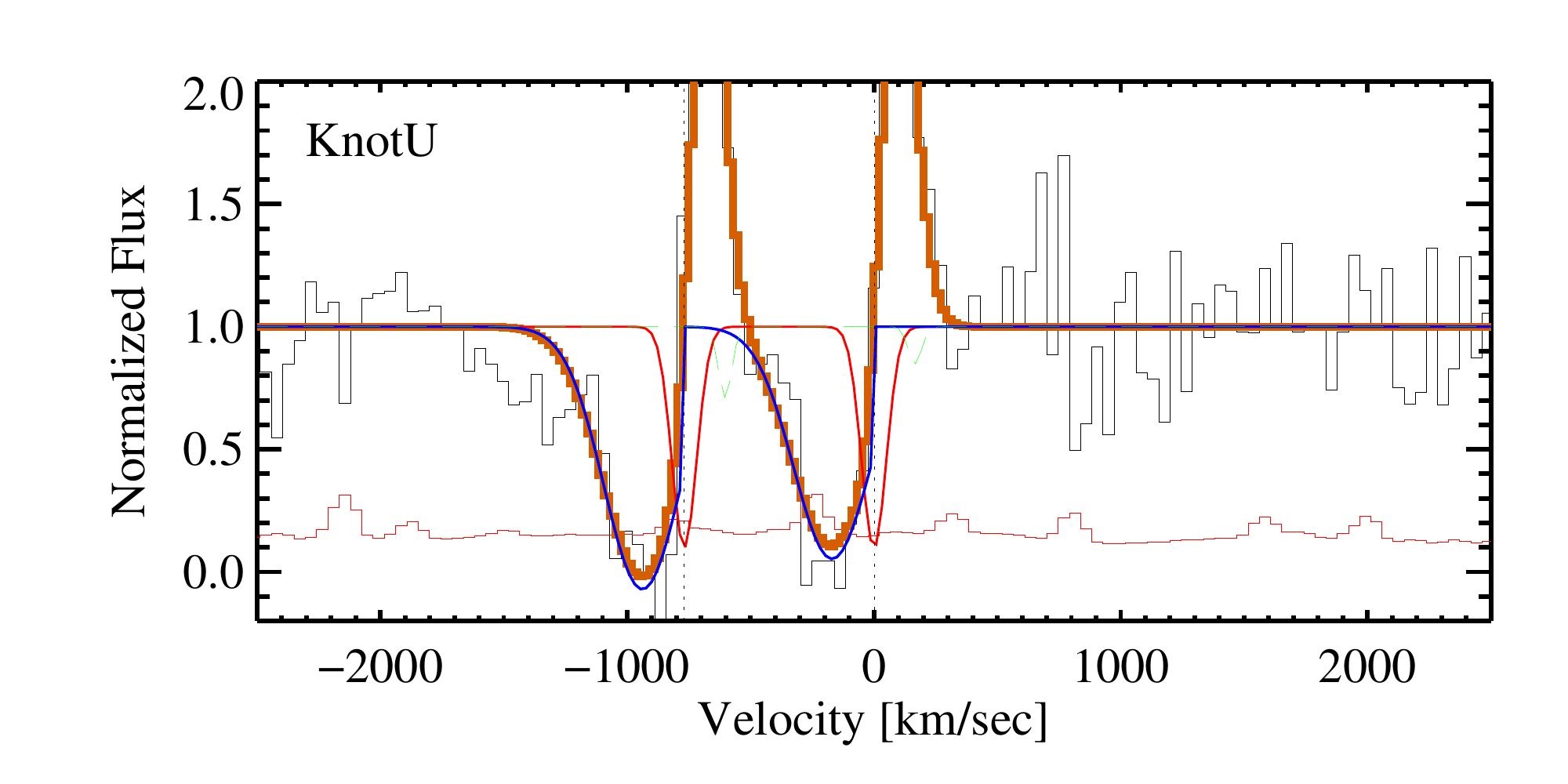}     
    \includegraphics[height=4.5cm,width=8.5cm]{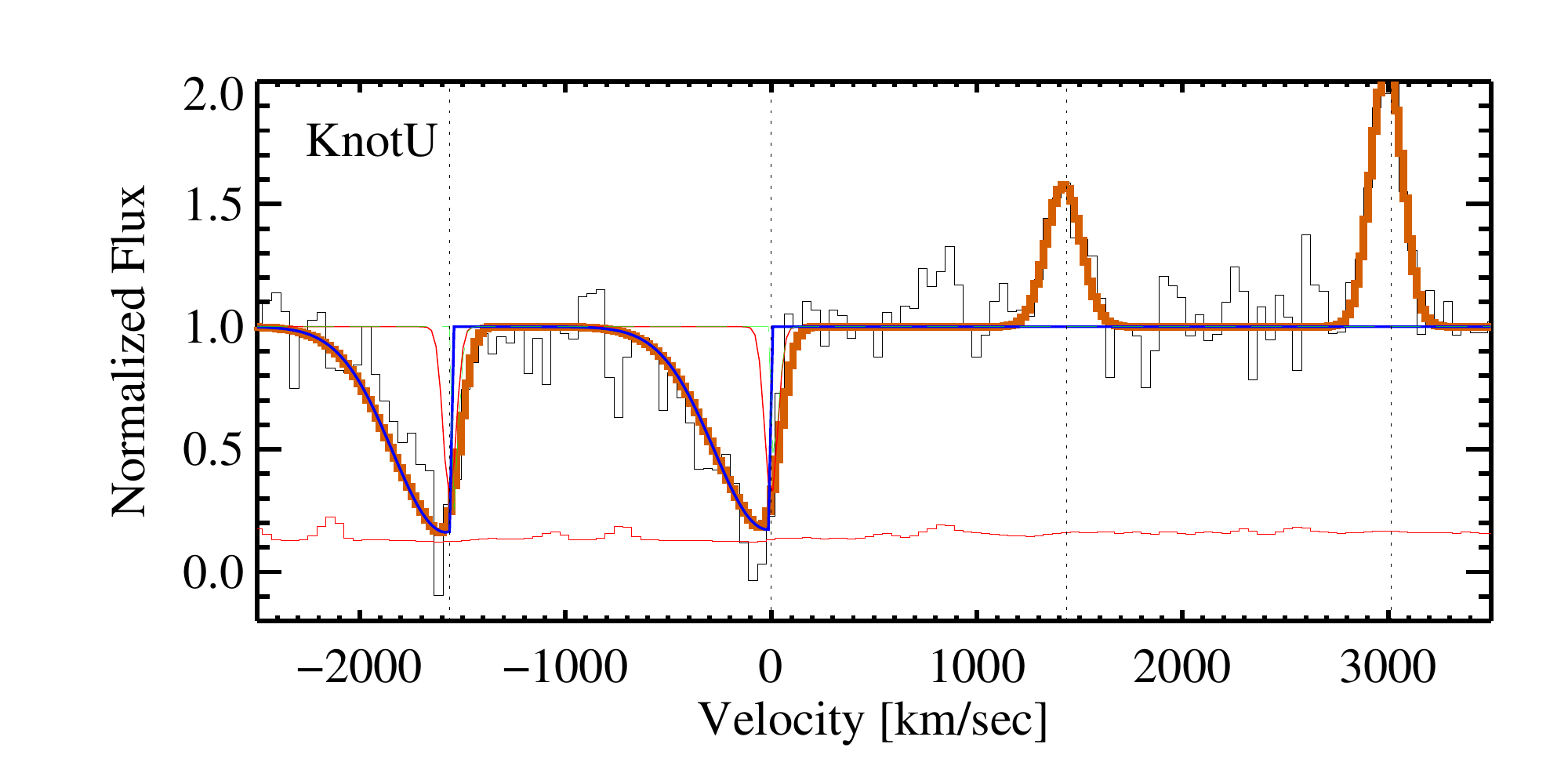}     

    \includegraphics[height=4.5cm,width=8.5cm]{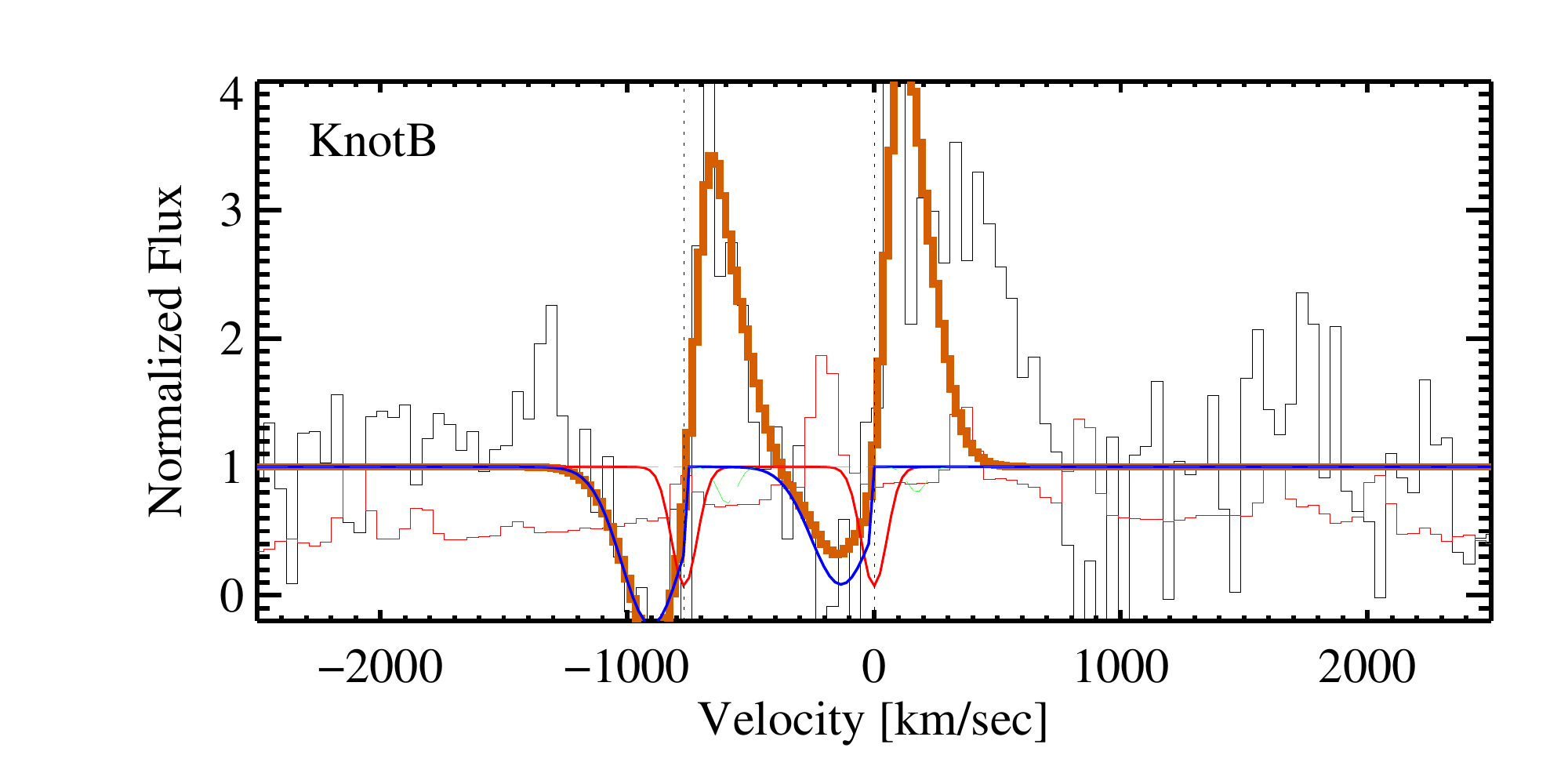}     
    \includegraphics[height=4.5cm,width=8.5cm]{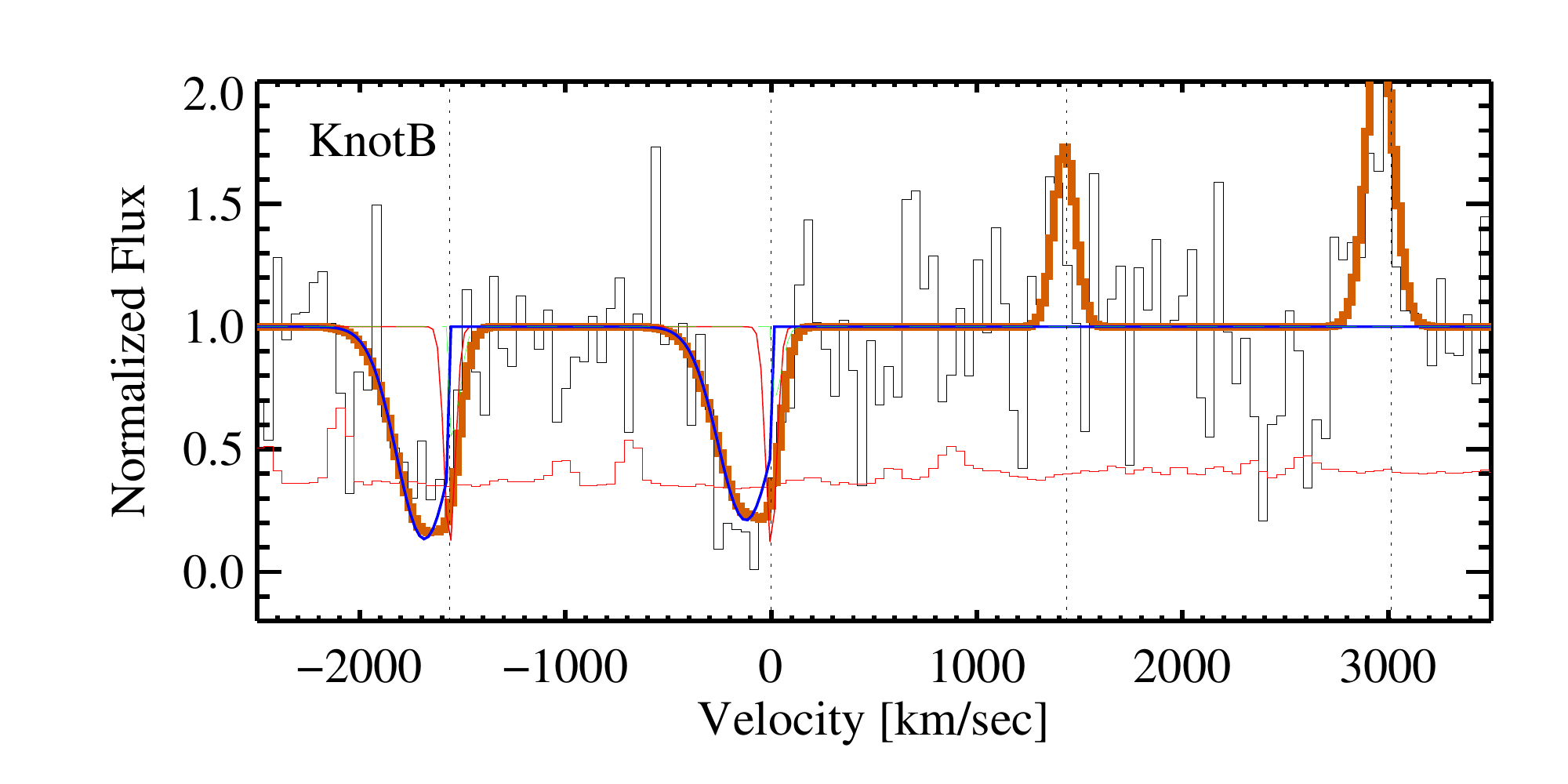}     

    \includegraphics[height=4.5cm,width=8.5cm]{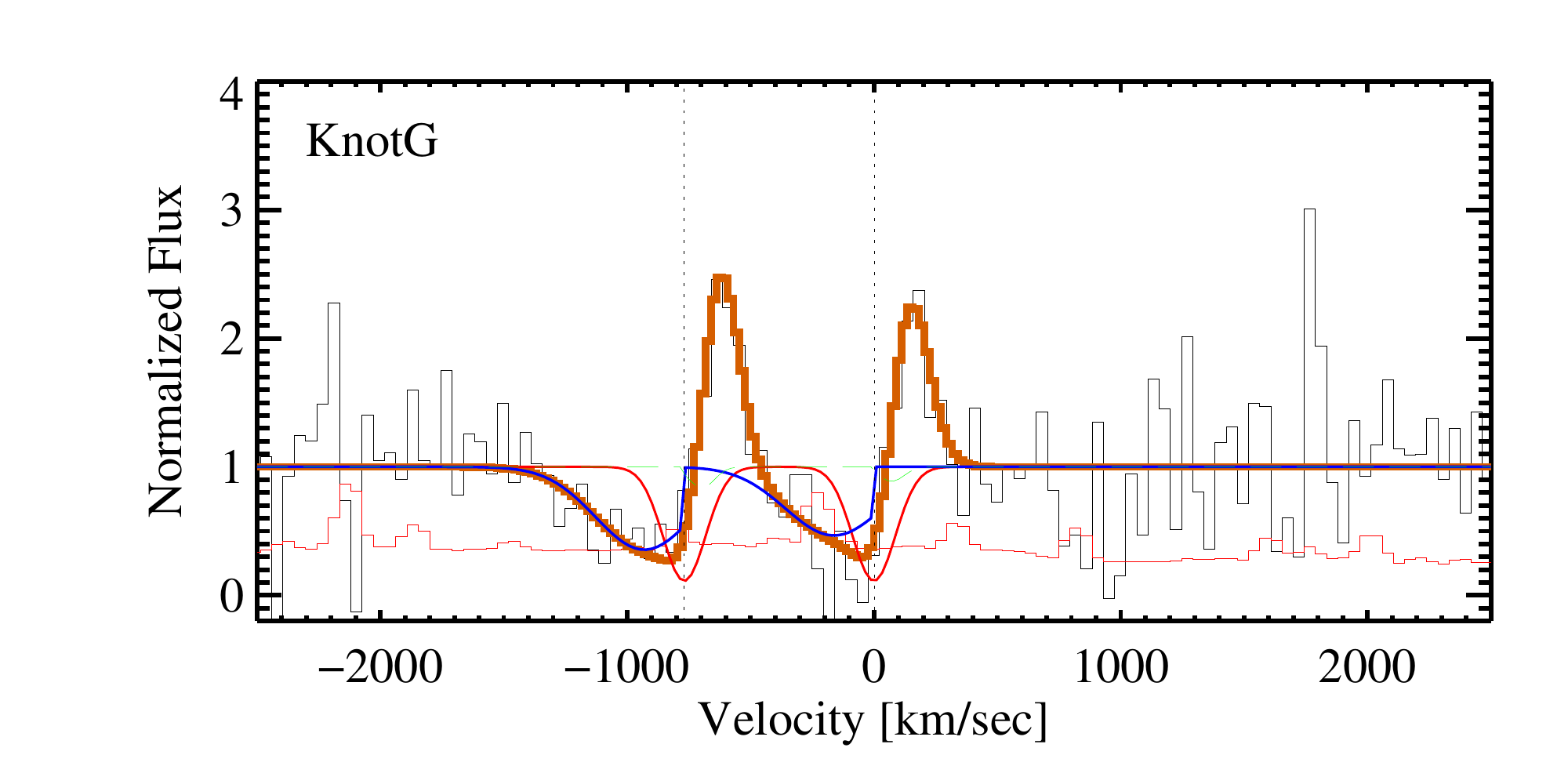}     
    \includegraphics[height=4.5cm,width=8.5cm]{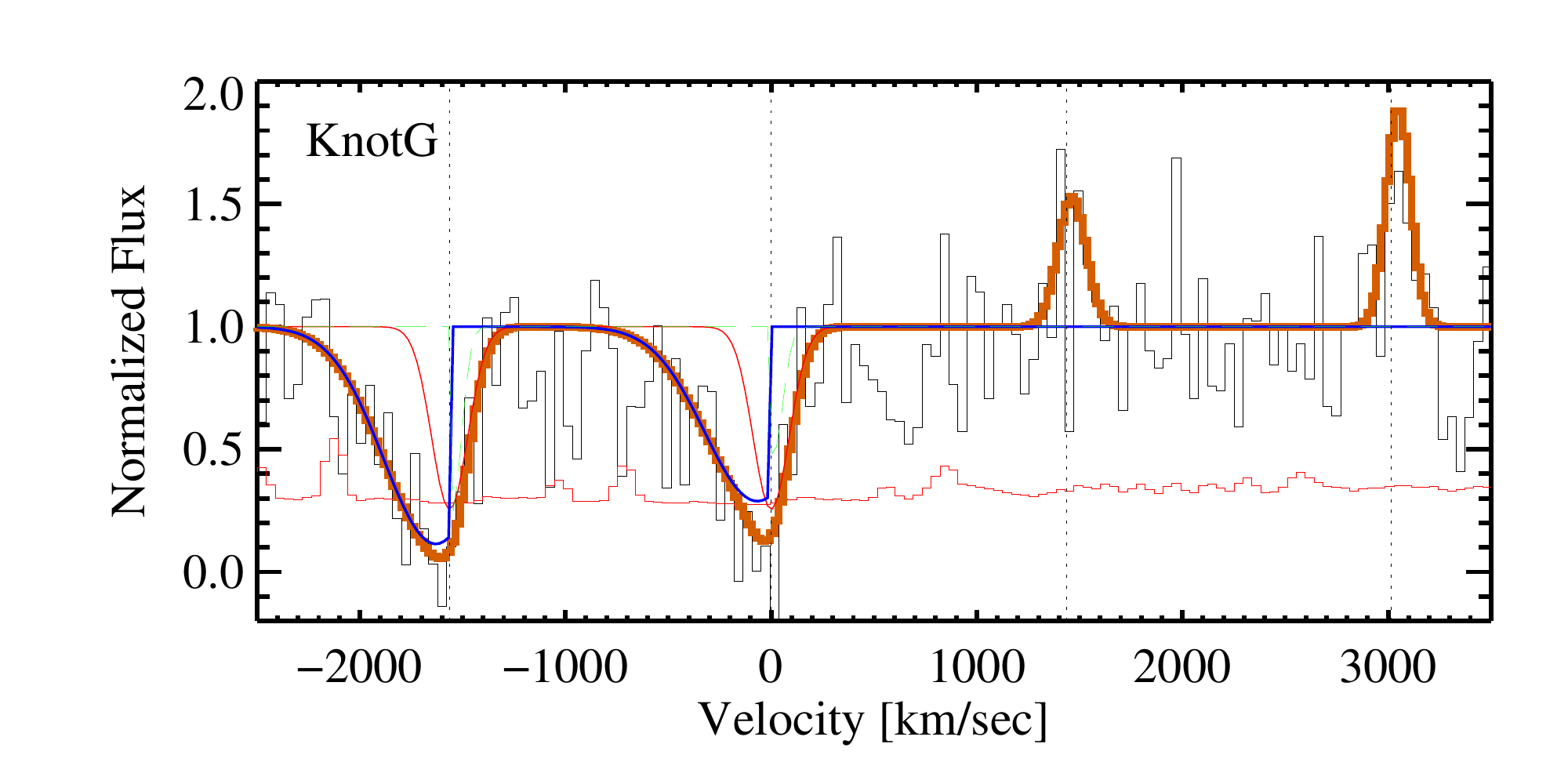}     
    \caption{{\MgII} and {\FeII} outflow profiles with fits. The black stairs represent the observer spectra and the red stairs represent the error spectra. The brown lines show the best fit model profile to the data, after convolving with the instrumental spread function. The blue lines show the outflowing components, and the red smooth line shows the systemic component. The dashed green line represents the inflowing component, which is very small. Knot B and Knot G were re-binned with 2 pixel windows for presentation purposes. }
\label{fig:MgII_fit}
\end{figure*}

\section{Modeling the observed M\lowercase{g}II \& F\lowercase{e}II line Profiles}
In this section we describe the models used to characterize the observed {\MgII} and {\FeII} doublets. The transitions used in this work are listed in Table \ref{table:Line_List}. We use these models to quantify the mean blueshifted absorption and redshifted emission profiles. We first describe the model used to characterize the {\MgII} doublet. We quantify the absorption due to the ISM of the host galaxy as the systemic ISM absorption component and model the outflowing and inflowing gas. The outflowing gas moving away from the galaxy will absorb photons at blueshifted (negative) velocities with respect to the systemic redshift of the galaxy and the inflowing gas will move towards the galaxy and will absorb photons at redshifted (positive) velocities. This outflowing gas could have some velocity dispersion (generally much lower than the thermal broadening of individual clouds) and even complex kinematic structures. However, due to moderate resolution of the spectra, such kinematic complexities can be neglected. The systemic redshift for each knot was obtained from H$\alpha$ emission lines in  NIRSPEC or OSIRIS spectra and are tabulated in Table \ref{table:Knot_properties}. The model that describes the systemic ISM component of the {\MgII} and {\FeII} absorption, centered on the systemic redshift of the galaxy is 

\begin{equation}
1 -  A_{sys} (\lambda) \; =\; 1-  (\tau_{1} G(v, b, \lambda_{1})+  \tau_{2} G(v, b, \lambda_{2})); \label{Equation1}
\end{equation}
where $\tau_{1}$ and $\tau_{2}$ are the optical depths at the center of the absorption profiles given by two Gaussians, $G(v, b, \lambda_{1})$ and  $G(v, b, \lambda_{2})$. These Gaussians are centered at the systemic zero velocity of the galaxy and have a velocity width given by $b$. For the systemic component, the individual absorption lines are likely to be optically thick as they are accounting for both the Galactic ISM as well as absorption from stellar atmospheres. Therefore we assume a doublet ratio of 1 ($\tau_{1} = \tau_{2}$) for all the systemic components. The final model for the observed absorption and emission line profile is given as
\begin{eqnarray}
F_{model} (\lambda) &=&   F_{c} (\lambda) [1- A_{sys} (\lambda)] [1 -A_{outflow}(\lambda)] \nonumber\\
   && {}[1-A_{inflow}(\lambda)] [1+F_{em}(\lambda)];
   \label{Equation2}
\end{eqnarray}
where $F_{model} (\lambda)$ is the model flux density, $F_{c} (\lambda)$ is the continuum level, $[1- A_{sys} (\lambda)]$ is the systemic ISM absorption component,  $[1- A_{outflow} (\lambda)]$ is the blueshifted outflowing gas, $[1- A_{inflow} (\lambda)]$ is the redshifted inflowing gas and $[1+F_{em} (\lambda)]$ is the emission component.  All the absorption components in the inflowing and outflowing models are taken to be the sum of two Gaussians centered at the rest frame wavelength of each component of the  doublet similar to  equation \ref{Equation1}. For the outflowing and inflowing components, we allow the doublet ratio to vary from 1 to 2 and the velocity centroids of the Gaussians are allowed to vary to represent outflowing or inflowing gas. 

For the {\MgII} transition we model the resonant emission lines with two Gaussians and allow their centers to vary to match the observed emission lines. Similarly we model the {\FeII} 2612 and {\FeII} 2626 fluorescent emission lines with two Gaussians. Finally, before comparing the model spectrum to the data, we convolve $F_{model} (\lambda)$ with a Gaussian having a FWHM equal to the velocity resolution of each spectrum (Table \ref{table:Knot_properties}). This gives the final model spectrum $F_{obs} (\lambda)$, which we compare to the data. Because the instrumental resolution is comparable to the expected flow velocities and widths (with FWHMs ranging between $\sim 108$ and $128$ \kms), this step significantly changes the shape of $F_{model} (\lambda)$, and this step is crucial to estimating robust model parameter constraints. There are four free parameters for each component: Doppler shift of the lines, optical depth at line center, Doppler width and doublet ratio. There are four independent model components: systemic, inflowing, outflowing and emission. This gives a total of 15 free parameters (the doublet ratio for the systemic component is fixed at 1). Because of the moderate spectral resolution and finite S/N of the individual spectrum, the Doppler shift and the optical depths are the best constrained and we will not discuss the other parameters in detail. 

 To constrain the best fit model parameters from the data, we developed a customized code in IDL, which samples the posterior probability density function (PPDF) for each model using the adaptive multiple chain Markov Chain Monte Carlo technique \citep{haario2001}. Figure \ref{fig:KnotE_MgII_estimate} top panel and Figure \ref{fig:MgII_fit} shows the best fit model profiles along with the observed spectra. We adopt a uniform prior over the allowed parameter intervals, adjusting slightly these intervals for each model and transition. Our code produces the marginalized PPDFs for each of the fitted parameters. We require each PPDF to be populated with 50,000 realizations, and we visually inspect the PPDFs for each fit to check that the MCMC chains have converged. 

For each realization, we compute the rest frame equivalent width of each model component given as $\rm{W\;=\;  \int (1- \tau_{1}G_{1}({\lambda})) d\lambda}$. This quantity is computed for both {\MgII} and {\FeII} outflow absorption and emission profiles. This yields a distribution of W measurements; we take the mode of the W distribution as the rest frame equivalent width for that transition, and the $\pm$ 34th percentile W around the mode as the lower and upper bounds on the uncertainty on the W measurements respectively. Similarly we estimate the absorption weighted outflow velocity for each transition in each realization given as $\rm{ \bar{v} \;= \;  \int v(1-  \tau_{1}G_{1}(v)) dv}$. We take the mode of the velocity distribution to parameterize the velocity for that transition and the $\pm$ 34th percentile velocities around the mode as the lower and upper bounds on the uncertainty on velocity respectively. For any absorption at the rest frame wavelength $\lambda_{rest}$, with an oscillator strength $f_{0}$, the estimated column density is given as 
\begin{equation}
\rm{  N\;[cm^{-2} ] \; =\;   \frac{\tau_{0} b_{D} \; [km\; sec^{-1}]} {1.497 \times 10^{-15}  \lambda_{rest}\;[\text{\normalfont\AA}]   f_{0}  } ;  }
\end{equation}
where $\tau_{0}$ is the optical depth at the center of the absorption line and $\rm{b_D}$ is the fitted doppler width of the absorption line (where $\rm{b_D = \sqrt{2} b}$).

The top panel of Figure \ref{fig:KnotE_MgII_estimate} shows the best fit {\MgII} model $F_{model}$ (solid brown line) with the observed the {\MgII} 2796, 2803 doublet for Knot E.  The lower panels show the marginalized PPDFs for the estimated outflow velocity ($V_{out}$), emission velocity ($V_{em}$), outflow absorption equivalent width ($W_{out}$), emission equivalent width ($W_{em}$) and {\MgII} column density estimate ($\log\; N$) respectively.  For the {\MgII} 2796, 2803 transition in Knot E we estimate the outflow velocity ($\rm{V_{out} }$) =  $\rm{-225^{+4}_{-7}\; (km/sec)}$,  outflow absorption equivalent width ($\rm{W_{out} }$) =  $\rm{3.37^{+0.10}_{-0.13} \; {\AA}}$, outflow emission velocity ($\rm{V_{em} }$) =  $\rm{158^{+27}_{-48} \; (km/sec)}$, outflow emission equivalent width ($\rm{W_{em} }$) =  $\rm{-4.87^{+0.87}_{-1.18}\; {\AA}}$, and  outflow {\MgII} column density ($\rm{\log N /cm^{-2}}$) = $\rm{14^{+0.02}_{-0.0.03}}$ respectively. These measurements for all four knots are listed in Table \ref{table:Result_summary}.

\begin{figure*}
\centering
    \includegraphics[height=6.5cm,width=7.5cm]{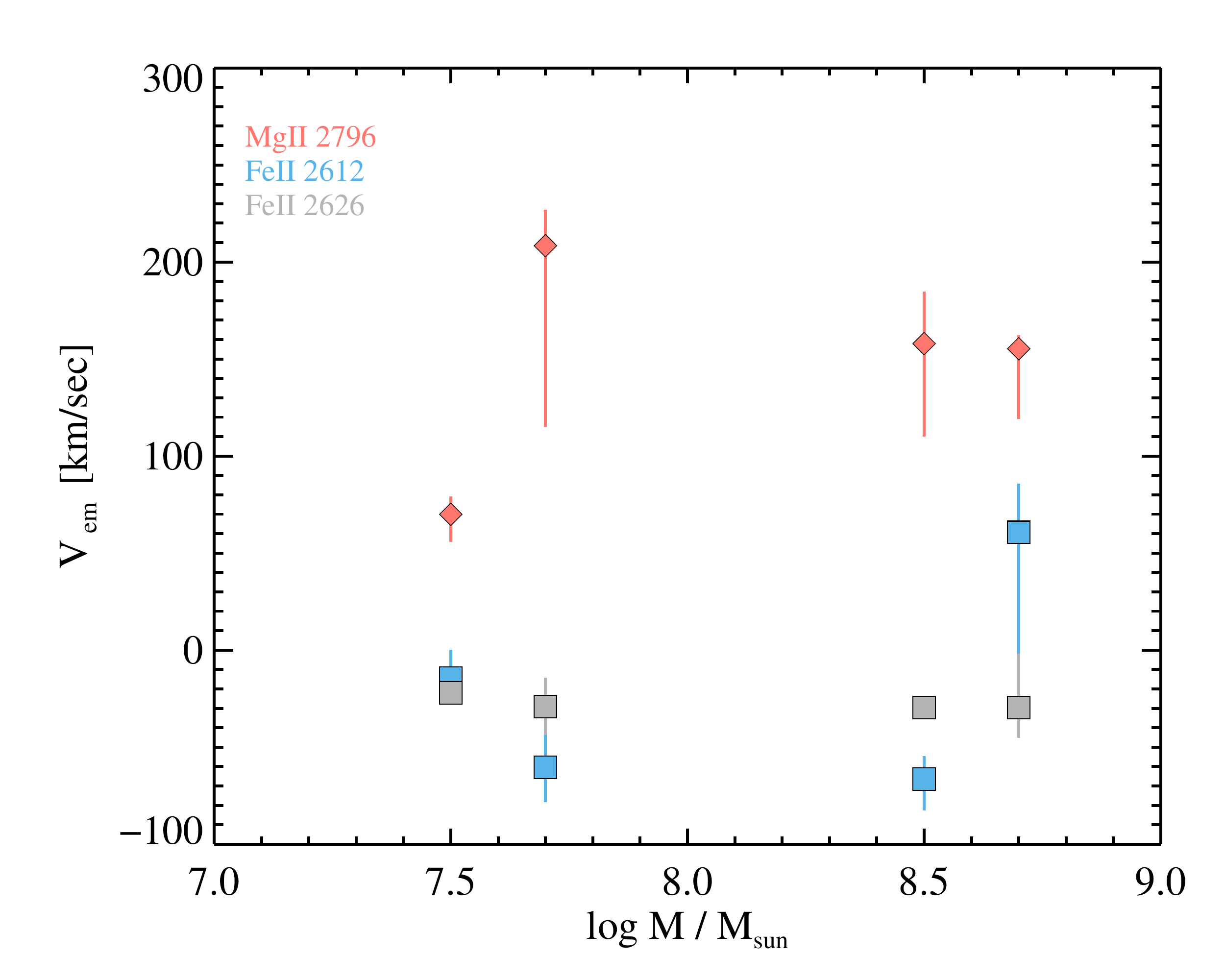}     
    \includegraphics[height=6.5cm,width=7.5cm]{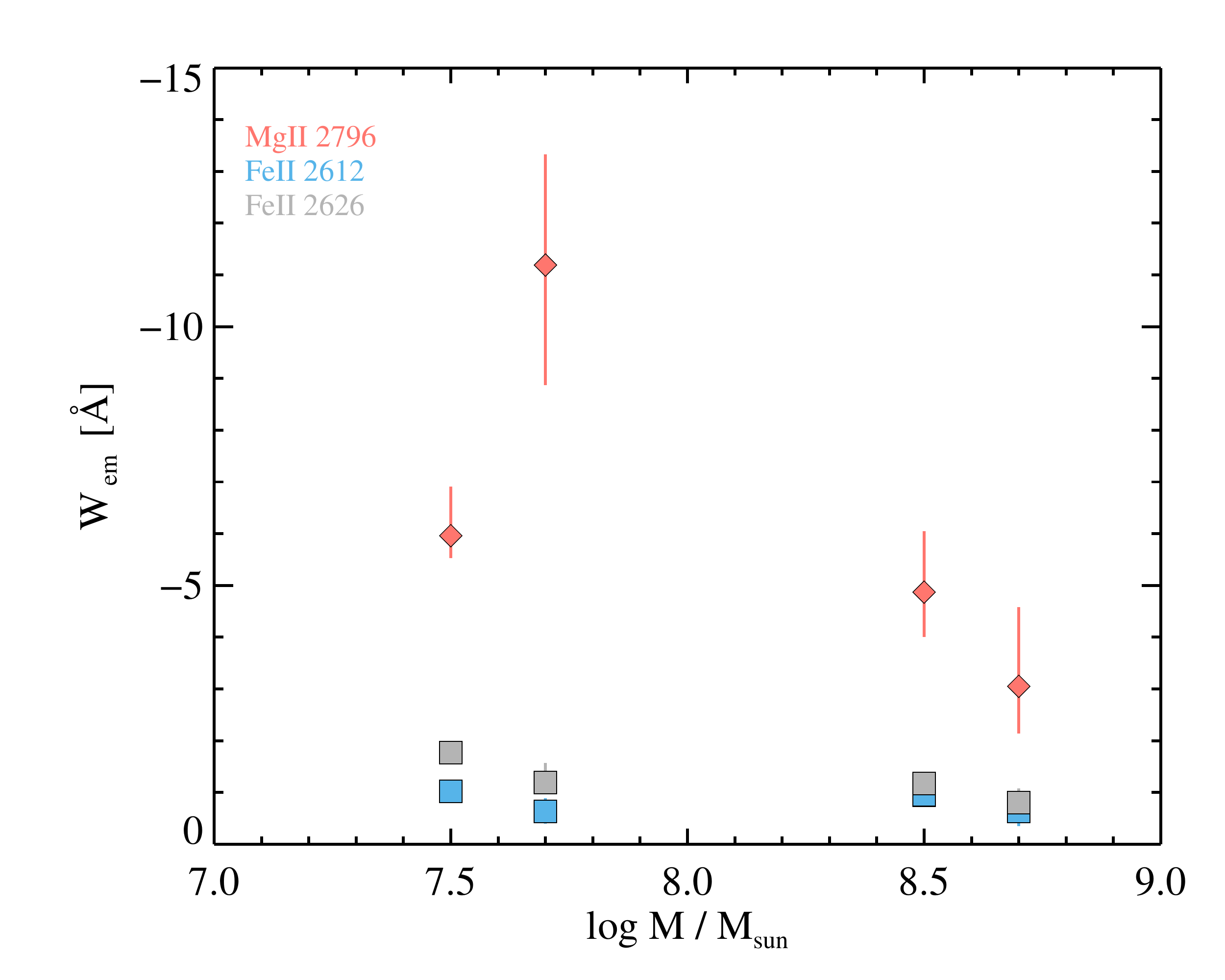}
    \includegraphics[height=6.5cm,width=7.5cm]{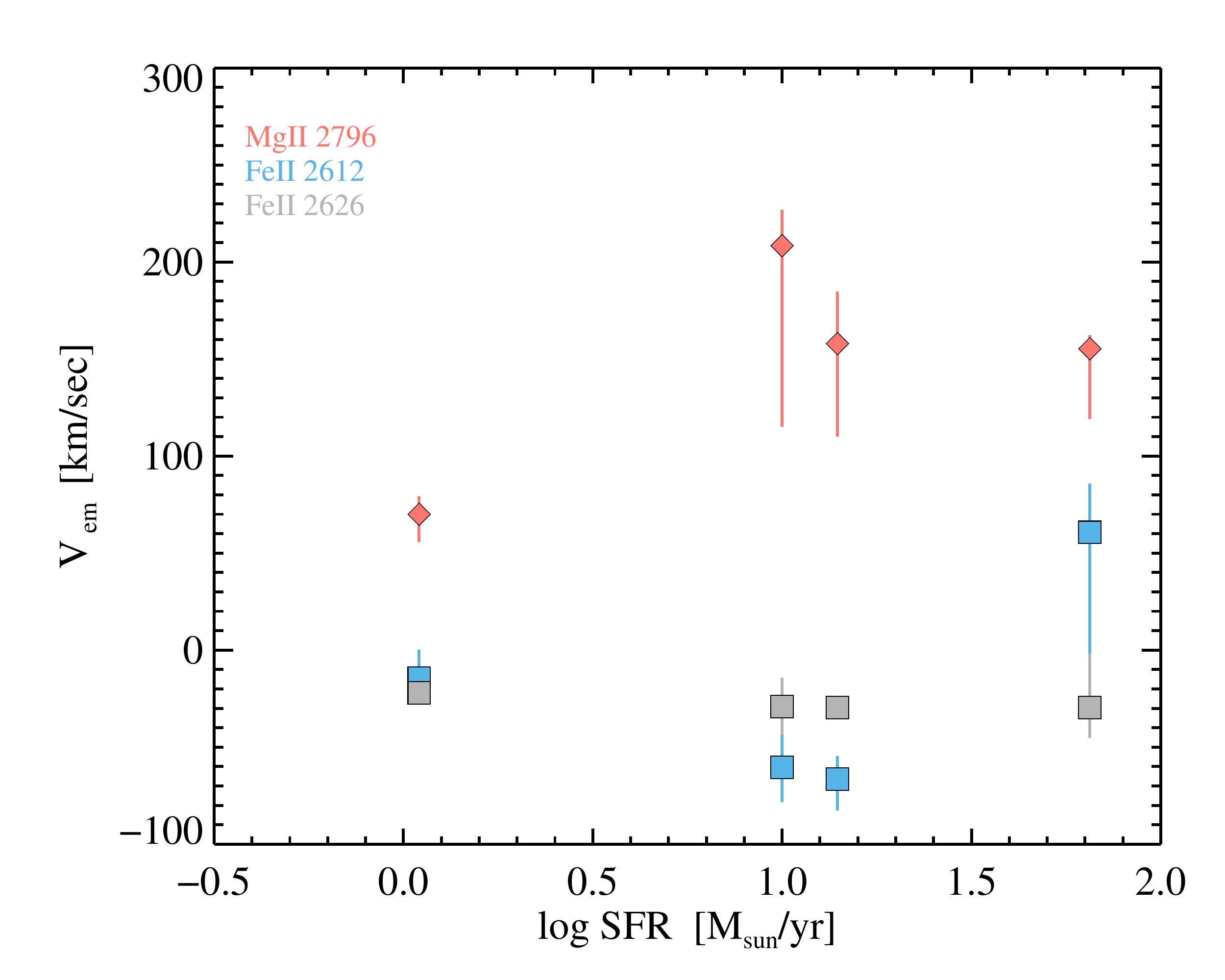}     
    \includegraphics[height=6.5cm,width=7.5cm]{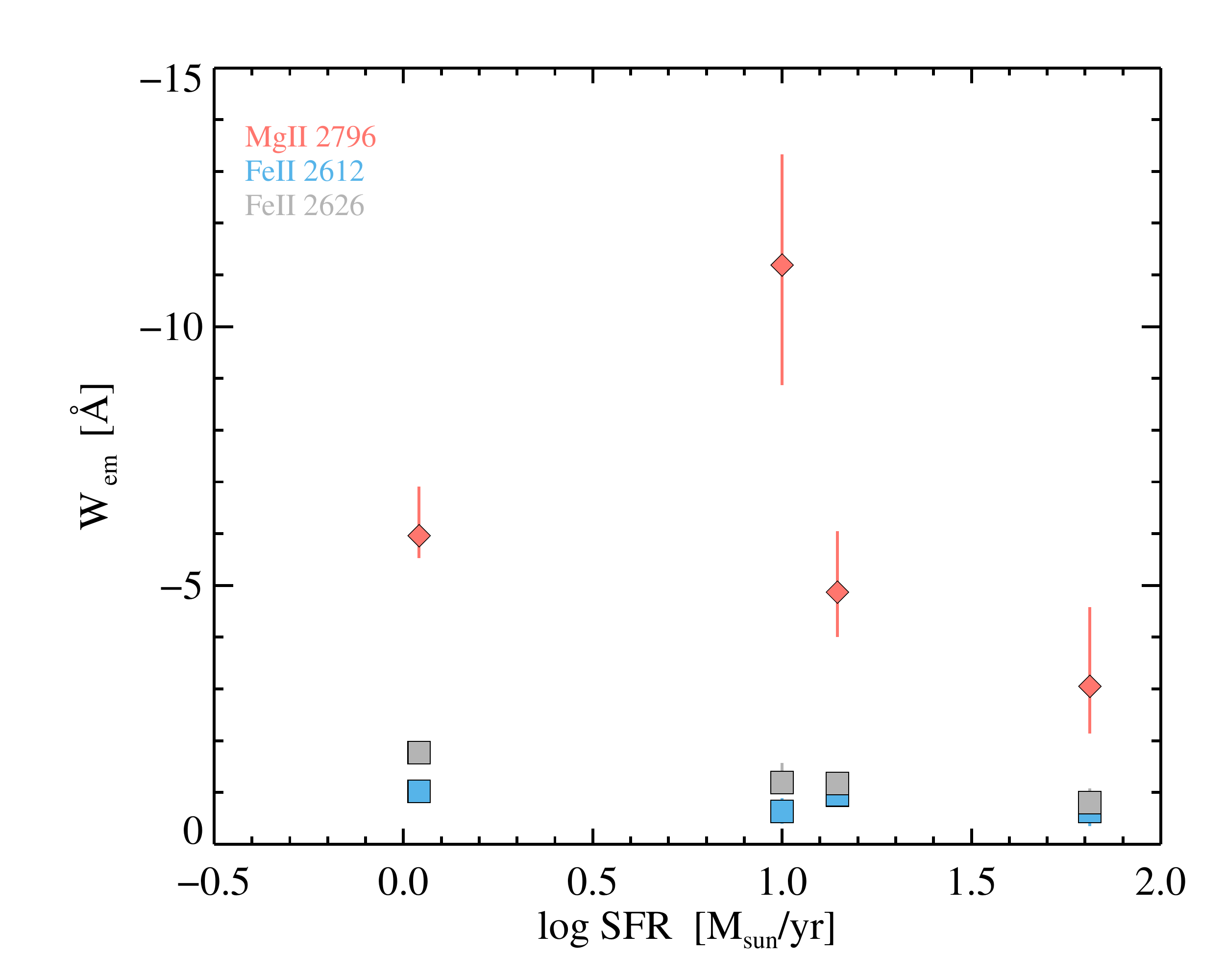}
     \includegraphics[height=6.5cm,width=7.5cm]{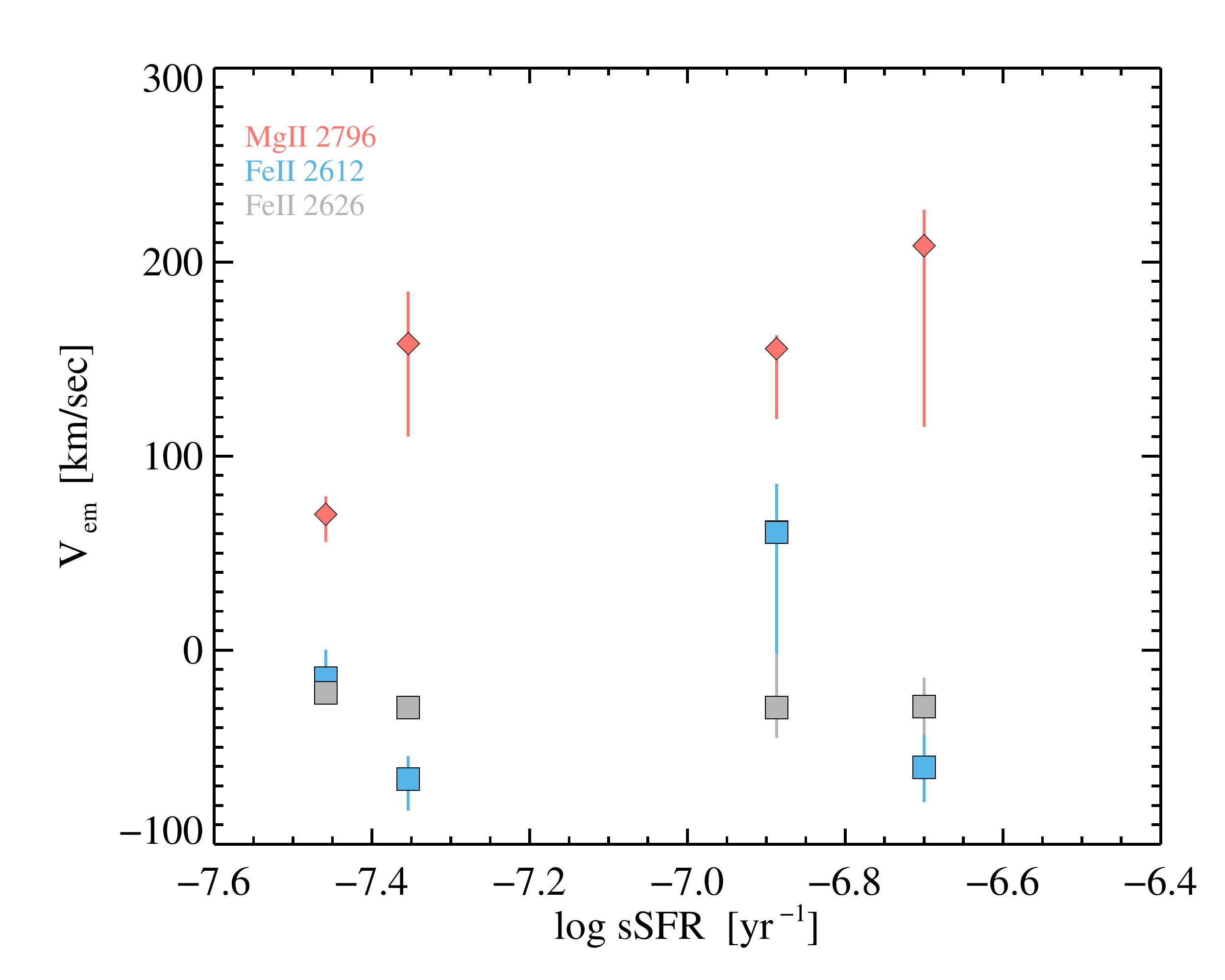}     
    \includegraphics[height=6.5cm,width=7.5cm]{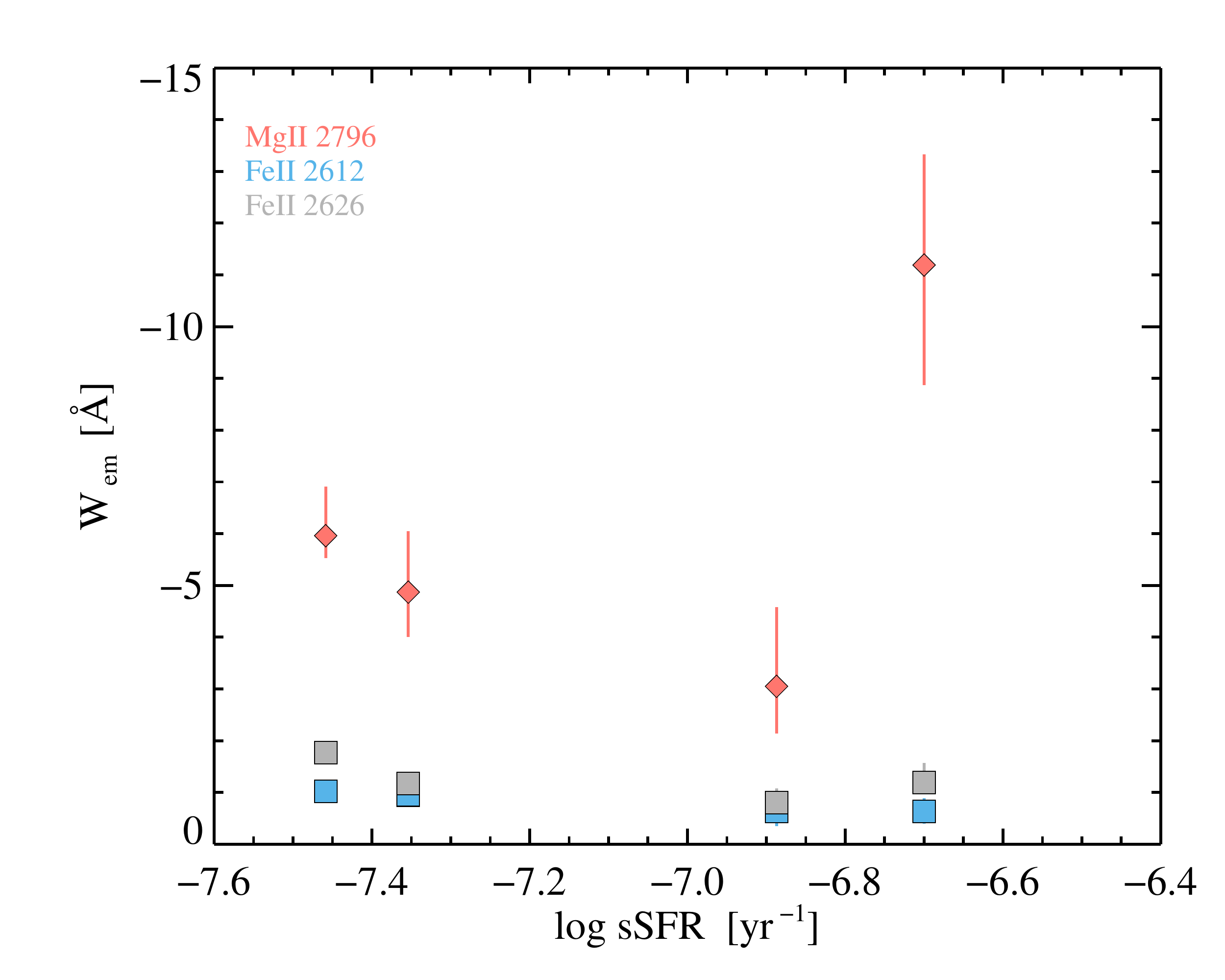} 
     \caption{Variation of {\MgII} and {\FeII} emission kinematics (left panels) and equivalent widths (right panels) as a function of stellar mass (top panels), SFR (middle panels) and sSFR (bottom panels) of the individual knots respectively. {\MgII} resonant emission (red diamonds) lines are always redshifted relative to the systemic zero velocity of the knots, whereas the fluorescent {\FeII} 2612 (blue squares) and {\FeII} 2626 (gray squares) emission lines are consistent with being at the systemic zero velocity of the knots. {\FeII} emission equivalent widths increase with decreasing SFR in the individual knots. }
\label{fig:emission_properties}
\end{figure*}

\begin{figure*}
\centering
     \includegraphics[height=6.5cm,width=7.5cm]{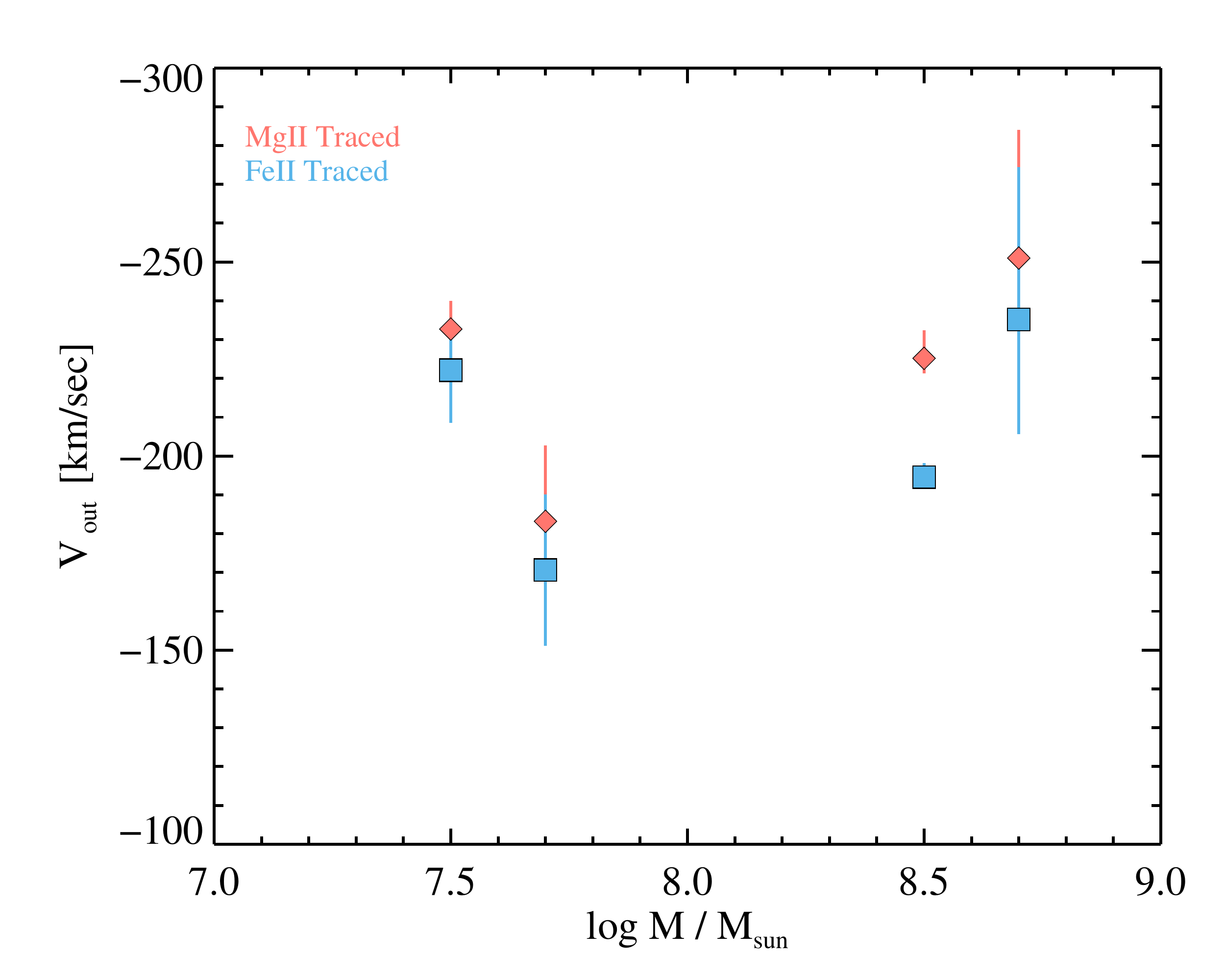}     
    \includegraphics[height=6.5cm,width=7.5cm]{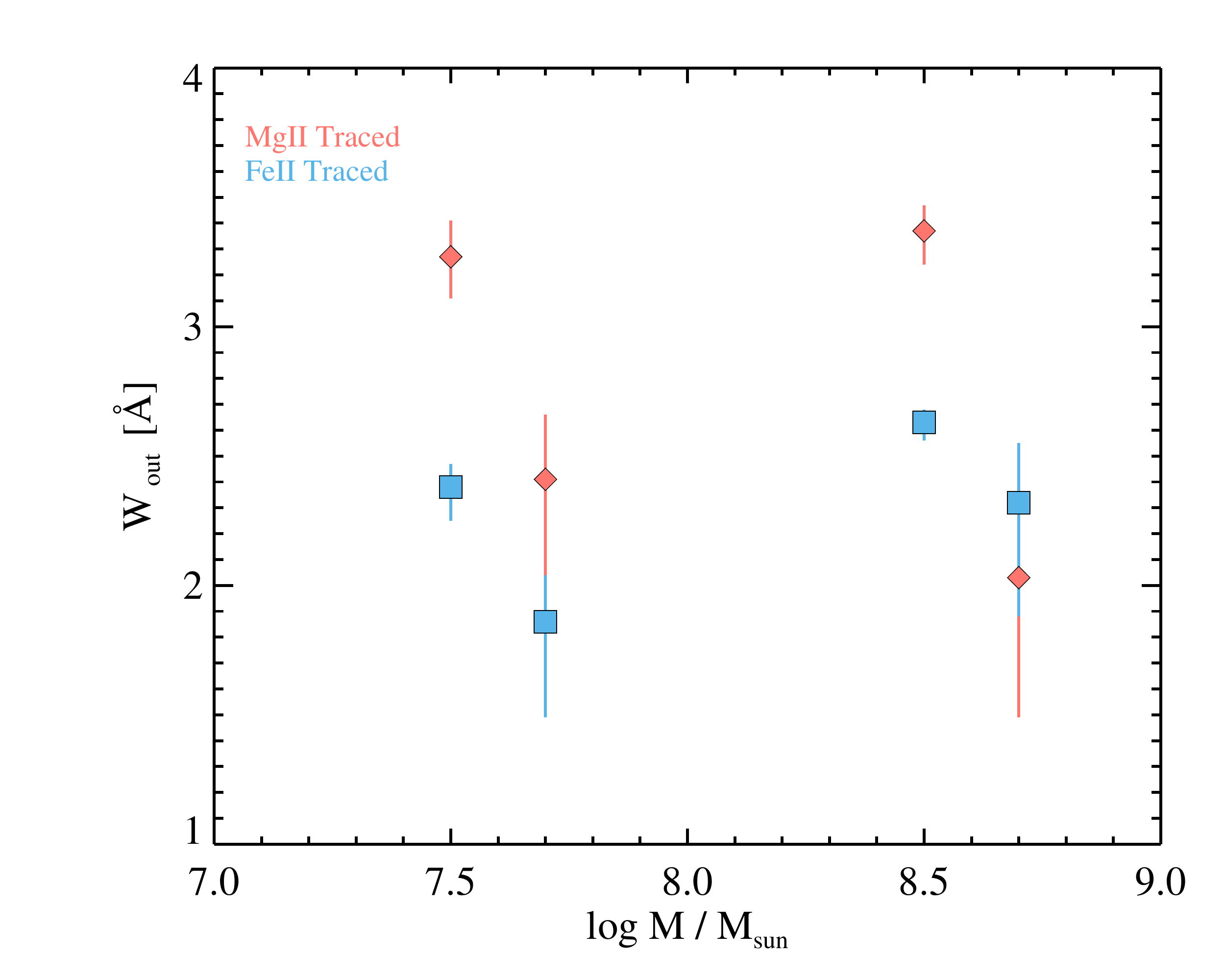}
    \includegraphics[height=6.5cm,width=7.5cm]{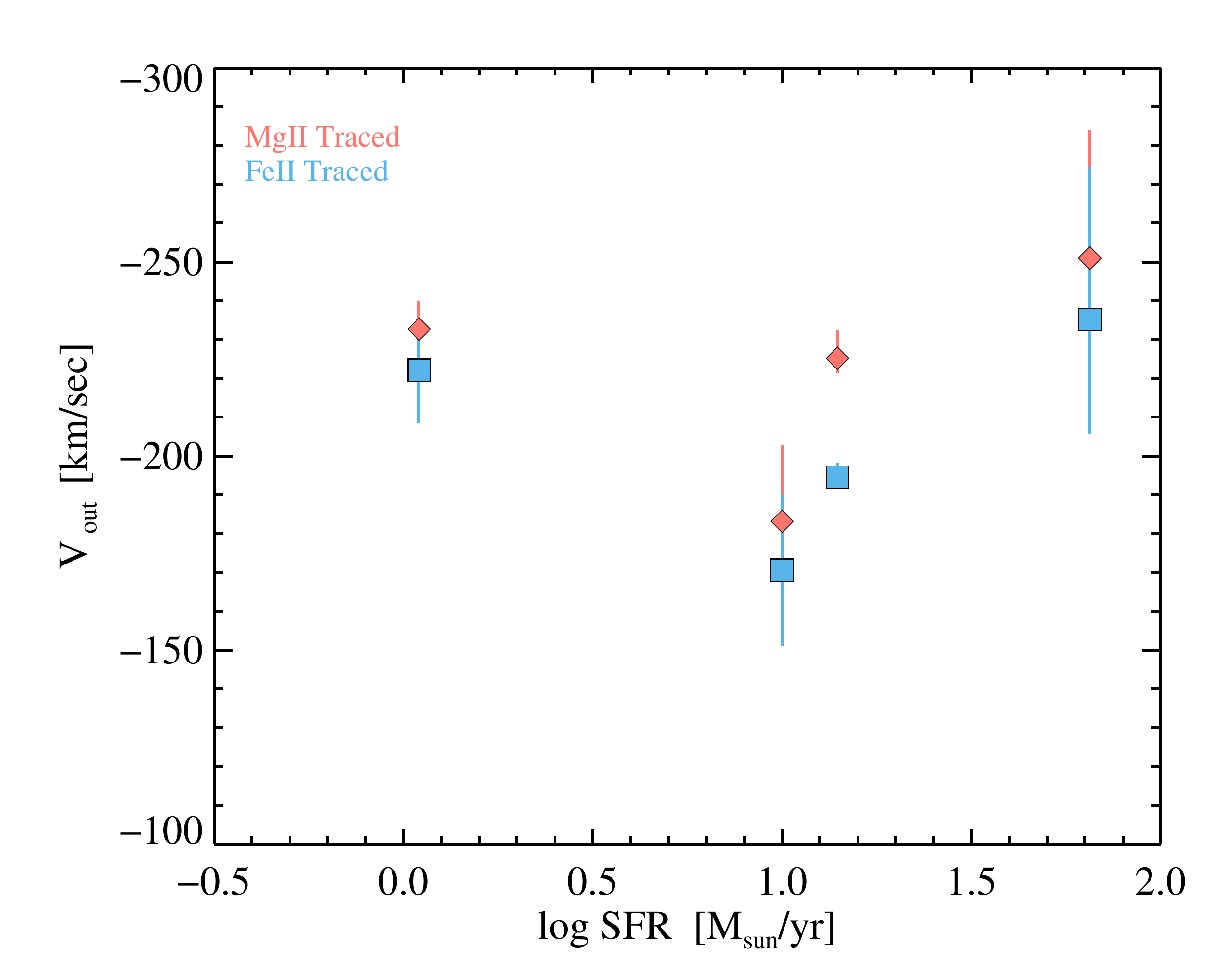}     
    \includegraphics[height=6.5cm,width=7.5cm]{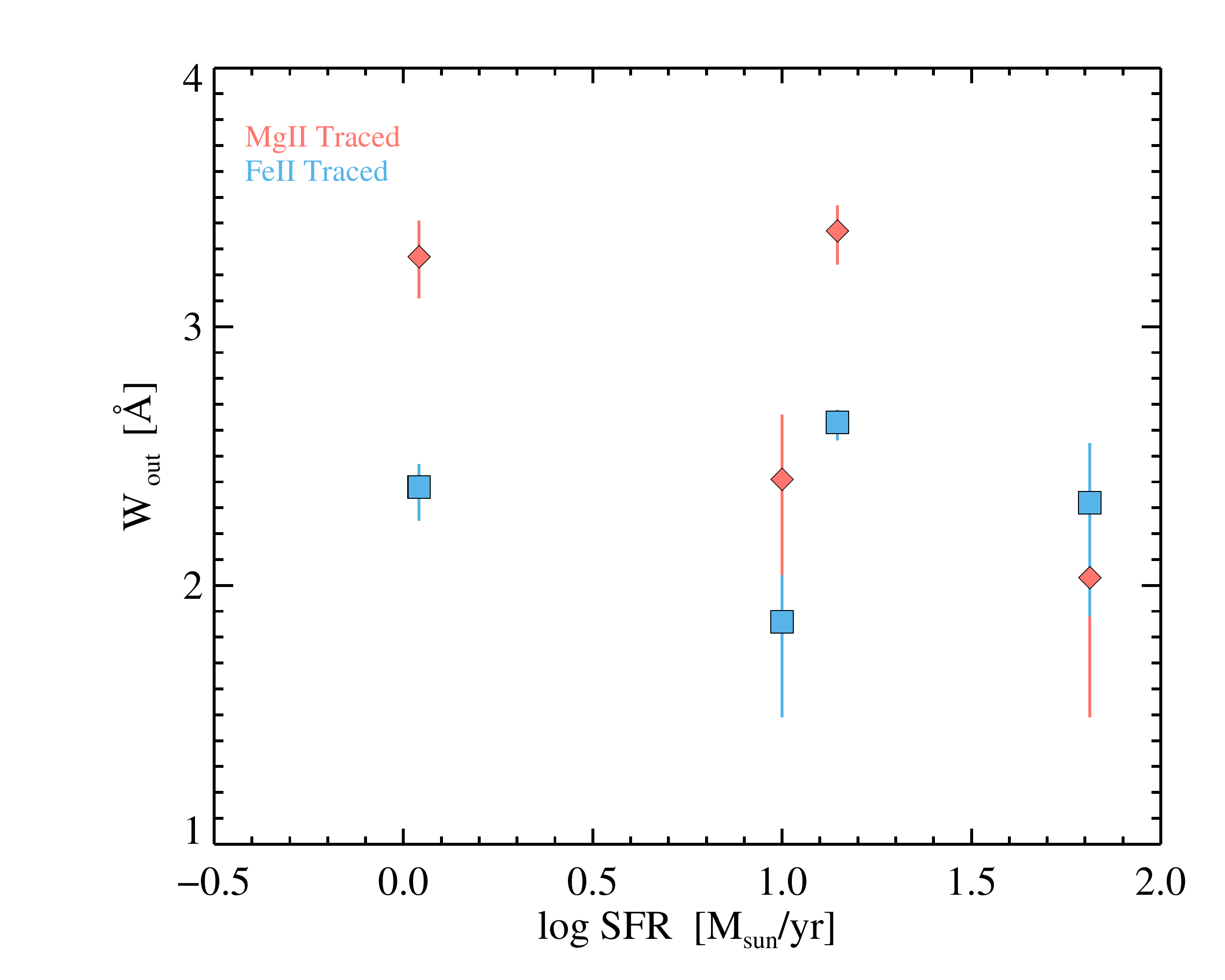}
     \includegraphics[height=6.5cm,width=7.5cm]{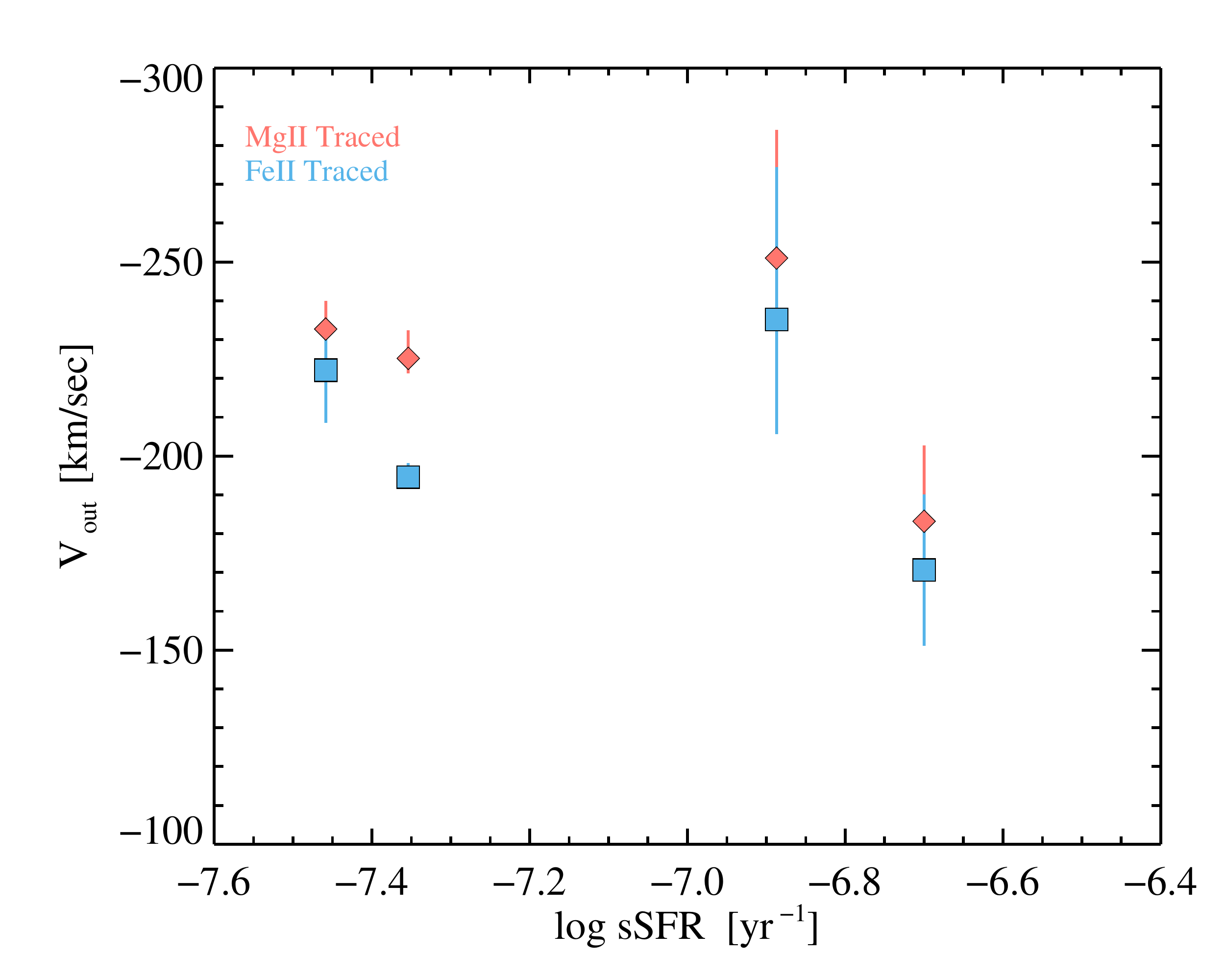}     
    \includegraphics[height=6.5cm,width=7.5cm]{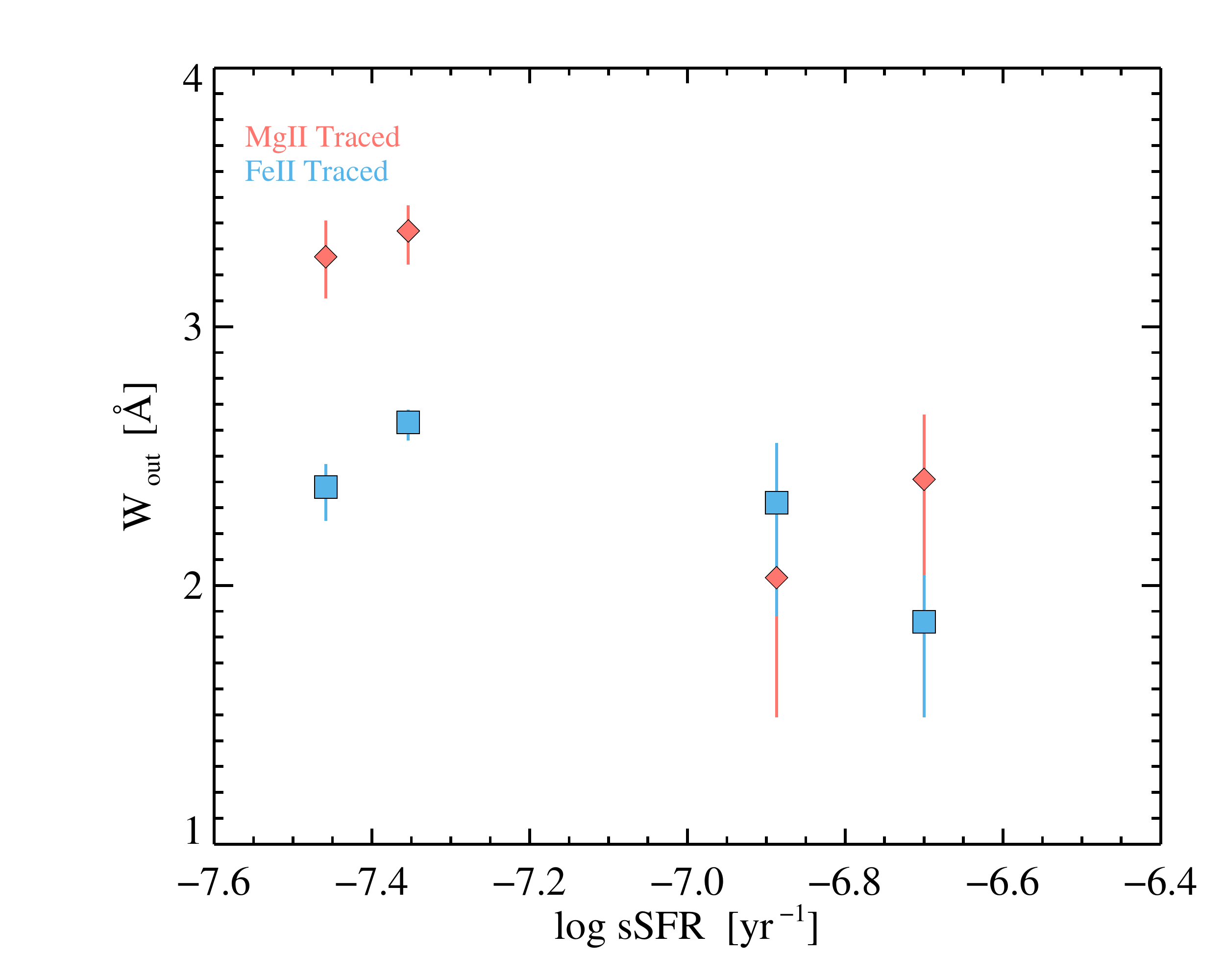} 
      \caption{Variation of blueshifted {\MgII} and {\FeII} absorption kinematics (left panels) and equivalent widths (right panels) as a function of stellar mass (top panels), SFR (middle panels) and sSFR (bottom panels) of the individual knots respectively. Outflow velocities traced by blueshifted {\MgII} absorption (red diamonds) and {\FeII} absorption (blue squares), show weak trends with increasing star formation rate and stellar mass of the knots. }
\label{fig:absorption_properties}
\end{figure*}

\section{Results}
In the following section we discuss the variation of emission/absorption strengths and kinematics of gas traced by {\MgII} and {\FeII} emission and absorption lines relative to their systemic redshift of the galaxy RCS0327. We also discuss the variation of outflow absorption and emission properties with the physical properties (stellar mass, star formation rate, specific star formation rate) of the individual knots. 

\subsection{Dependence with mass and star formation rates }

\subsubsection{{\FeII} emission}

We first focus on the emission strengths and kinematics of the {\FeII} transitions. The {\FeII} ion exhibits a complex and large set of transitions in the rest frame ultraviolet wavelength range. The Fe$^{+}$ transition alone is found to be associated with at least millions of recorded energy levels \citep{Kurucz2005}. The {\FeII} resonant absorption lines have been extensively used to study galactic outflows, as it redshifts into the optical wavelength range at high redshifts (\citealt{Rubin2011, Coil2011, Martin2012, Kornei2012, Kornei2013}). The absorption of a single resonant photon by Fe$^{+}$ can lead to either re-emission of another resonant photon to the ground state following an electronic transition to the ground state (resonant emission) or the emission of a photon to an excited ground state (fluorescence emission). In our data, we observe the {\FeII} 2586, 2600 resonant absorption doublet, but we do not observe resonant {\FeII} emission. We further observe fluorescence {\FeII} emission  at 2612 {\AA} and 2626 {\AA} respectively.  The absence of {\FeII} 2586, 2600 resonant emission is due to multiple scattering of the photons re-emitted with the resonant energy, as they interact with the same ions in the wind where they were created (see \citealt{Scarlata2015} for a detailed discussion). Such multiple scatterings will not change the shape of the absorption profile but will suppress the contribution of the re-emission in the resonant emission line, while simultaneously enhancing the re-emission strength in the florescence channel \citep{Scarlata2015}. In case of {\MgII} 2796, 2803 transition, the florescence channel is absent. Because of that, the multiple scattering of photons do not change the {\MgII} resonant emission profile and we observe the {\MgII} resonant emission lines but not the {\FeII} 2586, 2600 resonant emission lines (see Figure~\ref{fig:MgII_fit}).

Figure \ref{fig:emission_properties} shows how {\FeII} 2612, 2626 emission velocities and emission equivalent widths vary with the physical properties of the knots. The top row shows the variation with the stellar mass of the knots, the middle row shows the variation with star formation rates and the last panel shows the variation with the sSFR of these knots. We find that the {\FeII} emission velocity centroids (blue and gray squares) are more consistent with either being blueshifted or close to being at the systemic zero velocity of the individual knots. The mean velocity of the four individual measurements for {\FeII} 2612 emission line is $-20 \pm 29$ km/sec, and {\FeII} 2626 line is $-28 \pm 3$ km/sec respectively. We find no detectable correlation of {\FeII} emission velocities with stellar mass, SFR or sSFR of the individual knots.  

This absence of a kinematic shift relative to the systemic velocities does not necessarily imply that the associated gas is at rest with respect to the stars of the galaxy. Simple radiative transfer models for outflowing gas predict the net {\FeII} emission velocities to be close to the systemic zero velocity of the host galaxy for an optically thin wind \citep{Prochaska2011}. Such a model will transmit the scattered {\FeII} emission photons from both backside and frontside of the wind. Further, if the dust opacity increases, the {\FeII} emission velocities will be slightly blueshifted, because the redshifted photons scattering off the backside of the wind will be absorbed preferentially, due to the longer path lengths. This scenario is consistent with the measurements given in Figure \ref{fig:emission_properties}.

Looking at the {\FeII} 2612, 2626 emission equivalent widths, we see that knots with lower SFRs exhibit higher emission equivalent widths. A Pearson linear correlation test for the combined {\FeII} 2612, 2626 line equivalent widths show a 2.9$\sigma$ inverse correlation with log SFR of the individual knots. This result is consistent with \cite{Erb2012} and \cite{Kornei2013}, where amongst their {\FeII} emission line showing galaxies, the strongest {\FeII} emission lines were observed in the galaxies with smaller stellar masses and lower star formation rates. We do not detect any statistically significant correlation of {\FeII} emission line strength with stellar mass or specific star formation rates of the individual knots.

\subsubsection{{\MgII} emission}

We now focus on the emission kinematics and strength of the {\MgII} doublet transition. The strong resonant {\MgII}  absorption doublet transition is easily identifiable and extensively used as a tracer of galactic winds (e.g. \citealt{Weiner2009, Rubin2010,Martin2012, Bordoloi2014b}). While the {\MgII} doublet is primarily seen in absorption, it is also seen with a P-Cygni emission profile in a wide variety of objects. The P-Cygni {\MgII} emission profile has been seen in Seyferts \citep{1983ApJ...266...28W}, Ultra-Luminous Infrared galaxies (ULIRGS) \citep{Martin2009}, local star forming galaxies \citep{1993ApJS...86....5K}, high redshift star forming galaxies \citep{Weiner2009, Rubin2010, Bordoloi2014b} etc. The most viable hypothesis for the physical origin of {\MgII} emission doublet is photons scattering off the backside of galactic winds \citep{Prochaska2011}. This process has also been observed for Ly$\alpha$ emission in Lyman break galaxies \citep{Pettini2001}. Both the {\MgII} absorption lines can be strongly affected by resonant emission filling, as there are no excited ground states available for fluorescence.

Figure \ref{fig:emission_properties} shows how {\MgII} emission velocities and emission equivalent widths vary with the physical properties of the knots. The {\MgII} emission equivalent widths show a weak trend with SFR of the individual clumps. The {\MgII} emission velocities (red diamonds) are always redshifted with respect to the systemic redshift of the host star-forming knots. The mean {\MgII} emission line velocity for the four knots is $148 \pm 29$ km/sec. This finding with local star-forming knots is consistent with the other studies looking at global galaxy properites, where it was found that the typical {\MgII} emission velocities are $\approx$ 100 km/sec redward of the systemic redshift of the galaxies \citep{Weiner2009,Kornei2013,Rubin2014a}. Resonant  {\MgII} emission lines are not ubiquitously seen in down-the-barrel galaxy spectra: \cite{Kornei2013} reported that they detect {\MgII} emission in 15\% of their galaxies showing blueshifted {\MgII} absorption. \cite{Erb2012} studying a higher resolution sample reported the detection rate of {\MgII} emission in 30\% of their sample. \cite{Erb2012} found that {\MgII} emitters are typically drawn from bluer, higher star-forming galaxies. This might be owing to a combination of their lower resolution spectra and {\MgII} emission filling, making detection of any weak emission difficult.

\subsubsection{Outflows traced by {\MgII} \& {\FeII}}
In this section we study the outflow gas kinematics and absorption strengths traced by {\MgII} and {\FeII} resonant transitions. Figure \ref{fig:absorption_properties} shows the variation of {\MgII} and {\FeII} outflow velocities and equivalent widths with stellar mass, SFR and sSFR of the individual knots. The outflow velocities traced by {\MgII} and {\FeII} transitions are very similar, but the blueshifted outflow velocities traced by {\MgII} absorption are systematically higher  compared to that traced by {\FeII} absorption by an average of  $\sim$17$\pm$5 km/sec. We believe this  small offset is due to our inability to completely account for emission filling by resonant {\MgII} emission. However, the good agreement between the {\FeII} and {\MgII} traced outflow velocity estimates suggest that our model of {\MgII} emission is doing a reasonably good job in reproducing the total {\MgII} emission in these galaxies. 

The mean outflow velocities as traced by both {\FeII} and {\MgII} absorption are different from one knot to the other, after accounting for measurement uncertainties. The mean {\MgII} traced outflow velocity for the four knots is $-223 \pm 14$ km/sec, and the individual outflow velocities vary from -251 km/sec to -183 km/sec. Similarly the mean {\FeII} traced outflow velocity for the four knots is $ -206 \pm 15$ km/sec, and the individual outflow velocities vary from -235 km/sec to -170 km/sec. Hence for both independent tracers we see a maximum velocity difference $\approx$ 70 km/sec between individual knots. This variation is significantly larger than the differences in nebular redshift of the knots, which only differ by $\sim$ 15 km/sec (see Table \ref{table:Knot_properties}).

The outflow velocities show weak trends with SFR and stellar mass of the individual knots, but these trends are not statistically significant (Figure \ref{fig:absorption_properties}). The outflow velocities measured for individual star forming knots are similar to the average outflow velocities observed for whole galaxies by \cite{Weiner2009} and \cite{Bordoloi2014b} in their analysis of stacked down the barrel galaxy spectra.

\subsection{Dependence with $\Sigma_{SFR}$}
In this section we discuss how outflow properties and emission properties depend on the star formation rate surface density ($\Sigma_{SFR}$). We estimate $\Sigma_{SFR}$ for each knot self-consistently, using the extinction-corrected SFRs from the HST grism observations, and the apertures used to extract those spectra. $\Sigma_{SFR}$ is the demagnified SFR divided by the demagnified area of the aperture in the source plane. We employ this method so that the magnification used to compute the intrinsic star formation rate is the same magnification used to compute the area of the aperture in the source plane.

\subsubsection{{\FeII} \& {\MgII} emission}
In Figure \ref{fig: sigsfr_properties}, the top panels show the variation of {\MgII} and {\FeII} emission velocities and emission equivalent widths as a function of $\Sigma_{SFR}$. The {\MgII} emission equivalent width decreases with increasing $\Sigma_{SFR}$, while the {\FeII} emission equivalent widths show no trend with $\Sigma_{SFR}$. The decrease of {\MgII} emission equivalent width with increasing $\Sigma_{SFR}$ might be due to higher emission filling for weaker outflows.  We see no trend between emission velocities and $\Sigma_{SFR}$ of local knots.

\subsubsection{Outflows traced by {\MgII} \& {\FeII}}

There have been many studies that have looked into the correlation of $\langle \Sigma_{SFR} \rangle$ with the average blueshifted outflowing gas for the entire  galaxy.  At low redshifts, \cite{chen_outflow2010} used stacked NaD absorption lines of star-forming galaxies, and found that the outflow NaD equivalent width strongly depends on the average $\Sigma_{SFR}$ of their host galaxies.  At $z \sim 1$, \cite{Kornei2012} reported a 3.1$\sigma$ correlation of outflow velocity with $\Sigma_{SFR}$. \cite{Bordoloi2014b},  using stacked galaxy spectra found that on average below the canonical threshold ( $\Sigma_{SFR} \rm{\; =\; 0.1\; M_{\odot} yr^{-1} kpc^{-2}}$) star-forming galaxies exhibit little blueshifted absorption at  $z \sim 1$. At $z \sim 2$, \cite{Newman2012}  reported a $\Sigma_{SFR}$ threshold of 1 $\rm{M_{\odot} yr^{-1} kpc^{-2}}$, above which $\rm{M* > 10^{10} M_{\odot}}$ galaxies might have stronger outflows as compared to the low mass galaxies.

\begin{figure*}
\centering
   \includegraphics[height=6.5cm,width=7.5cm]{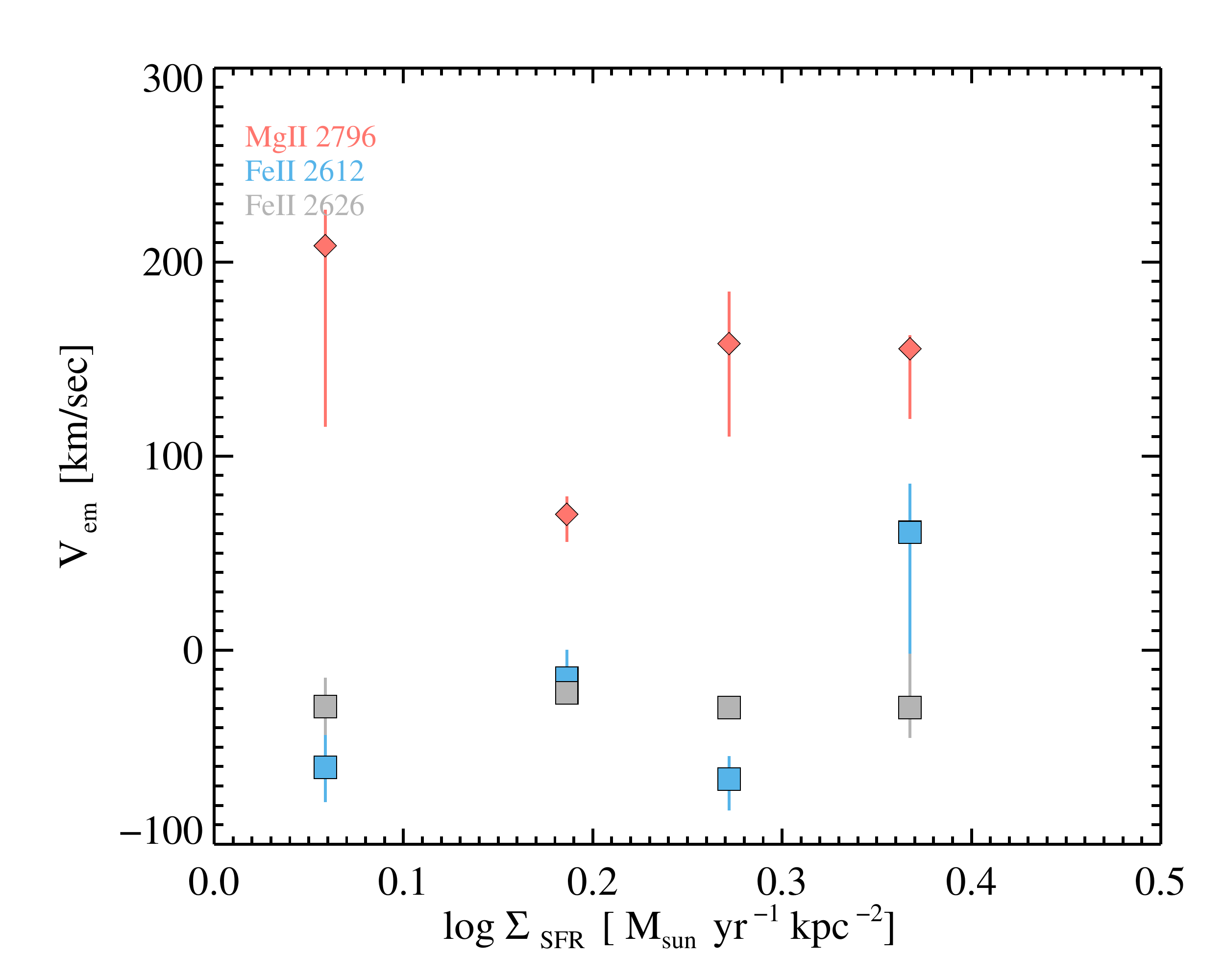}     
    \includegraphics[height=6.5cm,width=7.5cm]{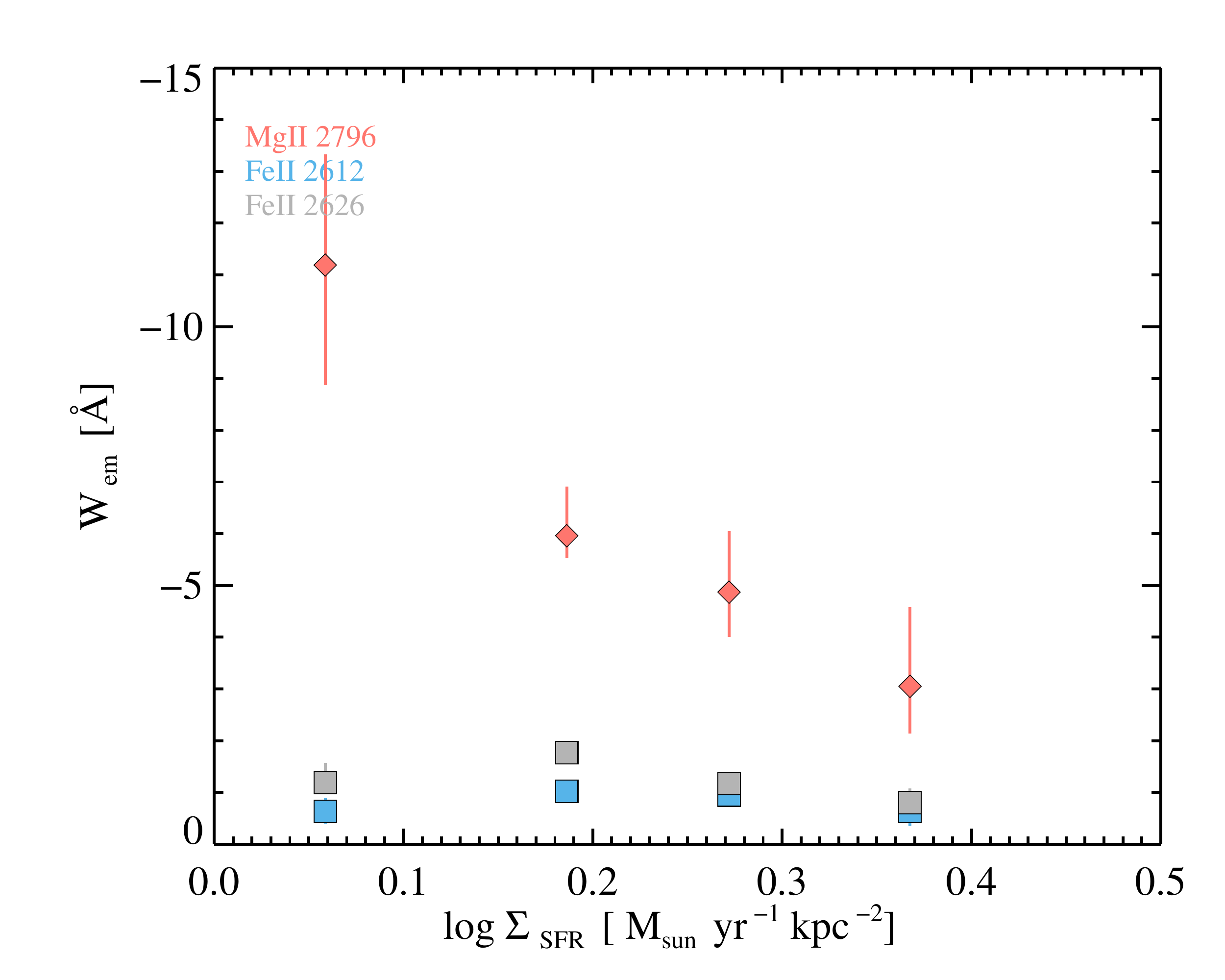}
    \includegraphics[height=6.5cm,width=7.5cm]{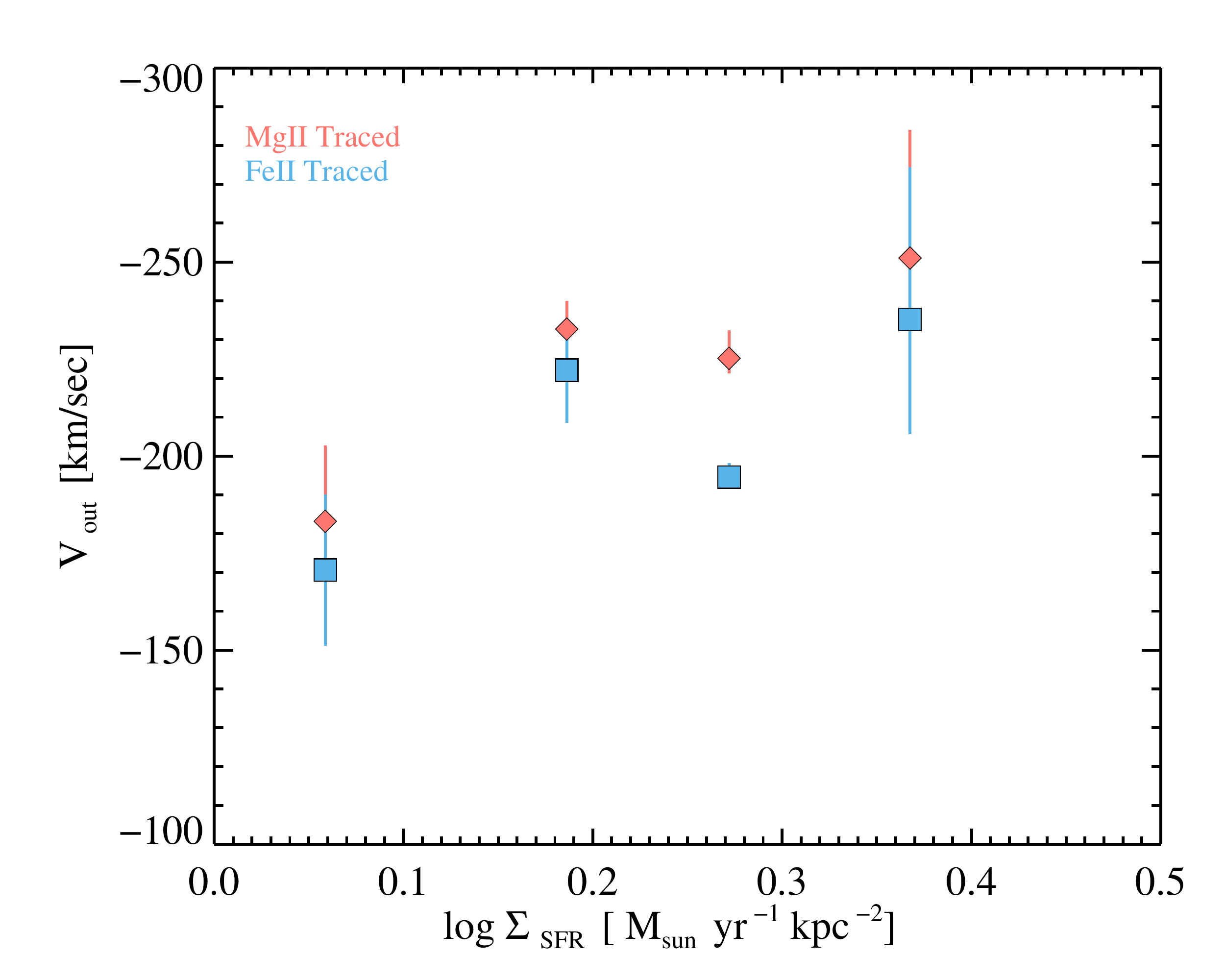}     
    \includegraphics[height=6.5cm,width=7.5cm]{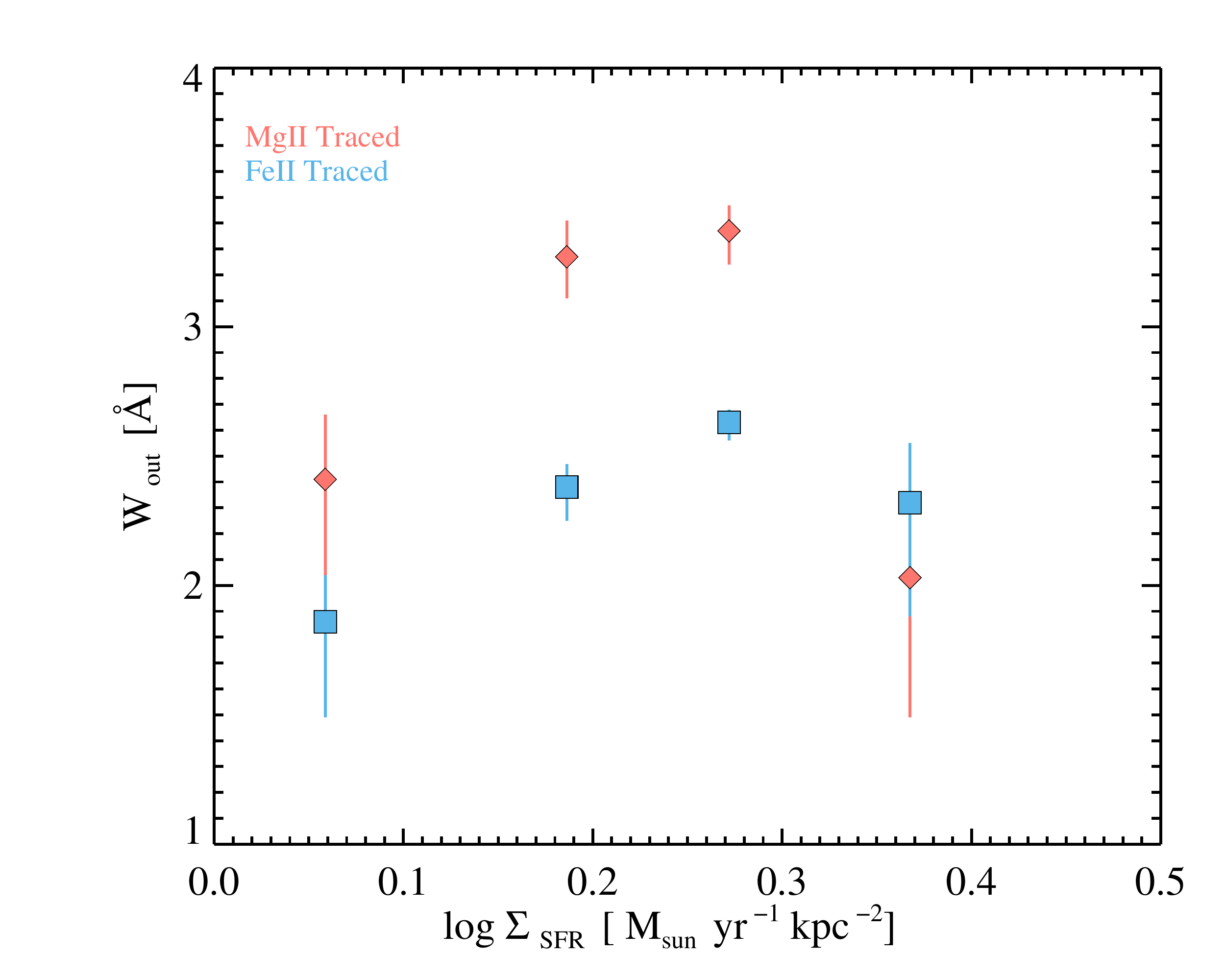}    
     \caption{Variation of outflow absorption and emission properties as a function of star-formation rate surface density ($\Sigma_{SFR}$) of the individual knots. Outflow velocity increases for increasing $\Sigma_{SFR}$. {\MgII} resonant emission equivalent width decreases with increasing $\Sigma_{SFR}$.}
\label{fig: sigsfr_properties}
\end{figure*}

 In this paper, we look at how outflow properties vary in the ``same'' galaxy as a function of the local $\Sigma_{SFR}$ of the star-forming regions. In Figure \ref{fig: sigsfr_properties}, the bottom panels show the variation of outflow velocities and equivalent widths with $\Sigma_{SFR}$. The red diamonds are measurement traced by {\MgII} absorption and the blue squares are measurements traced by {\FeII} absorption. The outflow velocities show a weak trend of increasing outflow velocities with increasing $\Sigma_{SFR}$.  A Pearson linear rank correlation test  shows a 2.4$\sigma$ correlation between $\log \; \Sigma_{SFR}$ and outflow velocities. This finding should be discussed in the context of the canonical star-formation rate surface density ``threshold'' of $\Sigma_{SFR} \rm{\; =\; 0.1\; M_{\odot} yr^{-1} kpc^{-2}}$. It has been argued that this is the minimum $\Sigma_{SFR}$ required for driving outflows in local starburst \citep{Heckman2002,Heckman2015}. Though we lack the dynamic range in $\Sigma_{SFR}$ to probe below this canonical threshold, we still notice a trend of increasing outflow velocities with localized  $\Sigma_{SFR}$. It should be noted that other studies observing the effect of $\Sigma_{SFR}$ on outflow properties have always used global flux-averaged $\Sigma_{SFR}$ measurements, as they could not resolve the individual star forming knots and the outflows separately. Here we are discussing the local $\Sigma_{SFR}$ properties of individual star-forming regions and the associated outflow velocities with these regions. Our findings are consistent with the  $z \sim 2$ \cite{Newman2012} results. However our current work is the first study that is observing the effects of $\Sigma_{SFR}$ and outflow velocities on sub-kpc scales within galaxies ($\leq$ 200 pc; \citealt{Wuyts2014}) scale.

\begin{figure*}
\centering
    \includegraphics[height=7.75cm,width=10.25cm]{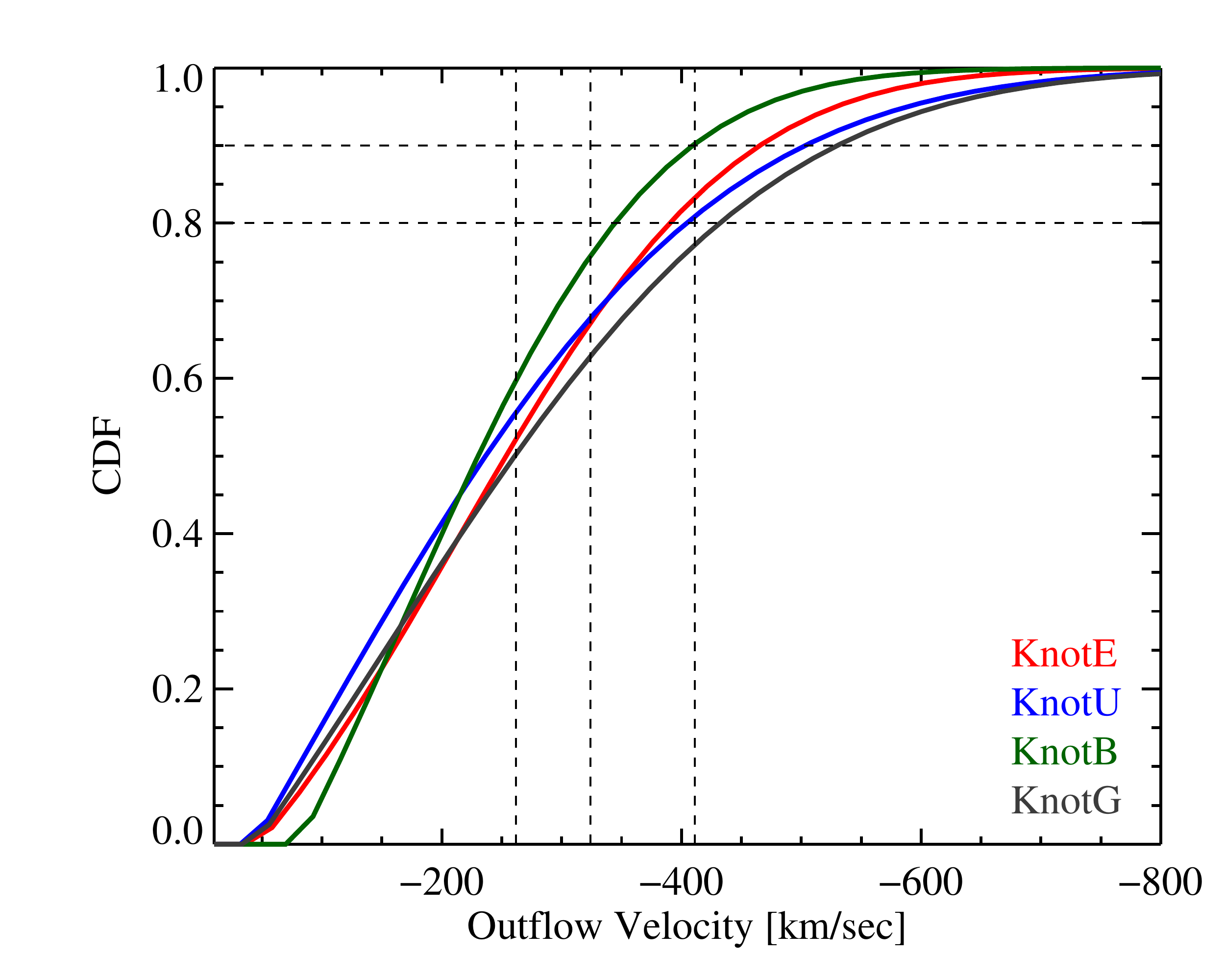}     
     \caption{Cumulative outflow pixel velocity distribution function examining how outflowing gas profiles vary. The horizontal dashed lines mark the velocities  which contain 80\% and 90\% of the blueshifted absorption, respectively. Bulk of the blueshifted outflowing absorption is confined within -400 km/sec for all cases. In case of Knot B, which exhibit the smallest $\Sigma_{SFR}$, 90\% of the blueshifted absorption occurs below -400 km/sec. The vertical dashed lines represent the escape velocity from the gravitational potential of the halo at R = 5 kpc, 20 kpc and 50 kpc respectively. Almost 20\% to 50\% of the blueshifted absorption has high enough velocity to escape.}
\label{fig: outflow_vel_pixel}
\end{figure*}

\begin{figure*}
\centering 
    \includegraphics[height=7.75cm,width=10.25cm]{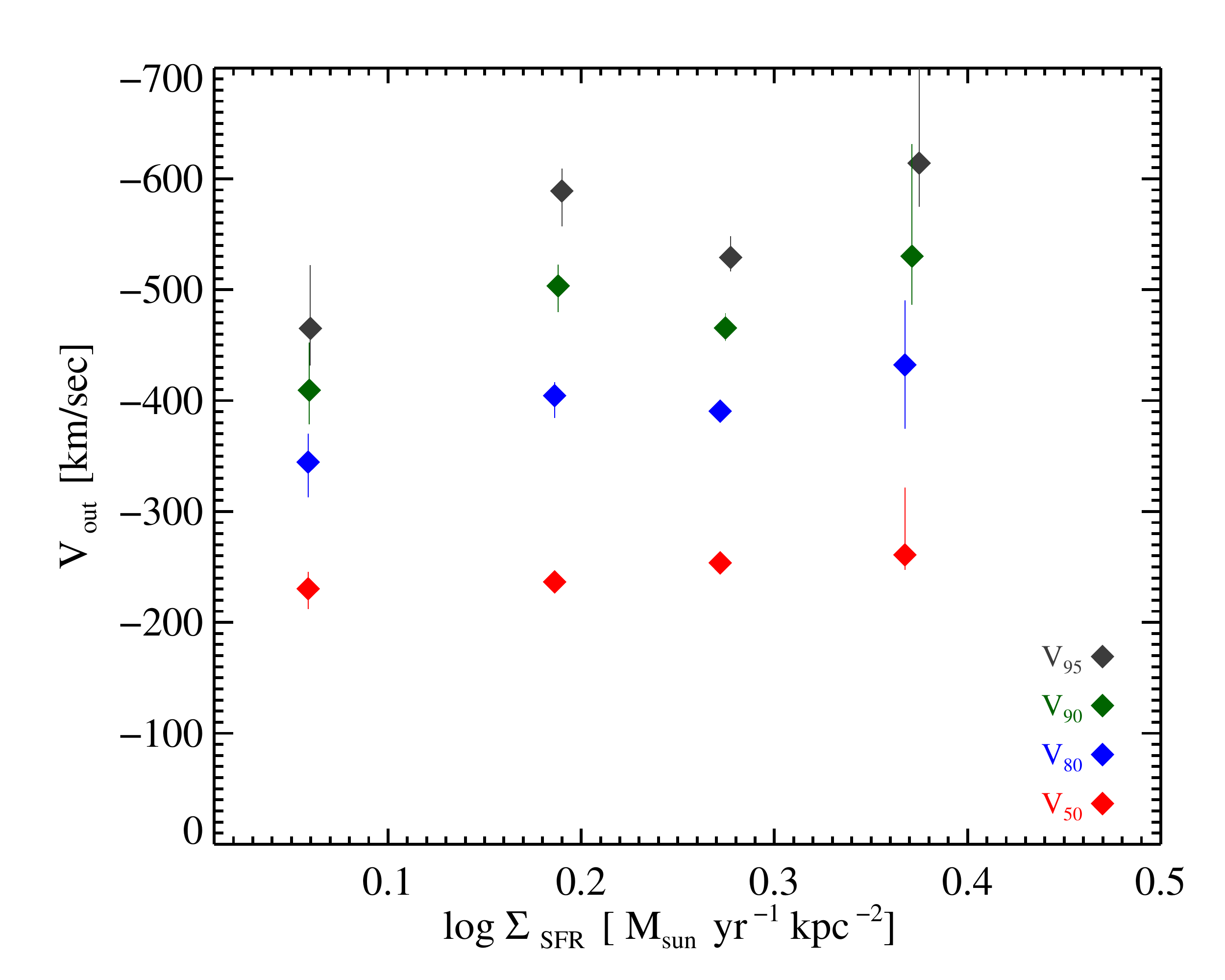}    
    \caption{Variation of cumulative outflow velocities with $\Sigma_{SFR}$ when the velocities account for 50\% blueshifted absorption (red diamonds), 80\% absorption (blue diamonds),90\% absorption (green diamonds), 95\% absorption (red diamonds). The dependence of outflow velocity with $\Sigma_{SFR}$ becomes more pronounced as we probe the high velocity tail of the outflowing gas. }
\label{fig: outflow_percentile}
\end{figure*}

\subsection{Pixel velocity distribution of outflowing gas}
In this section we examine the pixel velocity distribution of the blueshifed {\MgII} absorption. This approach allows us to probe the low column density, high velocity tail of the outflowing gas, as well as the bulk of the blueshifted gas as in Section 5. We use the fitted {\MgII} outflow absorption profile and compute the cumulative velocity distribution of the profile from 0 km/sec to -800 km/sec. Figure \ref{fig: outflow_vel_pixel} shows the cumulative pixel velocity distribution for the best fit {\MgII} profiles for each knot. This is an absorption weighted estimate of the amount of gas within a given outflowing velocity. The outflow velocity when the cumulative distribution function is 0.8 gives the maximum velocity below which 80\% of the observed {\MgII} absorption occurs (Figure \ref{fig: outflow_vel_pixel}, dashed line). 

We perform a two sample KS test, which rules out the null hypothesis that the pixel velocity distributions for all the knots are drawn from the same parent sample at 0.1\% significance level. This effectively tells us that the outflow velocity profiles are different for different star forming knots at 99.9\% confidence. This finding supports the picture that the properties of the galactic outflows, at least in this one $z=$ 1.7 galaxy, as being dominated by the properties of the nearest star-forming knots or star-clusters rather than the global properties of the galaxies. 

To estimate the absorption-weighted uncertainties on this cumulative velocity distribution, we compute the cumulative pixel velocity distribution for each realization of the MCMC chain. The mean velocity and the 1$\sigma$ velocity spread per pixel gives the estimated mean cumulative velocity and its uncertainty for each pixel. At the highest velocities, the uncertainties in mean velocity estimates rise, as they contain little absorption. Figure \ref{fig: outflow_percentile} shows the cumulative outflow velocities as a function of galaxy properties when these velocities account for 50\% (red diamonds), 80\% (blue diamonds), 90\% (green diamonds), and 95\% (black diamonds) of the total blueshifted absorption respectively. We see that the difference of outflow velocities between the knots with lowest and highest $\Sigma_{SFR}$ become more pronounced as we probe the higher velocity tail of the outflowing gas. All the knots have outflow velocities $>$ 300 km/sec for the top 20\% of the outflowing pixels (blue diamonds). For all the knots, the bulk of the absorption ($\approx$ 80\%), occurs within 400 km/sec of the systemic velocity of the galaxy. The escape velocities for the potential well of this galaxy at R = 5 kpc, 20 kpc and 50 kpc are given as  411 $\pm$ 74 {\kms}, 324 $\pm$ 59 {\kms} and 262 $\pm$ 48 {\kms}, respectively. These are shown as the vertical dashed lines in Figure \ref{fig: outflow_percentile}. We see that $\approx$ 20-50\% of the blueshifted absorption may escape the potential well of the halo.

\begin{figure*}
\centering
\includegraphics[width=7.25in,angle=0]{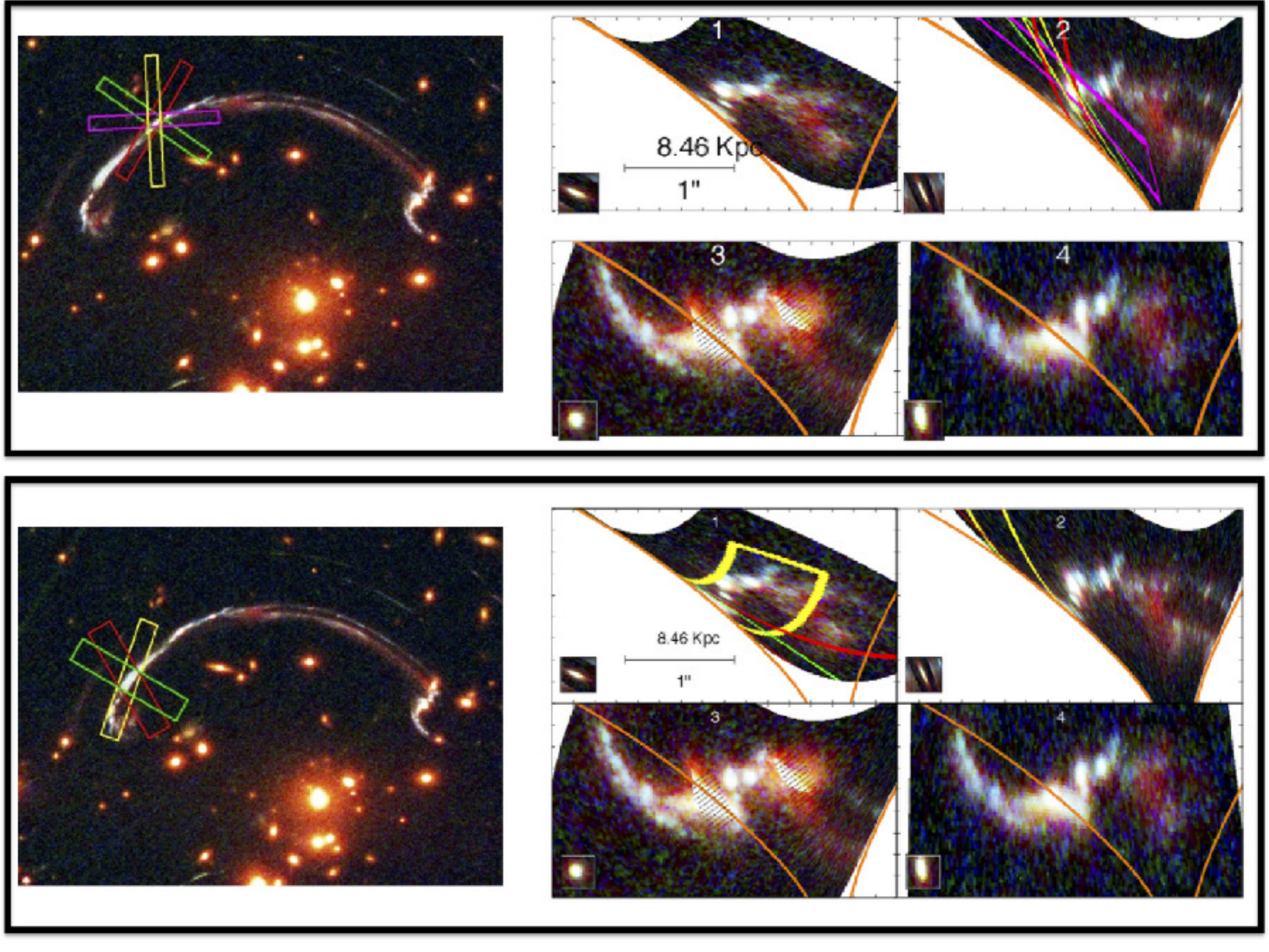}
\caption{Top left panel:  composite HST image of RCSGA 032727-132609, highlighting representative position angles for the MagE observations that targeted Knot E.  Top right panels:  source-plane reconstructions for each of the four magnified images of RCSGA 032727-132609.  Each sub panel is marked with the image number, which corresponds to the labels of the multiple images in Figure \ref{fig:finder_chart}. The representative MagE slits (pink, green, yellow, and red rectangles in the top left panel) are mapped to the source plane in the sub-panel for Image 2.  The hashed region in the source plane for image 3 marks contamination from two cluster galaxies, G1 and G2.  An orange line shows the source-plane caustic.
Bottom panels:  Same as top top, but for Knot U.  This figure is adapted from Figure 3 of \citet{Sharon2012}.}
\label{fig:knotE_slits}
\end{figure*}

\subsection{Spatial extent of M\lowercase{g}II emission region}

We now estimate the spatial extent of {\MgII} emission observed in these star-forming knots. We use the two-dimensional spectra to estimate how spatially extended are the observed {\MgII} emission regions. The galaxy {\arcname} is $\sim$ 15 kpc across along the long dimension (see Figure 5 of \citealt{Wuyts2014}).  As discussed in section 2 and shown in Figure~\ref{fig:finder_chart}, because the slit was rotated to track the parallactic angle, the observations cover a wide range of position angles in the image plane. Hence we cannot use the co-added two-dimensional spectrum of each knot to estimate spatial extent of {\MgII} emission. Instead, we perform this analysis on each exposure, since each exposure has a fixed position angle. Only in Knot U and Knot E, and only for the {\MgII} emission doublet, is the signal-to-noise high enough to meaningfully measure the spatial extent of the emission.  

In the two-dimensional spectrum of each exposure, we measure the full width at half maximum (FWHM) of the emission line along the spatial direction, summing over 5 pixels in the dispersion direction. We do the same for the nearby continuum emission, summing over 10 pixels in the dispersion direction. To translate these measurements from the image plane to the source plane we do the following.  For each exposure, we identify two points in the image plane that are located along the slit position angle, bracket the knot in question, and are located one FWHM apart as measured from the two dimensional spectrum.  We transform these image plane coordinates to the source plane, measure the distance between them, and call this the equivalent FWHM in the source plane. Figure \ref{fig:knotE_slits} illustrates that this linear approximation to the curved footprint of each slit in the source plane is adequate for the 
small sub-arcsecond scales over which the {\MgII} emission is spatially extended. Figure \ref{fig:knotE_slits} also shows that while the observations covered a wide range of position angles in the image plane, they are compressed to a much narrow range of position angles in the source plane.

\cite{Rubin2011a} chose to subtract the continuum emission from the 2D spectrum, then fit the residual line emission to a seeing-blurred model.  In our case, the continuum subtraction is negligible, as the line flux is much stronger per-pixel than the continuum flux. \cite{Rubin2011a} also dealt with a spatially--resolved galaxy. By contrast, our target star-forming knots are effectively unresolved in the slit spatial direction.  Therefore, we use the continuum FWHM as a proxy for the seeing, and by subtracting in quadrature the continuum FWHM from the observed {\MgII} emission line FWHM, we estimate the spatial extent of the {\MgII} emission.

For Knot E, in 9 of 10 exposures the FWHM of {\MgII} 2796, and in 7 of 9 cases for {\MgII} 2803, emission is larger than the FWHM of the continuum.  (In one additional exposure the emission profile is too noisy to fit.)  For Knot U, in 8 of 10 exposures the FWHM of {\MgII} 2796 emission is larger than that of
the continuum.  (Two other exposures have contaminating emission from other knots in the slit, and are therefore ignored.)  We find a mean FWHM
and error in the mean for {\MgII}~2796 of  $0.34 \pm 0.06$\arcsec\ ($2.9\pm 0.5$~kpc) for Knot E, and $0.30 \pm 0.05$\arcsec\ ($2.6 \pm 0.4$~kpc) for Knot U, both in the source plane. Hence the spatial extent up to which almost all the light from {\MgII} emission is seen (5$\sigma$ of the mean FWHM) is given as $R\;= 6.16 \pm 0.5$ kpc for Knot E  and $R\;= 5.52 \pm 0.4$ kpc for Knot U respectively. 

 Thus, in the spectra of Knot U and Knot E, the bulk of the Mg~II emission arises within 6~kpc of each knot, with the bulk of emission at smaller radii ($\sim 3$~kpc). These are smaller scales than those measured previously in {\MgII} emission. The field galaxies with the most spatially extended {\MgII} emission showed radii of  6-11  kpc (galaxy TKRS~4389, \citealt{{Rubin2011a}}); and $\sim11$~kpc (galaxy 32016857, \citealt{Martin2013}). The stacked spectrum of \cite{Erb2012} showed an excess at radii of 7~kpc.

 Thus, the {\MgII} emission in RCS0327 is extended on kpc scales, but the the emission is confined to smaller radii than seen previously in two galaxies and in a stack of 33 galaxies.  These results are not necessarily in conflict.  After all, the two galaxies of \cite{Rubin2011a} and \cite{Martin2013} were selected because they showed the most extended {\MgII} emission of their respective samples.  While the {\MgII} emission in RCS0327 is more compact than in the \cite{Erb2012} stack, the distribution of size scales within that stack is unknown; that is, after all, a limitation of stacking, and motivation for obtaining measurements in individual galaxies. 

\subsection{Minimum mass outflow rate}

In this section we report the minimum mass outflow rates for RCS0327. The mass outflow rate estimates are highly uncertain and depend on the assumed geometry of the outflowing gas. We always assume a this shell geometry to estimate the mass outflow rates.  As discussed in Section 3, our models fit the observations with typical outflow column densities of $N_{flow} \gtrsim 10^{14}\; \rm{cm^{-2}} $ for {\MgII} and $N_{flow} \gtrsim 10^{14.2}\; \rm{cm^{-2}} $ for {\FeII}. We estimate a conservative lower limit on the total hydrogen column density of the outflowing gas by assuming no ionization correction ($N_{MgII} = N_{Mg};\; N_{{\FeII}} = N_{Fe}$) and assume the minimum estimated column density for both {\MgII} and {\FeII} with no correction for line saturation. The nebular oxygen abundances for the clumps fall in the range of $8.02  \lesssim\; 12 + \log (O/H) \; \lesssim 8.28$ \citep{Wuyts2014}. This implies that the ISM of RCS0327 has abundances close to 27\% solar values (Solar abundance = 8.7; \citealt{2001ApJ...556L..63A}), with a spread of  $\approx$ 0.08 dex. Since the cool outflowing gas is likely to be entrained material swept up from the ISM, we adopt the ISM metallicity to be the metallicity of the cool outflowing gas. We assume a factor of -1.3 dex of Mg depletion onto dust and a factor of -1.0 dex of Fe depletion on dust \citep{Jenkins2009}. This allows us to estimate a lower limit of $N_{H}$ column densities for each knot. Using this we can compute the mass outflow rate for a wind with opening angle $\Omega_{w}$ and outflow velocity $v$ as 

\begin{equation}
\dot{M}_{out} \; =\;  \Omega_{w} C_{f} C_{\Omega} \mu m_{p} N_{H} R v.
\end{equation}

Here $C_{\Omega}$ is the angular covering fraction and $C_{f}$ is the clumpiness covering fraction, $\mu m_{p} $ is the mean atomic weight and $R$ is the spatial extent of the outflowing gas. Large scale galactic outflows are not spherical; but are preferentially observed along the minor axis of galaxies with an opening angle of $\approx$ 50-55 degrees \citep{Bordoloi2014b,Rubin2014a,Martin2012}. However, we are looking at local star-forming regions inside the galaxy, where it is safe to assume that gas is being swept up in spherical shells \citep{Heckman2002}. Hence we set the outflow angle $\Omega_{w} \;= 4\pi$ (thin shell approximation) and assume a smooth outflow ($C_{f}C_{\Omega} = 1$). Setting $\mu = 1.4$ we rewrite the mass outflow rate  for a thin shell geometry as (see \citealt{Weiner2009})

\begin{equation}
\rm{\dot{M}_{out} \;\gtrsim \; 22 M_{\odot} yr^{-1}  \frac{N_{H}}{10^{20} cm^{-2}}  \frac{R}{5 kpc}  \frac{v}{300 kms^{-1}}}  .
\end{equation}

The spatial extent ($R$) to be used to calculate the mass outflow rate depends upon the assumed model for the absorbing gas. In the idealized case of a thin shell-like structure, it is simply the radius of that shell. Constraining $R$ of the outflowing gas in absorption is not trivial. Down-the-barrel observation of galaxies can detect outflowing gas at any location along the line of sight to the target galaxy including material which is tens of kiloparsecs away from the host galaxy. Using stacked spectra of $\approx$ 4000 galaxies, \cite{Bordoloi2011a} have shown that outflowing gas traced by {\MgII} absorption can be detected out to $\gtrsim$ 50 kpc of the host galaxy at $z \sim 0.7$. However, most of such {\MgII} absorbers at impact parameters $>$ 20 kpc  have equivalent widths $<$ 2 {\AA}. Our outflow equivalent widths are typically $\gtrsim $ 2 {\AA}. Small numbers of such strong galaxy-absorber pairs are observed at impact parameters of $\approx$ 50 kpc  (e.g., \citealt{Nestor2007}). However, these surveys may misidentify the true galaxy counterparts if the emission from the host galaxy is blended with the bright QSO emission or is below the detection limit of the survey. Hence, we assume a conservative estimate for the spatial extent of the outflowing gas by assuming that the densest part of the outflowing gas can be traced by the spatial extent of the resonant {\MgII} emission lines. We assume the spatial extent of the {\MgII} emission lines estimated in Section 6 is the spatial extent of the outflowing gas. Since we only have information on Knot E ($R =$ 6.2 kpc) and Knot U ($R =$ 5.5 kpc), we assume that Knot B and Knot G have the same spatial extent as Knot E.

Given the uncertainties above, we note that the mass outflow rate calculations are highly uncertain and model dependent, and should only be taken as rough back of the envelope estimates, prone to systematic uncertainties. With that caveat in mind we report the minimum mass outflow rates below. The minimum mass outflow rate estimate for Knot E traced by the {\MgII} transition as $>$30$\rm{M_{\odot} yr^{-1}}$ and by the {\FeII} transition as $>$33$\rm{M_{\odot} yr^{-1}}$. The minimum mass outflow rate estimate for Knot G is given by the {\MgII} transition as $>$41$\rm{M_{\odot} yr^{-1}}$ and by the {\FeII} transition as $>$78$\rm{M_{\odot} yr^{-1}}$. The mass outflow rates are tabulated in Table \ref{table:mass_outflow}. Any such estimate of mass outflow rate is strictly a lower limit as we are neglecting any ionization correction and underestimating any effects of saturation. The true mass outflow rate could be easily a factor of ten higher than these observed values. These values are consistent with the values derived by \cite{Martin2012}. We find that the mass outflow rates are greater than the typical star formation rates of the individual knots and we can put lower limits on the mass loading factor $\rm{\eta}$  ($\rm{\dot{M}_{out} \; =\; \eta \;SFR}$). As an example; we find that $\rm{\eta_{\MgII}}$ for Knot E is $>$ 2.1 and for Knot G is $>$1. These mass loading factor estimates are consistent with the findings of \cite{Newman2012} at $z \sim 2$. They estimate a mass loading factor $\approx$ 2 for their $\rm{M* > 10^{10} M_{\odot}}$ galaxies.  All the mass outflow rate estimates are tabulated in Table \ref{table:mass_outflow}.
  
\begin{figure*}
    \includegraphics[height=8.25cm,width=8.8cm]{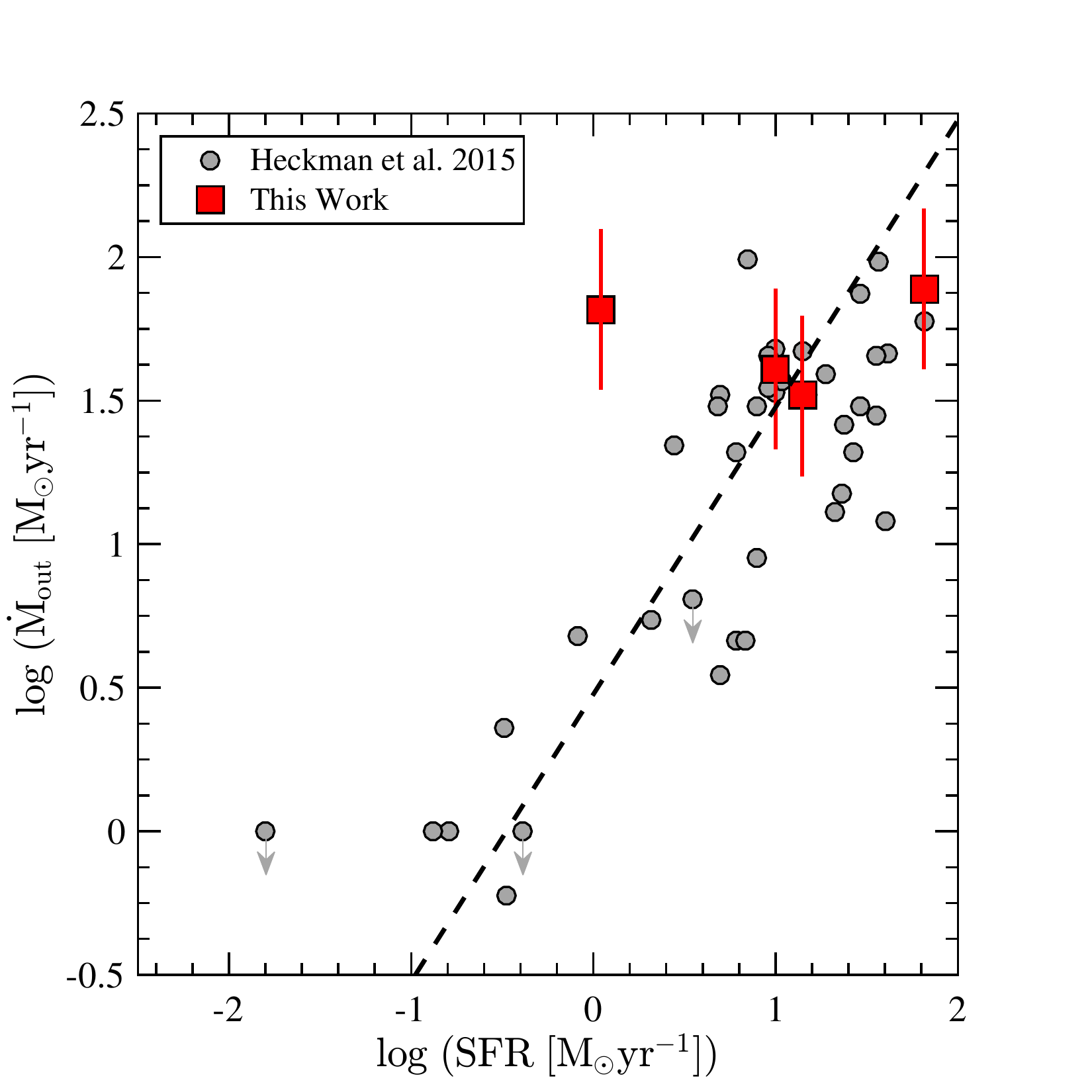}     
    \includegraphics[height=8.25cm,width=8.8cm]{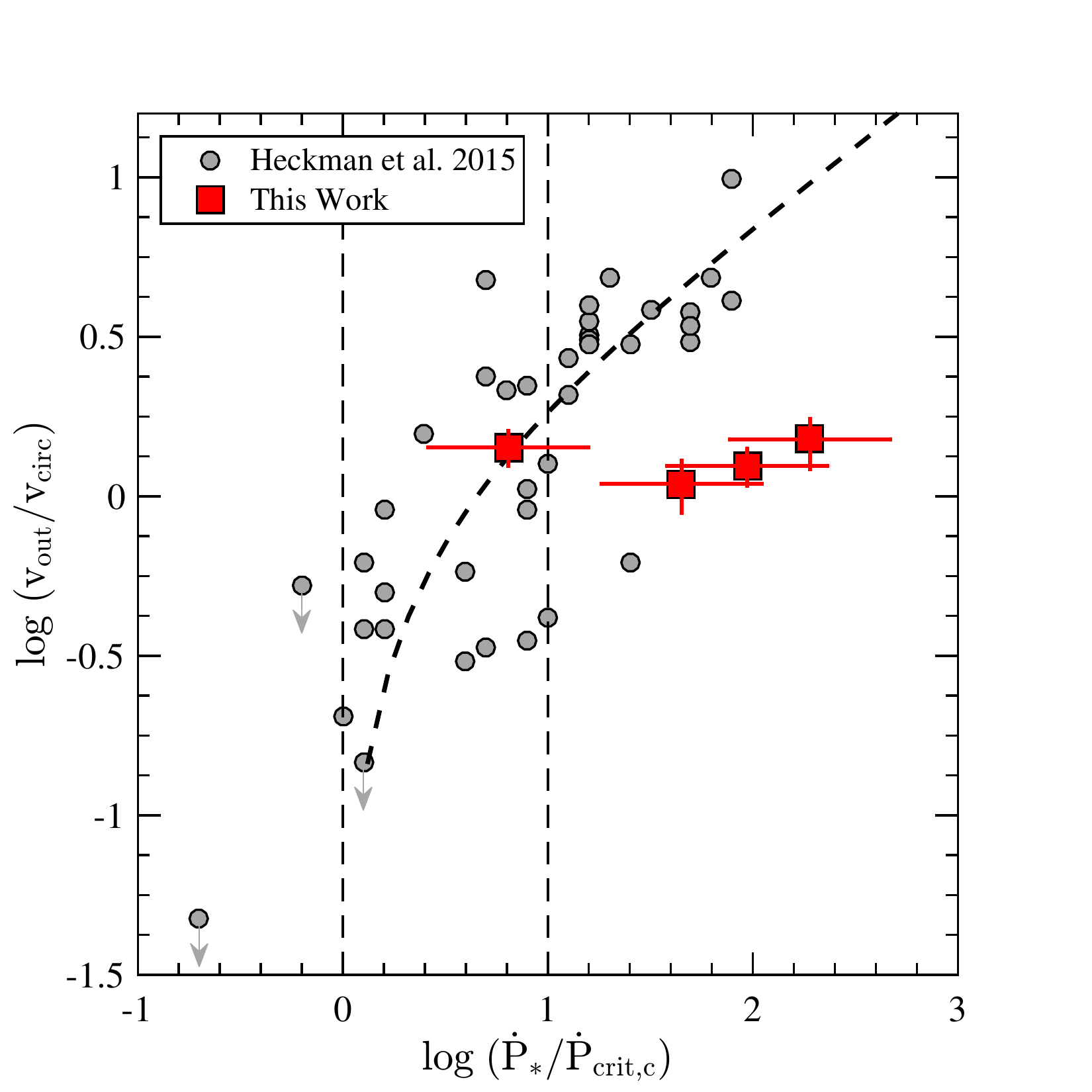}     
    \includegraphics[height=8.25cm,width=8.8cm]{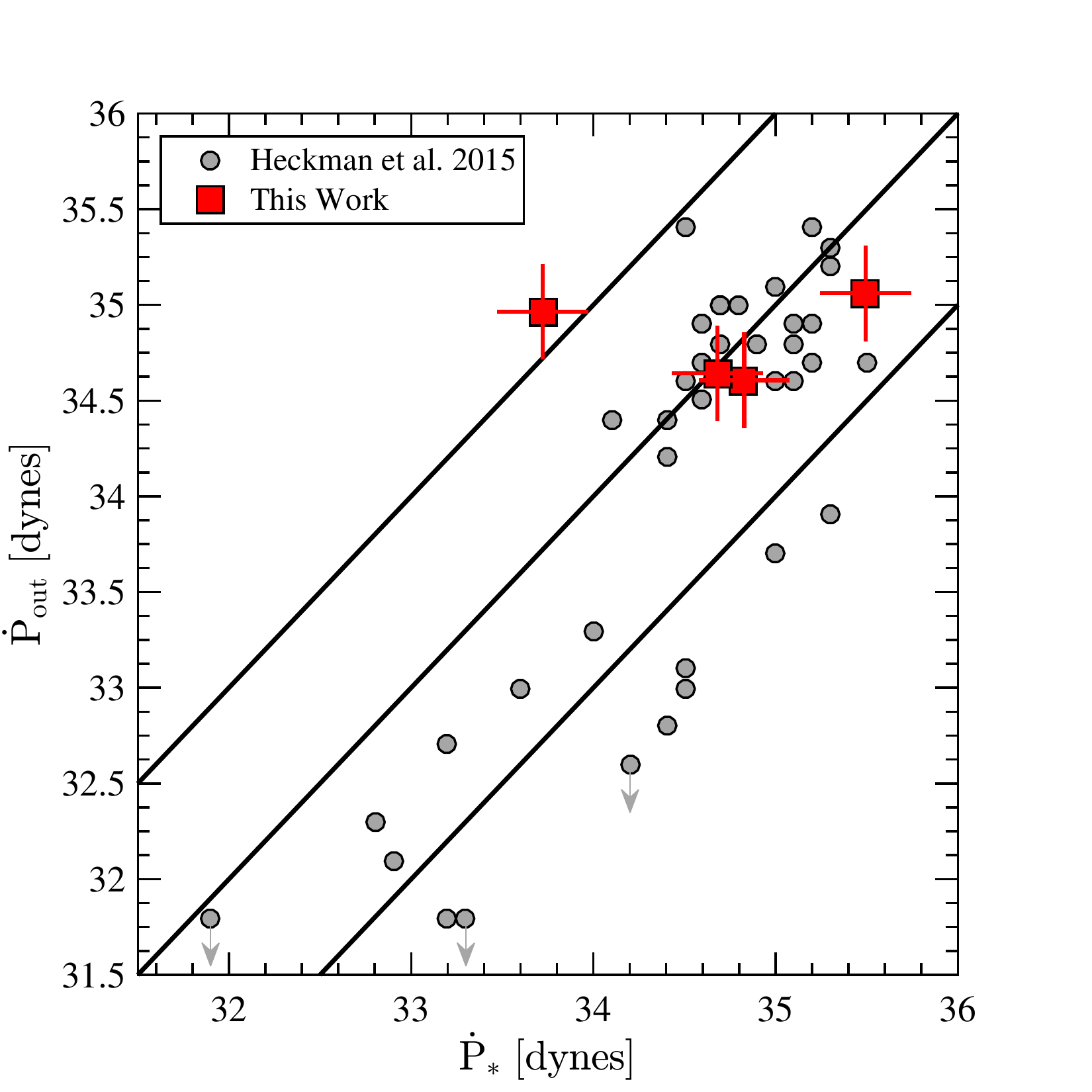}     
    
\caption{Top Left Panel: Mass outflow rate estimates plotted as a function of star formation rates. The gray points show the mass outflow rates for local star-burst galaxies from \citealt{Heckman2015}. The red points are mass outflow rates estimated in this work for individual star-forming knots in one $z$=1.7 galaxy. The dashed line shows the mass outflow rate when the mass loading factor $\eta$ =3. Top Right Panel: The observed normalized outflow velocities ($v_{out}/v_{circ}$) as a function of the ratio of the amount of momentum flux supplied by the starburst to the critical value needed to overcome gravity and drive an outflow. The local starburst galaxies (gray points) exhibit a range of momentum flux ratio driving outflows. The individual star-forming knots of a single high $z$ galaxy (red squares) have a range of momentum fluxes driving outflows, comparable to a set of star-burst galaxies. The dashed line is from the equation of motion for a momentum driven cloud model. The vertical dashed lines  show the ratio of $\dot{P}_{*}/\dot{P}_{crit}$ = 1 and 10.   Bottom Panel: Outflow momentum flux estimates as a function of rate of momentum supplied by the starburst. The diagonal lines show the ratios of $\dot{P}_{out}/\dot{P}_{*}$ = 10,1.0 and 0.1. The outflows driven by individual star-forming knots (red squares) are carrying the full momentum supplied by the starburst (see text). }
\label{fig: model_compare}
\end{figure*}

\subsection{Comparison with Local Starburst galaxies}
In this section we will compare the observed outflow properties in RCS0327 with that for local starburst galaxies and compare them with a simple idealized  analytical momentum driven wind model from \cite{Heckman2015}. They studied a sample of 39 local starburst galaxies, and used HST/COS and FUSE observations to estimate the outflow velocities and mass outflow rates for these galaxies. The gray points in Figure \ref{fig: model_compare} show these measurements. It should be noted that the mass outflow rate calculations are highly uncertain and should only be taken as roughly indicative estimates. We are assuming a statistical uncertainty on $\log \dot{M}_{out}$ of 0.3 dex, similar to \cite{Heckman2015} for comparison.

Figure \ref{fig: model_compare}, top left panel shows the relationship between mass outflow rates and SFR of these starburst galaxies (gray points). The mass outflow rates estimated in different regions of one $z=$1.7 galaxy (red squares) shows similar variation as seen in local starburst galaxies. For these estimates, we find that the typical $\rm{\eta}$ within a galaxy can vary from 1 to 4 times the local SFR. 

We estimate the momentum flux carried out by the outflowing gas as $\dot{P}_{out} = \dot{M}_{out} v_{out}$ dynes. These values are listed in Table \ref{table:momentum_flux}. We further estimate the total momentum flux supplied by a typical starburst due to a combination of hot wind fluid driven by the ejecta of massive stars and the radiation pressure. We follow \cite{Heckman2015}, and adopt an expression based on Starburst99 \citep{Starburst99} and write the total momentum flux in a starburst as $\dot{P}_{*}\; =\; SFR  \times \;4.8 \times 10^{33}$ dynes. Figure  \ref{fig: model_compare}, bottom panel, shows the distribution of momentum flux carried by the outflow and the total momentum flux in a starburst, for local starbursts (gray points) and this work (red squares). This shows that almost all of the momentum flux ($\sim$ 100\%) supplied by the star forming knots are being carried out by the observed outflows. Outflows driven by individual star-forming knots are as forceful as the outflows driven by local starbursts ($\dot{P}_{out}\; >\; 10^{34} $ dynes, \citealt{Heckman2015}). 

We now compare the data to a simple analytical model described in \cite{Heckman2015}. This model assumes that absorption lines are produced by a distribution of clouds that are being driven out of the galaxy by the momentum supplied by a starburst as described above. Following \cite{Heckman2015} we define a critical momentum flux which is needed for gas clouds to escape the local star forming regions of radius $r_{*}$ given as $\dot{p}_{crit}\; =\; 4\pi r_{*} N_{c} \langle m \rangle v_{circ}^{2}$. Here $r_{*}$ is the size of the star forming knots \citep{Wuyts2014} and the values are listed in Table \ref{table:momentum_flux}. $N_{c}$ is the cloud neutral hydrogen column density. 

Figure \ref{fig: model_compare}, top right panel shows the normalized outflow velocities ($v_{out}/v_{circ}$) as a function of the ratio of  the amount of momentum flux supplied by the starburst to the critical value needed to overcome gravity and drive an outflow ($\dot{p}_{*}/\dot{p}_{crit}$) for local starbursts (gray points) and this work (red squares). We estimate the outflow velocity for this model from equation 5 of \cite{Heckman2015}, and this is shown as the dashed line. As compared to the local starbursts, outflows driven from this one $z =1.7$ galaxy produces stronger outflows ($\dot{p}_{*}/\dot{p}_{crit} \gtrsim 10.$). However the normalized outflow velocities saturate for the strongest outflows. Bulk of the mass observed in these stronger outflows at $z=$ 1.7 exceed the escape velocity of the halo. If we underestimate $\dot{p}_{crit}$, as the values of  $N_{c}$ is uncertain, the ratio $\dot{p}_{*}/\dot{p}_{crit}$ will change and the red squares in Figure \ref{fig: model_compare}, top right panel will be more consistent with the local observations.

In RCS0327 the knot-to-knot variations in mass outflow rates, momentum flux and $\dot{p}_{*}/\dot{p}_{crit}$ are comparable to that observed in a sample of starburst galaxies in the local Universe. This certainly need not have been the case; if the observed outflows were well-mixed and large ($ >>1$ kpc), we would have measured the same properties for each knot. We thus conclude that, at least in this one galaxy, we are seeing a ``locally sourced outflows'', i.e.  the outflow properties are determined locally by nearby star-forming regions, and are not uniform on kiloparsec scales.

\begin{table*}
 \begin{minipage}{175mm}
 \begin{center}
            Properties of the individual knots of RCS0327 \\
            \renewcommand\tabcolsep{4.pt}
  \begin{tabular}{crrrrrrr}

  \hline
  Knot &   $\log M*/M_{\odot}$     &     SFR                             &  $\rm{ \Sigma_{SFR}}$                      & $z_{sys} \pm z_{error}$ & FWHM   & Instrument & Offset\\
 &                                     & [$\rm{M_{\odot}\; yr^{-1}}$] &  $[\rm{ \;M_{\odot} yr^{-1}kpc^{-2}}$] &                  & [km/sec] & & [km/sec]  \\
 \hline
Knot E    &   $\rm{8.5^{+0.1}_{-0.4}}$ &  $\rm{14 \pm 0.4}  $  &     $\rm{1.87 \pm 0.06}  $       & $1.7037455 \pm  0.000005$       &   109    & NIRSPEC      &     0    \\
Knot U\footnotetext{$^1$ The magnification of knot U is extremely high and therefore extremely uncertain.  This uncertainty is not fully captured by the errorbar.}  &  $\rm{ 7.5}^{1}$ &  $\rm{1.1 \pm 0.1} $  &     $\rm{1.54 \pm 0.24}  $       &$1.703884 0 \pm  0.000007$       &    128     &    NIRSPEC    &     15  \\
Knot B    &  $\rm{ 7.7^{+0.5}_{-0.2}}$ &  $\rm{10 \pm 0.4}  $  &      $\rm{1.14 \pm 0.28}  $      &$1.7036000  \pm 0.000070$      &     109     &   OSIRIS    &    -16 \\
Knot G    &  $\rm{ 8.7^{+0.5}_{-0.1}}$ &  $\rm{65 \pm 3.0}  $  &      $\rm{2.33 \pm 0.11}  $      & $1.7038500   \pm  0.000050$    &     108      &    OSIRIS   &   12   \\
\end{tabular}
\caption{ The physical properties of the individual knots of RCS0327. Nebular redshifts from the H$\alpha$ emission line, from the following references:  1) the Keck/NIRSPEC spectra of \citet{Rigby11} re-analyzed by \citet{Wuyts2014}; and 2) the Keck/OSIRIS spectra of \citet{Wuyts2014}.  For convenience, the last column gives the velocity offset with respect to Knot E. SFR is measured from H$\beta$ luminosity, following \citet{Kennicutt1998}, assuming a Chabrier IMF, corrected for extinction and outflows, and demagnified for lensing. The associated uncertainty is the standard deviation when images of multiple knots could be measured. Otherwise the uncertainty quoted is the statistical uncertainty, which mostly comes from the line flux uncertainty.}
\label{table:Knot_properties}
\end{center}
\renewcommand\tabcolsep{6pt}

\end{minipage}
\end{table*}

\begin{table*}
\centering
{\MgII} and {\FeII} Transitions studied \\
  \begin{tabular}{crrrrr}
  \hline
  Transition &   $\lambda^{a}$     &     $J_{high}^{b}$  &   $J_{low}^{b}$ &                $A^{c}$ & Feature\\
                   &  [{\AA}]                  &                               &                         &         [$sec^{-1}$]   &\\
       \hline
{\MgII}                   &  2796.351   &       3/2                    &   0                    &        2.63E+08     &  Resonant abs/em   \\
                             &  2803.528   &       1/2                    &   0                    &        2.60E+08     &  Resonant abs/em   \\
    \hline
{\FeII}                   &  2600.173   &       9/2                    &   9/2                  &        2.36E+08     &  Resonant abs   \\
                             &  2586.650   &      7/2                    &   9/2                    &        8.61E+07     &   Resonant abs   \\
                             &  2626.451   &      9/2                    &   7/2                    &        3.41E+07     &   Fine-structure em   \\
                             &  2612.654   &      7/2                    &   7/2                    &        1.23E+07     &   Fine-structure em   \\    
\hline
\end{tabular}
\caption{$^{a}$ Vacuum wavelengths from \citet{Morton2003}. $^b$ Orbital angular momenta. $^{c}$ Einstein $A$-coefficients.}
\label{table:Line_List}
\end{table*}

%
 \begin{table*}
 \begin{center}
  Measured absorption and emission kinematics and strengths of the individual knots of RCS0327 \\
  \begin{tabular}{crrrrr}
  \hline
Transition & Measurement           &                 Knot E             &              Knot U                 &                Knot B             & Knot G \\
\hline
{\MgII}    & $\rm{V_{out} (km/sec)}$ &  $\rm{-225^{+4}_{-7}}$        & $\rm{-233^{+10}_{-7}}$         & $\rm{-183^{+16}_{-20}}$   & $\rm{-251^{+33}_{-33}}$    \\
{\MgII}    & $\rm{W_{out} /{\AA}}$ &  $\rm{3.37^{+0.10}_{-0.13}}$ & $\rm{3.27^{+0.14}_{-0.16}}$ & $\rm{2.41^{+0.25}_{-0.46}}$   & $\rm{2.03^{+0.44}_{-0.54}}$    \\
{\MgII}    & $\rm{\log N /cm^{-2}}$ &  $\rm{14^{+0.02}_{-0.0.03}}$ & $\rm{14.16^{+0.03}_{-0.03}}$ & $\rm{13.9^{+0.11}_{-0.08}}$   & $\rm{13.9^{+0.12}_{-0.20}}$    \\
{\MgII}    & $\rm{V_{em} (km/sec)}$ &  $\rm{158^{+27}_{-48}}$        & $\rm{70^{+9}_{-14}}$         & $\rm{208^{+18}_{-93}}$   & $\rm{155^{+7}_{-36}}$    \\
{\MgII}    & $\rm{W_{em} /{\AA}}$ &  $\rm{-4.87^{+0.87}_{-1.18}}$ & $\rm{-5.96^{+0.43}_{-0.95}}$ & $\rm{-11.19^{+2.32}_{-2.14}}$   & $\rm{-3.05^{+0.91}_{-1.53}}$    \\
{\FeII}    & $\rm{V_{out} (km/sec)}$ &  $\rm{-195^{+3}_{-4}}$        & $\rm{-222^{+14}_{-13}}$         & $\rm{-171^{+20}_{-20}}$   & $\rm{-235^{+30}_{-39}}$    \\
{\FeII}    & $\rm{W_{out} /{\AA}}$ &  $\rm{2.63^{+0.05}_{-0.07}}$ & $\rm{2.38^{+0.09}_{-0.13}}$ & $\rm{1.86^{+0.18}_{-0.37}}$   & $\rm{2.32^{+0.23}_{-0.44}}$    \\
{\FeII}    & $\rm{\log N /cm^{-2}}$ &  $\rm{14.34^{+0.02}_{-0.02}}$ & $\rm{14.52^{+0.02}_{-0.04}}$ & $\rm{14.23^{+0.12}_{-0.14}}$   & $\rm{14.42^{+0.11}_{-0.10}}$    \\
{\FeII}2612    & $\rm{V_{em} (km/sec)}$ &  $\rm{-66^{+12}_{-16}}$        & $\rm{-14^{+15}_{-12}}$         & $\rm{-60^{+32}_{-18}}$   & $\rm{61^{+25}_{-89}}$    \\
{\FeII}2612    & $\rm{W_{em} /{\AA}}$ &  $\rm{-0.95^{+0.09}_{-0.08}}$ & $\rm{-1.02^{+0.14}_{-0.13}}$ & $\rm{-0.63^{+0.23}_{-0.26}}$   & $\rm{-0.63^{+0.28}_{-0.30}}$    \\
{\FeII}2626    & $\rm{V_{em} (km/sec)}$ &  $\rm{-30^{+5}_{-5}}$        & $\rm{-22^{+7}_{-5}}$         & $\rm{-29^{+15}_{-15}}$   & $\rm{-30^{+28}_{-16}}$    \\
{\FeII}2626    & $\rm{W_{em} /{\AA}}$ &  $\rm{-1.17^{+0.08}_{-0.10}}$ & $\rm{-1.77^{+0.15}_{-0.13}}$ & $\rm{-1.19^{+0.20}_{-0.38}}$   & $\rm{-0.80^{+0.19}_{-0.28}}$    \\
\end{tabular}
\caption{ The measured {\MgII} and {\FeII} outflow kinematics, equivalent widths and column densities as well as the {\MgII} resonant emission and {\FeII} 2612, 2626 fluorescence emission kinematics and strengths. }
\label{table:Result_summary}
\end{center}
\end{table*}

\begin{table}
 \begin{center}
            Mass outflow rates  [$\rm{M_{\odot} \;yr^{-1}}$ ]\\
  \begin{tabular}{crrrr}
  \hline
  Knot &   {\MgII} Traced     &     {\FeII} Traced   &  $\eta_{{\MgII}}$& $\eta_{{\FeII}}$ \\
       \hline
Knot E    & 30          & 33        & 2.1   & 2.4 \\
Knot U    & 51          & 66        & 46  & 60. \\
Knot B    & 34          & 41        & 3.4  & 4.1 \\
Knot G    & 41         & 78        & 0.64  & 1.2 \\
\end{tabular}
\caption{Minimum mass outflow rates for each star-forming knots. $\eta$ gives the lower limits on the mass loading factor for each knot as traced by {\MgII} and {\FeII} transitions respectively.}
\label{table:mass_outflow}
\end{center}
\end{table}

\begin{table}
 \begin{center}
 Momentum Flux \\
  \begin{tabular}{crrr}
  \hline
  Knot &   $\log \; \dot{P}_{out}^{1}$     &     $\log \; \dot{P}_{*}^{2}$  &   $\log \; \dot{P}_{crit,c}^{3}$\\
          &  [$\log$ dynes]      &     [$\log$ dynes]   &   [$\log$ dynes] \\
       \hline
Knot E    & 34.6          & 34.8        & 32.9   \\
Knot U    & 34.9          & 33.7        & 32.9  \\
Knot B    & 34.6         & 34.7        & 33.0   \\
Knot G    & 35.0         & 35.5        & 32.2  \\
\end{tabular}
\caption{$^{1}$ The momentum flux in the outflows. The uncertainties will be dominated by mass outflow rate estimates. $^2$ The momentum flux from starburst proportional to the star formation rate (see text). The uncertainties are hard to quantify \citep{Starburst99}. $^{3}$ The critical momentum flux in a starburst that is required to escape gravity for a cloud at the radius of the star-forming knots.}
\label{table:momentum_flux}
\end{center}
\end{table}

\section{Conclusions}
In this paper we present the rest-frame ultraviolet spectra of  {\MgII} and {\FeII} emission and absorption lines along four lines of sight separated by up to 6~kpc in a gravitationally lensed $z=1.70$ galaxy, targeting four bright regions.  Such observations are challenging, and only possible with current telescopes with the assistance of strong gravitational lensing and a favorable lensing configuration. While RCS0327 is one of the brightest lensed galaxies known, there are hundreds of fainter lensed galaxies in which this experiment could be repeated with upcoming 20-30~m ground-based optical telescopes. The main finding of this work are as follows:

\begin{center}
\begin{itemize}

\item All four individual star forming knots are driving outflows traced by blueshifted {\MgII} and {\FeII} absorption profiles. The outflow velocities traced by blueshifted {\MgII} absorption range from $\approx$ -183 km/sec to -251 km/sec; and outflow velocities traced by blueshifted {\FeII} absorption range from $\approx$ -171 km/sec to -235 km/sec. The rest frame {\MgII} outflow equivalent widths range from 2.03 {\AA} to 3.37 {\AA}; and the rest frame {\FeII} outflow equivalent widths range from 1.86 {\AA} to 2.63 {\AA} respectively.

\item Individual star forming knots with higher local star-formation rate surface densities ($\Sigma_{SFR}$) exhibit higher outflow velocities. This is consistent with the picture of outflows being driven due to local star-formation activity in star clusters. We find a 2.4$\sigma$ correlation between outflow velocities and the $\Sigma_{SFR}$ of the individual knots. 

\item We detect resonant {\MgII} emission in all the four spectra studied in this paper. The {\MgII} emission velocity centroids are always redshifted relative to the systemic zero velocity of the individual knots. This is consistent with the scenario that the physical origin of the {\MgII} emission doublet is in photons scattering off the backside of galactic winds. 

\item Star forming knots with smaller $\Sigma_{SFR}$ exhibit the strongest {\MgII} emission equivalent widths. Knot G with the highest $\Sigma_{SFR}$, exhibits a {\MgII} emission equivalent width  $\sim$ 3.7 times smaller than Knot B; which has the lowest  $\Sigma_{SFR}$ of all the four knots.

\item We detect {\FeII} 2612, 2626 fluorescence emission lines for all the four knots whereas we do not detect any resonant {\FeII} emission lines. The {\FeII} 2612, 2626 emission lines are either slightly blueshifted or are consistent with the systemic zero velocities of the individual clumps. 

\item We estimate the spatial extent of the {\MgII} emission lines from the extended two-dimensional emission profile. We estimate the FWHM of the extended {\MgII} 2796 emission at the source plane to be $0.34 \pm 0.06$\arcsec\ ($2.9\pm 0.5$~kpc) for Knot E, and $0.30 \pm 0.05$\arcsec\ ($2.6 \pm 0.4$~kpc) for Knot U respectively.

\item Assuming a thin shell geometry, we estimate the minimum mass outflow rates ($\dot{M}_{out}$), which range from $>$ 30 to $>$51 $M_{\odot} yr^{-1}$ as estimated from {\MgII} absorption and  $>$ 33 to $>$78 $M_{\odot} yr^{-1}$ as estimated from {\FeII} absorption. These are conservative lower limits as we are not accounting for line saturation and ionization corrections. Assuming $\dot{M}_{out} \; \propto $ SFR, we find that the mass loading factor $\eta$ is a few times the SFR of the individual star forming knots. This is the first time that $\eta$s have been measured for multiple individual regions within the same galaxy in the distant Universe, and is a crucial observable to models of how galaxies process their gas.  

\item  Given the low circular velocity inferred for two galaxies with the stellar mass of the merging pair of galaxies RCS0327; almost  20-50\% of the blueshifted absorption may escape the gravitational potential of the galaxy.

\item The knot-to-knot variations in mass outflow rates, momentum flux and $\dot{p}_{*}/\dot{p}_{crit}$ in RCS0327 are comparable to starburst galaxies in local Universe. This is the first comparison of mass-loading factor and wind momentum in outflows at z$\sim$2 to outflows in the local Universe. The observed knot to knot variation in these quantities argues for the importance of localized sources in the galaxies in driving outflows in the Universe.

The spectra for RCS0327 show that {\MgII} emission is occurring across the galaxy, but that the properties of the outflow are heavily influenced by the nearest star forming regions.  {\MgII} emission is seen in each of the four knots we targeted, separated by up to 6 kpc. In the two knots with high signal-to-noise spectra, the {\MgII} emission is more spatially extended than the continuum, out to 6~kpc scales.  It is clearly not the case that only one region is producing all the {\MgII} emission in this galaxy.  However, (section 4.2) the outflow properties correlate with the nearest star forming region, and (section 4.3) the outflow velocity profile varies from knot to knot, across separations of 0.4 to 6 kpc in the source plane (see Figure 5 of \citealt{Wuyts2014}). Thus, while the emission in RCS0327 extends out to $\sim$6 kpc from knots E and U, the bulk of the emission arises from $<$3~kpc, on the same scale as the $\sim$2~kpc spacing between knots.  Thus, though the outflow in RCS0327 does extend to galaxy-wide scales,  the properties of the outflow appear to be shaped most strongly by the closest star-forming complex.

From knot to knot, we find large variations in the strength of the {\MgII} emission, the velocity peak of that emission, and in the velocity profiles of the inferred outflows. In RCS0327, the outflowing winds appear quite different on lines of sight separated by only a few kiloparsecs. This did not have to be the case; if the outflow were well-mixed and large, $>>$1 kpc, we would have measured the same outflow properties for each knot. That is very different from what we observed. We thus conclude that, at least in this one galaxy, we are seeing ``locally sourced outflows''---the outflows must be close to the star-forming regions, and still bear the velocity signatures of the gas flowing out of those regions, such that the different sight lines encounter quite different outflows.  At least in this one galaxy, the outflow properties are determined locally by nearby star-forming regions, and are not uniform on kiloparsec scales.  Repeating this experiment in other lensed galaxies with multiple sight lines will test the universality of this result. If they also show significant line of sight variation, then it will be clear that studies of outflows in distant galaxies must spatially resolve them.
\end{itemize}
\end{center}

\section{Acknowledgments}
RB would like to thank the Aspen Center for Physics (supported by NSF grant 1066293) for hospitality and productive atmosphere and organizers of the workshop ``Physics of Accretion and Feedback in the Circum-Galactic Medium''  in June 2015, for opportunities to discuss results presented here during completion of this paper. RB would like to thank Timothy Heckman for stimulating discussions on this work. RB would also like to thank Peter Behroozi for providing mock catalogues to estimate halo mass of the host galaxy. Partial support for this work was provided by NASA through Hubble Fellowship grant $\#$51354 awarded by the Space Telescope Science Institute, which is operated by the Association of Universities for Research in Astronomy, Inc., for NASA, under contract NAS 5-26555. This paper includes data gathered with the 6.5 meter Magellan Telescopes located at Las Campanas Observatory, Chile. Magellan observing time for this program was granted by the telescope allocation committees of the Carnegie Observatories, the University of Chicago, the University of Michigan, and Harvard University.

\bibliographystyle{mnras}
\bibliography{mybibliography}

\begin{thebibliography}{}
\makeatletter
\relax
\def\mn@urlcharsother{\let\do\@makeother \do\$\do\&\do\#\do\^\do\_\do\%\do\~}
\def\mn@doi{\begingroup\mn@urlcharsother \@ifnextchar [ {\mn@doi@}
  {\mn@doi@[]}}
\def\mn@doi@[#1]#2{\def\@tempa{#1}\ifx\@tempa\@empty \href
  {http://dx.doi.org/#2} {doi:#2}\else \href {http://dx.doi.org/#2} {#1}\fi
  \endgroup}
\def\mn@eprint#1#2{\mn@eprint@#1:#2::\@nil}
\def\mn@eprint@arXiv#1{\href {http://arxiv.org/abs/#1} {{\tt arXiv:#1}}}
\def\mn@eprint@dblp#1{\href {http://dblp.uni-trier.de/rec/bibtex/#1.xml}
  {dblp:#1}}
\def\mn@eprint@#1:#2:#3:#4\@nil{\def\@tempa {#1}\def\@tempb {#2}\def\@tempc
  {#3}\ifx \@tempc \@empty \let \@tempc \@tempb \let \@tempb \@tempa \fi \ifx
  \@tempb \@empty \def\@tempb {arXiv}\fi \@ifundefined
  {mn@eprint@\@tempb}{\@tempb:\@tempc}{\expandafter \expandafter \csname
  mn@eprint@\@tempb\endcsname \expandafter{\@tempc}}}

\bibitem[\protect\citeauthoryear{{Adelberger}, {Shapley}, {Steidel}, {Pettini},
  {Erb}  \& {Reddy}}{{Adelberger} et~al.}{2005}]{Adelberger2005}
{Adelberger} K.~L.,  {Shapley} A.~E.,  {Steidel} C.~C.,  {Pettini} M.,  {Erb}
  D.~K.,   {Reddy} N.~A.,  2005, \mn@doi [ApJ] {10.1086/431753}, \href
  {http://adsabs.harvard.edu/abs/2005ApJ...629..636A} {629, 636}

\bibitem[\protect\citeauthoryear{{Allende Prieto}, {Lambert}  \&
  {Asplund}}{{Allende Prieto} et~al.}{2001}]{2001ApJ...556L..63A}
{Allende Prieto} C.,  {Lambert} D.~L.,   {Asplund} M.,  2001, \mn@doi [ApJL]
  {10.1086/322874}, \href {http://adsabs.harvard.edu/abs/2001ApJ...556L..63A}
  {556, L63}

\bibitem[\protect\citeauthoryear{{Asplund}, {Grevesse}, {Sauval}  \&
  {Scott}}{{Asplund} et~al.}{2009}]{Asplund09}
{Asplund} M.,  {Grevesse} N.,  {Sauval} A.~J.,   {Scott} P.,  2009, \mn@doi
  [ARA\&A] {10.1146/annurev.astro.46.060407.145222}, \href
  {http://adsabs.harvard.edu/abs/2009ARA%26A..47..481A} {47, 481}

\bibitem[\protect\citeauthoryear{{Behroozi}, {Wechsler}  \& {Wu}}{{Behroozi}
  et~al.}{2013}]{Behroozi2013}
{Behroozi} P.~S.,  {Wechsler} R.~H.,   {Wu} H.-Y.,  2013, \mn@doi [ApJ]
  {10.1088/0004-637X/762/2/109}, \href
  {http://adsabs.harvard.edu/abs/2013ApJ...762..109B} {762, 109}

\bibitem[\protect\citeauthoryear{{Bell}, {McIntosh}, {Katz}  \&
  {Weinberg}}{{Bell} et~al.}{2003}]{Bell2003}
{Bell} E.~F.,  {McIntosh} D.~H.,  {Katz} N.,   {Weinberg} M.~D.,  2003, \mn@doi
  [ApJS] {10.1086/378847}, \href
  {http://adsabs.harvard.edu/abs/2003ApJS..149..289B} {149, 289}

\bibitem[\protect\citeauthoryear{{Binney} \& {Tremaine}}{{Binney} \&
  {Tremaine}}{1987}]{Galactic_Dynamics}
{Binney} J.,  {Tremaine} S.,  1987, {Galactic dynamics}.
Princeton University Press

\bibitem[\protect\citeauthoryear{{Blanton} et~al.,}{{Blanton}
  et~al.}{2003}]{Blanton2003}
{Blanton} M.~R.,  et~al., 2003, \mn@doi [ApJ] {10.1086/375776}, \href
  {http://adsabs.harvard.edu/abs/2003ApJ...592..819B} {592, 819}

\bibitem[\protect\citeauthoryear{{Bordoloi} et~al.,}{{Bordoloi}
  et~al.}{2011}]{Bordoloi2011a}
{Bordoloi} R.,  et~al., 2011, \mn@doi [ApJ] {10.1088/0004-637X/743/1/10}, \href
  {http://adsabs.harvard.edu/abs/2011ApJ...743...10B} {743, 10}

\bibitem[\protect\citeauthoryear{{Bordoloi}, {Lilly}, {Kacprzak}  \&
  {Churchill}}{{Bordoloi} et~al.}{2014a}]{Bordoloi2012a}
{Bordoloi} R.,  {Lilly} S.~J.,  {Kacprzak} G.~G.,   {Churchill} C.~W.,  2014a,
  \mn@doi [ApJ] {10.1088/0004-637X/784/2/108}, \href
  {http://adsabs.harvard.edu/abs/2014ApJ...784..108B} {784, 108}

\bibitem[\protect\citeauthoryear{{Bordoloi} et~al.,}{{Bordoloi}
  et~al.}{2014b}]{Bordoloi2014b}
{Bordoloi} R.,  et~al., 2014b, \mn@doi [ApJ] {10.1088/0004-637X/794/2/130},
  \href {http://adsabs.harvard.edu/abs/2014ApJ...794..130B} {794, 130}

\bibitem[\protect\citeauthoryear{{Bordoloi} et~al.,}{{Bordoloi}
  et~al.}{2014c}]{Bordoloi2014c}
{Bordoloi} R.,  et~al., 2014c, ApJ, \href
  {http://adsabs.harvard.edu/abs/2014arXiv1406.0509B} {796, 136}

\bibitem[\protect\citeauthoryear{{Brammer} et~al.,}{{Brammer}
  et~al.}{2011}]{Brammer2011}
{Brammer} G.~B.,  et~al., 2011, \mn@doi [ApJ] {10.1088/0004-637X/739/1/24},
  \href {http://adsabs.harvard.edu/abs/2011ApJ...739...24B} {739, 24}

\bibitem[\protect\citeauthoryear{{Calzetti}, {Kinney}  \&
  {Storchi-Bergmann}}{{Calzetti} et~al.}{1994}]{1994ApJ...429..582C}
{Calzetti} D.,  {Kinney} A.~L.,   {Storchi-Bergmann} T.,  1994, \mn@doi [ApJ]
  {10.1086/174346}, \href {http://adsabs.harvard.edu/abs/1994ApJ...429..582C}
  {429, 582}

\bibitem[\protect\citeauthoryear{{Chabrier}}{{Chabrier}}{2003}]{Chabrier03}
{Chabrier} G.,  2003, \mn@doi [PASP] {10.1086/376392}, \href
  {http://adsabs.harvard.edu/abs/2003PASP..115..763C} {115, 763}

\bibitem[\protect\citeauthoryear{{Chen}, {Tremonti}, {Heckman}, {Kauffmann},
  {Weiner}, {Brinchmann}  \& {Wang}}{{Chen} et~al.}{2010}]{chen_outflow2010}
{Chen} Y.-M.,  {Tremonti} C.~A.,  {Heckman} T.~M.,  {Kauffmann} G.,  {Weiner}
  B.~J.,  {Brinchmann} J.,   {Wang} J.,  2010, \mn@doi [Astronomical Journal]
  {10.1088/0004-6256/140/2/445}, \href
  {http://adsabs.harvard.edu/abs/2010AJ....140..445C} {140, 445}

\bibitem[\protect\citeauthoryear{{Coil}, {Weiner}, {Holz}, {Cooper}, {Yan}  \&
  {Aird}}{{Coil} et~al.}{2011}]{Coil2011}
{Coil} A.~L.,  {Weiner} B.~J.,  {Holz} D.~E.,  {Cooper} M.~C.,  {Yan} R.,
  {Aird} J.,  2011, \mn@doi [ApJ] {10.1088/0004-637X/743/1/46}, \href
  {http://adsabs.harvard.edu/abs/2011ApJ...743...46C} {743, 46}

\bibitem[\protect\citeauthoryear{{Cooksey}, {Thom}, {Prochaska}  \&
  {Chen}}{{Cooksey} et~al.}{2010}]{Cooksey2010}
{Cooksey} K.~L.,  {Thom} C.,  {Prochaska} J.~X.,   {Chen} H.-W.,  2010, \mn@doi
  [ApJ] {10.1088/0004-637X/708/1/868}, \href
  {http://adsabs.harvard.edu/abs/2010ApJ...708..868C} {708, 868}

\bibitem[\protect\citeauthoryear{{Daddi} et~al.,}{{Daddi}
  et~al.}{2007}]{Daddi2007}
{Daddi} E.,  et~al., 2007, \mn@doi [ApJ] {10.1086/521818}, \href
  {http://adsabs.harvard.edu/abs/2007ApJ...670..156D} {670, 156}

\bibitem[\protect\citeauthoryear{{Dav{\'e}}, {Finlator}  \&
  {Oppenheimer}}{{Dav{\'e}} et~al.}{2012}]{Dave2012}
{Dav{\'e}} R.,  {Finlator} K.,   {Oppenheimer} B.~D.,  2012, \mn@doi [MNRAS]
  {10.1111/j.1365-2966.2011.20148.x}, \href
  {http://adsabs.harvard.edu/abs/2012MNRAS.421...98D} {421, 98}

\bibitem[\protect\citeauthoryear{{Erb}, {Quider}, {Henry}  \& {Martin}}{{Erb}
  et~al.}{2012}]{Erb2012}
{Erb} D.~K.,  {Quider} A.~M.,  {Henry} A.~L.,   {Martin} C.~L.,  2012, \mn@doi
  [ApJ] {10.1088/0004-637X/759/1/26}, \href
  {http://adsabs.harvard.edu/abs/2012ApJ...759...26E} {759, 26}

\bibitem[\protect\citeauthoryear{{Faber} et~al.,}{{Faber}
  et~al.}{2007}]{Faber2007}
{Faber} S.~M.,  et~al., 2007, \mn@doi [ApJ] {10.1086/519294}, \href
  {http://adsabs.harvard.edu/abs/2007ApJ...665..265F} {665, 265}

\bibitem[\protect\citeauthoryear{{F{\"o}rster Schreiber} et~al.,}{{F{\"o}rster
  Schreiber} et~al.}{2014}]{ForsterSchreiber2014}
{F{\"o}rster Schreiber} N.~M.,  et~al., 2014, \mn@doi [ApJ]
  {10.1088/0004-637X/787/1/38}, \href
  {http://adsabs.harvard.edu/abs/2014ApJ...787...38F} {787, 38}

\bibitem[\protect\citeauthoryear{{Gabor}, {Dav{\'e}}, {Oppenheimer}  \&
  {Finlator}}{{Gabor} et~al.}{2011}]{Gabor2011}
{Gabor} J.~M.,  {Dav{\'e}} R.,  {Oppenheimer} B.~D.,   {Finlator} K.,  2011,
  \mn@doi [MNRAS] {10.1111/j.1365-2966.2011.19430.x}, \href
  {http://adsabs.harvard.edu/abs/2011MNRAS.417.2676G} {417, 2676}

\bibitem[\protect\citeauthoryear{{Genzel} et~al.,}{{Genzel}
  et~al.}{2014}]{Genzel2014}
{Genzel} R.,  et~al., 2014, \mn@doi [ApJ] {10.1088/0004-637X/796/1/7}, \href
  {http://adsabs.harvard.edu/abs/2014ApJ...796....7G} {796, 7}

\bibitem[\protect\citeauthoryear{{Giavalisco} et~al.,}{{Giavalisco}
  et~al.}{2011}]{Giavalisco2011}
{Giavalisco} M.,  et~al., 2011, \mn@doi [ApJ] {10.1088/0004-637X/743/1/95},
  \href {http://adsabs.harvard.edu/abs/2011ApJ...743...95G} {743, 95}

\bibitem[\protect\citeauthoryear{{Gilbank}, {Gladders}, {Yee}  \&
  {Hsieh}}{{Gilbank} et~al.}{2011}]{Gilbank2011}
{Gilbank} D.~G.,  {Gladders} M.~D.,  {Yee} H.~K.~C.,   {Hsieh} B.~C.,  2011,
  \mn@doi [AJ] {10.1088/0004-6256/141/3/94}, \href
  {http://adsabs.harvard.edu/abs/2011AJ....141...94G} {141, 94}

\bibitem[\protect\citeauthoryear{Haario, Saksman  \& Tamminen}{Haario
  et~al.}{2001}]{haario2001}
Haario H.,  Saksman E.,   Tamminen J.,  2001, Bernoulli, 7, 223

\bibitem[\protect\citeauthoryear{{Heckman}}{{Heckman}}{2002}]{Heckman2002}
{Heckman} T.~M.,  2002, in {Mulchaey} J.~S.,  {Stocke} J.~T.,  eds,
  Astronomical Society of the Pacific Conference Series Vol. 254, Extragalactic
  Gas at Low Redshift. p.~292 (\mn@eprint {} {arXiv:astro-ph/0107438})

\bibitem[\protect\citeauthoryear{{Heckman}, {Lehnert}, {Strickland}  \&
  {Armus}}{{Heckman} et~al.}{2000}]{Heckman2000}
{Heckman} T.~M.,  {Lehnert} M.~D.,  {Strickland} D.~K.,   {Armus} L.,  2000,
  \mn@doi [ApJS] {10.1086/313421}, \href
  {http://adsabs.harvard.edu/abs/2000ApJS..129..493H} {129, 493}

\bibitem[\protect\citeauthoryear{{Heckman}, {Alexandroff}, {Borthakur},
  {Overzier}  \& {Leitherer}}{{Heckman} et~al.}{2015}]{Heckman2015}
{Heckman} T.~M.,  {Alexandroff} R.~M.,  {Borthakur} S.,  {Overzier} R.,
  {Leitherer} C.,  2015, preprint, \href
  {http://adsabs.harvard.edu/abs/2015arXiv150705622H} {} (\mn@eprint {arXiv}
  {1507.05622})

\bibitem[\protect\citeauthoryear{{Jenkins}}{{Jenkins}}{2009}]{Jenkins2009}
{Jenkins} E.~B.,  2009, \mn@doi [ApJ] {10.1088/0004-637X/700/2/1299}, \href
  {http://adsabs.harvard.edu/abs/2009ApJ...700.1299J} {700, 1299}

\bibitem[\protect\citeauthoryear{{Kassin} et~al.,}{{Kassin}
  et~al.}{2012}]{Kassin2012}
{Kassin} S.~A.,  et~al., 2012, \mn@doi [ApJ] {10.1088/0004-637X/758/2/106},
  \href {http://adsabs.harvard.edu/abs/2012ApJ...758..106K} {758, 106}

\bibitem[\protect\citeauthoryear{{Kennicutt}}{{Kennicutt}}{1998}]{Kennicutt1998}
{Kennicutt} Jr. R.~C.,  1998, \mn@doi [ARAA] {10.1146/annurev.astro.36.1.189},
  \href {http://adsabs.harvard.edu/abs/1998ARA%26A..36..189K} {36, 189}

\bibitem[\protect\citeauthoryear{{Kinney}, {Bohlin}, {Calzetti}, {Panagia}  \&
  {Wyse}}{{Kinney} et~al.}{1993}]{1993ApJS...86....5K}
{Kinney} A.~L.,  {Bohlin} R.~C.,  {Calzetti} D.,  {Panagia} N.,   {Wyse}
  R.~F.~G.,  1993, \mn@doi [ApJS] {10.1086/191771}, \href
  {http://adsabs.harvard.edu/abs/1993ApJS...86....5K} {86, 5}

\bibitem[\protect\citeauthoryear{{Klypin}, {Trujillo-Gomez}  \&
  {Primack}}{{Klypin} et~al.}{2011}]{Klypin2011}
{Klypin} A.~A.,  {Trujillo-Gomez} S.,   {Primack} J.,  2011, \mn@doi [ApJ]
  {10.1088/0004-637X/740/2/102}, \href
  {http://adsabs.harvard.edu/abs/2011ApJ...740..102K} {740, 102}

\bibitem[\protect\citeauthoryear{{Kornei}, {Shapley}, {Martin}, {Coil}, {Lotz},
  {Schiminovich}, {Bundy}  \& {Noeske}}{{Kornei} et~al.}{2012}]{Kornei2012}
{Kornei} K.~A.,  {Shapley} A.~E.,  {Martin} C.~L.,  {Coil} A.~L.,  {Lotz}
  J.~M.,  {Schiminovich} D.,  {Bundy} K.,   {Noeske} K.~G.,  2012, \mn@doi
  [ApJ] {10.1088/0004-637X/758/2/135}, \href
  {http://adsabs.harvard.edu/abs/2012ApJ...758..135K} {758, 135}

\bibitem[\protect\citeauthoryear{{Kornei}, {Shapley}, {Martin}, {Coil}, {Lotz}
  \& {Weiner}}{{Kornei} et~al.}{2013}]{Kornei2013}
{Kornei} K.~A.,  {Shapley} A.~E.,  {Martin} C.~L.,  {Coil} A.~L.,  {Lotz}
  J.~M.,   {Weiner} B.~J.,  2013, \mn@doi [ApJ] {10.1088/0004-637X/774/1/50},
  \href {http://adsabs.harvard.edu/abs/2013ApJ...774...50K} {774, 50}

\bibitem[\protect\citeauthoryear{{Kurucz}}{{Kurucz}}{2005}]{Kurucz2005}
{Kurucz} R.~L.,  2005, Memorie della Societa Astronomica Italiana Supplementi,
  \href {http://adsabs.harvard.edu/abs/2005MSAIS...8...14K} {8, 14}

\bibitem[\protect\citeauthoryear{{Lehnert} \& {Heckman}}{{Lehnert} \&
  {Heckman}}{1996}]{Lehnert1996}
{Lehnert} M.~D.,  {Heckman} T.~M.,  1996, \mn@doi [ApJ] {10.1086/177180}, \href
  {http://adsabs.harvard.edu/abs/1996ApJ...462..651L} {462, 651}

\bibitem[\protect\citeauthoryear{{Leitherer} et~al.,}{{Leitherer}
  et~al.}{1999}]{Starburst99}
{Leitherer} C.,  et~al., 1999, \mn@doi [ApJS] {10.1086/313233}, \href
  {http://adsabs.harvard.edu/abs/1999ApJS..123....3L} {123, 3}

\bibitem[\protect\citeauthoryear{{Leitherer}, {Tremonti}, {Heckman}  \&
  {Calzetti}}{{Leitherer} et~al.}{2011}]{Leitherer2011}
{Leitherer} C.,  {Tremonti} C.~A.,  {Heckman} T.~M.,   {Calzetti} D.,  2011,
  \mn@doi [AJ] {10.1088/0004-6256/141/2/37}, \href
  {http://adsabs.harvard.edu/abs/2011AJ....141...37L} {141, 37}

\bibitem[\protect\citeauthoryear{{Marshall} et~al.,}{{Marshall}
  et~al.}{2008}]{magespie}
{Marshall} J.~L.,  et~al., 2008, in Society of Photo-Optical Instrumentation
  Engineers (SPIE) Conference Series. p.~54 (\mn@eprint {arXiv} {0807.3774}),
  \mn@doi{10.1117/12.789972}

\bibitem[\protect\citeauthoryear{{Martin}}{{Martin}}{2005}]{Martin2005}
{Martin} C.~L.,  2005, \mn@doi [ApJ] {10.1086/427277}, \href
  {http://adsabs.harvard.edu/abs/2005ApJ...621..227M} {621, 227}

\bibitem[\protect\citeauthoryear{{Martin} \& {Bouch{\'e}}}{{Martin} \&
  {Bouch{\'e}}}{2009}]{Martin2009}
{Martin} C.~L.,  {Bouch{\'e}} N.,  2009, \mn@doi [ApJ]
  {10.1088/0004-637X/703/2/1394}, \href
  {http://adsabs.harvard.edu/abs/2009ApJ...703.1394M} {703, 1394}

\bibitem[\protect\citeauthoryear{{Martin}, {Shapley}, {Coil}, {Kornei},
  {Bundy}, {Weiner}, {Noeske}  \& {Schiminovich}}{{Martin}
  et~al.}{2012}]{Martin2012}
{Martin} C.~L.,  {Shapley} A.~E.,  {Coil} A.~L.,  {Kornei} K.~A.,  {Bundy} K.,
  {Weiner} B.~J.,  {Noeske} K.~G.,   {Schiminovich} D.,  2012, \mn@doi [ApJ]
  {10.1088/0004-637X/760/2/127}, \href
  {http://adsabs.harvard.edu/abs/2012ApJ...760..127M} {760, 127}

\bibitem[\protect\citeauthoryear{{Martin}, {Shapley}, {Coil}, {Kornei},
  {Murray}  \& {Pancoast}}{{Martin} et~al.}{2013}]{Martin2013}
{Martin} C.~L.,  {Shapley} A.~E.,  {Coil} A.~L.,  {Kornei} K.~A.,  {Murray} N.,
    {Pancoast} A.,  2013, \mn@doi [ApJ] {10.1088/0004-637X/770/1/41}, \href
  {http://adsabs.harvard.edu/abs/2013ApJ...770...41M} {770, 41}

\bibitem[\protect\citeauthoryear{{Morton}}{{Morton}}{2003}]{Morton2003}
{Morton} D.~C.,  2003, \mn@doi [ApJS] {10.1086/377639}, \href
  {http://adsabs.harvard.edu/abs/2003ApJS..149..205M} {149, 205}

\bibitem[\protect\citeauthoryear{{Murray}, {M{\'e}nard}  \&
  {Thompson}}{{Murray} et~al.}{2011}]{Murray2011}
{Murray} N.,  {M{\'e}nard} B.,   {Thompson} T.~A.,  2011, \mn@doi [ApJ]
  {10.1088/0004-637X/735/1/66}, \href
  {http://adsabs.harvard.edu/abs/2011ApJ...735...66M} {735, 66}

\bibitem[\protect\citeauthoryear{{Nestor}, {Turnshek}, {Rao}  \&
  {Quider}}{{Nestor} et~al.}{2007}]{Nestor2007}
{Nestor} D.~B.,  {Turnshek} D.~A.,  {Rao} S.~M.,   {Quider} A.~M.,  2007,
  \mn@doi [ApJ] {10.1086/511411}, \href
  {http://adsabs.harvard.edu/abs/2007ApJ...658..185N} {658, 185}

\bibitem[\protect\citeauthoryear{{Newman} et~al.,}{{Newman}
  et~al.}{2012}]{Newman2012}
{Newman} S.~F.,  et~al., 2012, \mn@doi [ApJ] {10.1088/0004-637X/761/1/43},
  \href {http://adsabs.harvard.edu/abs/2012ApJ...761...43N} {761, 43}

\bibitem[\protect\citeauthoryear{{Noeske} et~al.,}{{Noeske}
  et~al.}{2007}]{Noeske2007}
{Noeske} K.~G.,  et~al., 2007, \mn@doi [ApJL] {10.1086/517927}, \href
  {http://adsabs.harvard.edu/abs/2007ApJ...660L..47N} {660, L47}

\bibitem[\protect\citeauthoryear{{Patel} et~al.,}{{Patel}
  et~al.}{2013}]{Patel2013}
{Patel} S.~G.,  et~al., 2013, \mn@doi [ApJ] {10.1088/0004-637X/766/1/15}, \href
  {http://adsabs.harvard.edu/abs/2013ApJ...766...15P} {766, 15}

\bibitem[\protect\citeauthoryear{{Pettini}, {Shapley}, {Steidel}, {Cuby},
  {Dickinson}, {Moorwood}, {Adelberger}  \& {Giavalisco}}{{Pettini}
  et~al.}{2001}]{Pettini2001}
{Pettini} M.,  {Shapley} A.~E.,  {Steidel} C.~C.,  {Cuby} J.-G.,  {Dickinson}
  M.,  {Moorwood} A.~F.~M.,  {Adelberger} K.~L.,   {Giavalisco} M.,  2001,
  \mn@doi [ApJ] {10.1086/321403}, \href
  {http://adsabs.harvard.edu/abs/2001ApJ...554..981P} {554, 981}

\bibitem[\protect\citeauthoryear{{Prochaska}, {Kasen}  \& {Rubin}}{{Prochaska}
  et~al.}{2011}]{Prochaska2011}
{Prochaska} J.~X.,  {Kasen} D.,   {Rubin} K.,  2011, \mn@doi [ApJ]
  {10.1088/0004-637X/734/1/24}, \href
  {http://adsabs.harvard.edu/abs/2011ApJ...734...24P} {734, 24}

\bibitem[\protect\citeauthoryear{{Rigby}, {Wuyts}, {Gladders}, {Sharon}  \&
  {Becker}}{{Rigby} et~al.}{2011}]{Rigby11}
{Rigby} J.~R.,  {Wuyts} E.,  {Gladders} M.~D.,  {Sharon} K.,   {Becker} G.~D.,
  2011, \mn@doi [ApJ] {10.1088/0004-637X/732/1/59}, \href
  {http://adsabs.harvard.edu/abs/2011ApJ...732...59R} {732, 59}

\bibitem[\protect\citeauthoryear{{Rubin}, {Weiner}, {Koo}, {Martin},
  {Prochaska}, {Coil}  \& {Newman}}{{Rubin} et~al.}{2010}]{Rubin2010}
{Rubin} K.~H.~R.,  {Weiner} B.~J.,  {Koo} D.~C.,  {Martin} C.~L.,  {Prochaska}
  J.~X.,  {Coil} A.~L.,   {Newman} J.~A.,  2010, \mn@doi [ApJ]
  {10.1088/0004-637X/719/2/1503}, \href
  {http://adsabs.harvard.edu/abs/2010ApJ...719.1503R} {719, 1503}

\bibitem[\protect\citeauthoryear{{Rubin}, {Prochaska}, {M{\'e}nard}, {Murray},
  {Kasen}, {Koo}  \& {Phillips}}{{Rubin} et~al.}{2011a}]{Rubin2011a}
{Rubin} K.~H.~R.,  {Prochaska} J.~X.,  {M{\'e}nard} B.,  {Murray} N.,  {Kasen}
  D.,  {Koo} D.~C.,   {Phillips} A.~C.,  2011a, \mn@doi [ApJ]
  {10.1088/0004-637X/728/1/55}, \href
  {http://adsabs.harvard.edu/abs/2011ApJ...728...55R} {728, 55}

\bibitem[\protect\citeauthoryear{{Rubin}, {Prochaska}, {M{\'e}nard}, {Murray},
  {Kasen}, {Koo}  \& {Phillips}}{{Rubin} et~al.}{2011b}]{Rubin2011b}
{Rubin} K.~H.~R.,  {Prochaska} J.~X.,  {M{\'e}nard} B.,  {Murray} N.,  {Kasen}
  D.,  {Koo} D.~C.,   {Phillips} A.~C.,  2011b, \mn@doi [ApJ]
  {10.1088/0004-637X/728/1/55}, \href
  {http://adsabs.harvard.edu/abs/2011ApJ...728...55R} {728, 55}

\bibitem[\protect\citeauthoryear{{Rubin}, {Prochaska}, {Koo}  \&
  {Phillips}}{{Rubin} et~al.}{2012}]{Rubin2011}
{Rubin} K.~H.~R.,  {Prochaska} J.~X.,  {Koo} D.~C.,   {Phillips} A.~C.,  2012,
  \mn@doi [ApJL] {10.1088/2041-8205/747/2/L26}, \href
  {http://adsabs.harvard.edu/abs/2012ApJ...747L..26R} {747, L26}

\bibitem[\protect\citeauthoryear{{Rubin}, {Prochaska}, {Koo}, {Phillips},
  {Martin}  \& {Winstrom}}{{Rubin} et~al.}{2014}]{Rubin2014a}
{Rubin} K.~H.~R.,  {Prochaska} J.~X.,  {Koo} D.~C.,  {Phillips} A.~C.,
  {Martin} C.~L.,   {Winstrom} L.~O.,  2014, \mn@doi [ApJ]
  {10.1088/0004-637X/794/2/156}, \href
  {http://adsabs.harvard.edu/abs/2014ApJ...794..156R} {794, 156}

\bibitem[\protect\citeauthoryear{{Rupke}, {Veilleux}  \& {Sanders}}{{Rupke}
  et~al.}{2005}]{Rupke2005b}
{Rupke} D.~S.,  {Veilleux} S.,   {Sanders} D.~B.,  2005, \mn@doi [ApJS]
  {10.1086/432889}, \href {http://adsabs.harvard.edu/abs/2005ApJS..160..115R}
  {160, 115}

\bibitem[\protect\citeauthoryear{{Scarlata} \& {Panagia}}{{Scarlata} \&
  {Panagia}}{2015}]{Scarlata2015}
{Scarlata} C.,  {Panagia} N.,  2015, \mn@doi [ApJ]
  {10.1088/0004-637X/801/1/43}, \href
  {http://adsabs.harvard.edu/abs/2015ApJ...801...43S} {801, 43}

\bibitem[\protect\citeauthoryear{{Shapley}, {Steidel}, {Pettini}  \&
  {Adelberger}}{{Shapley} et~al.}{2003}]{Shapley2003}
{Shapley} A.~E.,  {Steidel} C.~C.,  {Pettini} M.,   {Adelberger} K.~L.,  2003,
  \mn@doi [ApJ] {10.1086/373922}, \href
  {http://adsabs.harvard.edu/abs/2003ApJ...588...65S} {588, 65}

\bibitem[\protect\citeauthoryear{{Sharon}, {Gladders}, {Rigby}, {Wuyts},
  {Koester}, {Bayliss}  \& {Barrientos}}{{Sharon} et~al.}{2012}]{Sharon2012}
{Sharon} K.,  {Gladders} M.~D.,  {Rigby} J.~R.,  {Wuyts} E.,  {Koester} B.~P.,
  {Bayliss} M.~B.,   {Barrientos} L.~F.,  2012, \mn@doi [ApJ]
  {10.1088/0004-637X/746/2/161}, \href
  {http://adsabs.harvard.edu/abs/2012ApJ...746..161S} {746, 161}

\bibitem[\protect\citeauthoryear{{Tremonti} et~al.,}{{Tremonti}
  et~al.}{2004}]{Tremonti2004}
{Tremonti} C.~A.,  et~al., 2004, \mn@doi [ApJ] {10.1086/423264}, \href
  {http://adsabs.harvard.edu/abs/2004ApJ...613..898T} {613, 898}

\bibitem[\protect\citeauthoryear{{Tumlinson} et~al.,}{{Tumlinson}
  et~al.}{2011}]{Tumlinson2011a}
{Tumlinson} J.,  et~al., 2011, \mn@doi [Science] {10.1126/science.1209840},
  \href {http://adsabs.harvard.edu/abs/2011Sci...334..948T} {334, 948}

\bibitem[\protect\citeauthoryear{{Veilleux}, {Cecil}  \&
  {Bland-Hawthorn}}{{Veilleux} et~al.}{2005}]{Veilleux2005}
{Veilleux} S.,  {Cecil} G.,   {Bland-Hawthorn} J.,  2005, \mn@doi [Annual
  Review of Astronomy and Astrophysics]
  {10.1146/annurev.astro.43.072103.150610}, \href
  {http://adsabs.harvard.edu/abs/2005ARA%26A..43..769V} {43, 769}

\bibitem[\protect\citeauthoryear{{Wark} \& {Mercer}}{{Wark} \&
  {Mercer}}{1965}]{WarkMercer}
{Wark} D.~Q.,  {Mercer} D.~M.,  1965, \mn@doi [AO] {10.1364/AO.4.000839}, \href
  {http://adsabs.harvard.edu/abs/1965ApOpt...4..839W} {4, 839}

\bibitem[\protect\citeauthoryear{{Weiner} et~al.,}{{Weiner}
  et~al.}{2009}]{Weiner2009}
{Weiner} B.~J.,  et~al., 2009, \mn@doi [ApJ] {10.1088/0004-637X/692/1/187},
  \href {http://adsabs.harvard.edu/abs/2009ApJ...692..187W} {692, 187}

\bibitem[\protect\citeauthoryear{{Werk} et~al.,}{{Werk}
  et~al.}{2014}]{Werk2014}
{Werk} J.~K.,  et~al., 2014, \mn@doi [ApJ] {10.1088/0004-637X/792/1/8}, \href
  {http://adsabs.harvard.edu/abs/2014ApJ...792....8W} {792, 8}

\bibitem[\protect\citeauthoryear{{Whitaker}, {Rigby}, {Brammer}, {Gladders},
  {Sharon}, {Teng}  \& {Wuyts}}{{Whitaker} et~al.}{2014}]{Whitaker2014}
{Whitaker} K.~E.,  {Rigby} J.~R.,  {Brammer} G.~B.,  {Gladders} M.~D.,
  {Sharon} K.,  {Teng} S.~H.,   {Wuyts} E.,  2014, \mn@doi [ApJ]
  {10.1088/0004-637X/790/2/143}, \href
  {http://adsabs.harvard.edu/abs/2014ApJ...790..143W} {790, 143}

\bibitem[\protect\citeauthoryear{{Wu}, {Boggess}  \& {Gull}}{{Wu}
  et~al.}{1983}]{1983ApJ...266...28W}
{Wu} C.-C.,  {Boggess} A.,   {Gull} T.~R.,  1983, \mn@doi [ApJ]
  {10.1086/160756}, \href {http://adsabs.harvard.edu/abs/1983ApJ...266...28W}
  {266, 28}

\bibitem[\protect\citeauthoryear{{Wuyts} et~al.,}{{Wuyts}
  et~al.}{2010}]{Wuyts2010}
{Wuyts} E.,  et~al., 2010, \mn@doi [ApJ] {10.1088/0004-637X/724/2/1182}, \href
  {http://adsabs.harvard.edu/abs/2010ApJ...724.1182W} {724, 1182}

\bibitem[\protect\citeauthoryear{{Wuyts}, {Rigby}, {Gladders}  \&
  {Sharon}}{{Wuyts} et~al.}{2014}]{Wuyts2014}
{Wuyts} E.,  {Rigby} J.~R.,  {Gladders} M.~D.,   {Sharon} K.,  2014, \mn@doi
  [ApJ] {10.1088/0004-637X/781/2/61}, \href
  {http://adsabs.harvard.edu/abs/2014ApJ...781...61W} {781, 61}

\bibitem[\protect\citeauthoryear{{Zhu} \& {M{\'e}nard}}{{Zhu} \&
  {M{\'e}nard}}{2013}]{zhu2013a}
{Zhu} G.,  {M{\'e}nard} B.,  2013, \mn@doi [ApJ] {10.1088/0004-637X/770/2/130},
  \href {http://adsabs.harvard.edu/abs/2013ApJ...770..130Z} {770, 130}

\makeatother
\end{thebibliography}

   \label{lastpage}

\end{document}